\documentclass[12pt]{article}
\usepackage{graphicx}
\usepackage{amsmath,amssymb,amsfonts}
\usepackage{color}
\usepackage{amsmath}
\usepackage{amsfonts}
\usepackage{amssymb}
\usepackage{float}
\usepackage{cancel}
\usepackage{caption}
\usepackage{slashed}            
\usepackage{subfig}             
\usepackage{bbm}                
\usepackage{collref}
\usepackage{setspace}

\usepackage[hypertexnames=false]{hyperref}

\numberwithin{equation}{section} 
\newcommand{\meg}{\mu\to e \gamma}

\newcommand{\co}{\text{c}}
\newcommand{\si}{\text{s}}
\newcommand{\dbar}{d\mkern-6mu\mathchar'26}
\newcommand{\tF}[5]{\tilde{F}^{#2 #4 #5}_{#1,#3}}
\newcommand{\tDF}[5]{\tilde{D}_{#1} \tilde{F}^{#2 #4 #5}_{#1,#3}}

\usepackage{tikz}
\usetikzlibrary{arrows,shapes}
\usetikzlibrary{trees}
\usetikzlibrary{matrix,arrows} 				
\usetikzlibrary{positioning}				
\usetikzlibrary{calc,through}				
\usetikzlibrary{decorations.pathreplacing}  
\usetikzlibrary{backgrounds}  				
\usepackage{pgffor}							

\usetikzlibrary{decorations.pathmorphing}	
\usetikzlibrary{decorations.markings}
\usetikzlibrary{snakes}
\tikzset{
    vector/.style={decorate, decoration={snake}, draw},
	provector/.style={decorate, decoration={snake,amplitude=2.5pt}, draw},
	antivector/.style={decorate, decoration={snake,amplitude=-2.5pt}, draw},
    fermion/.style={draw=black, postaction={decorate},
        decoration={markings,mark=at position .55 with {\arrow[draw=black]{>}}}},
    fermionbar/.style={draw=black, postaction={decorate},
        decoration={markings,mark=at position .55 with {\arrow[draw=black]{<}}}},
    fermionnoarrow/.style={draw=black},
    gluon/.style={decorate, draw=black,
        decoration={coil,amplitude=4pt, segment length=5pt}},
    scalar/.style={dashed,draw=black, postaction={decorate},
        decoration={markings,mark=at position .55 with {\arrow[draw=black]{>}}}},
    scalarbar/.style={dashed,draw=black, postaction={decorate},
        decoration={markings,mark=at position .55 with {\arrow[draw=black]{<}}}},
    scalarnoarrow/.style={dash pattern = on 6 pt off 3 pt,draw=black},
    electron/.style={draw=black, postaction={decorate},
        decoration={markings,mark=at position .55 with {\arrow[draw=black]{>}}}},
	bigvector/.style={decorate, decoration={snake,amplitude=4pt}, draw},
	vectorscalar/.style={loosely dotted,draw=black, postaction={decorate}},
}

\begin{document}
\begin{titlepage}
\begin{flushright}
\end{flushright}

\begin{center}
{\huge\bf Warped Penguins}\\

\medskip
\bigskip\color{black}\vspace{0.6cm}
{
{\large\bf Csaba Cs\'aki, Yuval Grossman,\\Philip Tanedo, and Yuhsin Tsai}
}
\\[7mm]
{\it Institute for High Energy Phenomenology,\\ Newman Laboratory of Elementary Particle Physics,\\
Cornell University, Ithaca, NY 14853, USA}
\\
\vspace*{0.3cm}
{\it E-mail: \rm{\href{mailto:csaki@cornell.edu}{csaki@cornell.edu}, \href{mailto:yg73@cornell.edu}{yg73@cornell.edu}, \href{mailto:pt267@cornell.edu}{pt267@cornell.edu}, \href{mailto:yt237@cornell.edu}{yt237@cornell.edu}}}
\bigskip\bigskip\bigskip

{
\centerline{\large\bf Abstract}
\begin{quote}

We present an analysis of the loop-induced magnetic dipole operator in the Randall-Sundrum model of a warped extra dimension with anarchic bulk fermions and an IR brane-localized Higgs. 
These operators are finite at one-loop order and we explicitly calculate the branching ratio for $\mu\to e\gamma$ using the mixed position/momentum space formalism.
The particular bound on the anarchic Yukawa and Kaluza-Klein (KK) scales can depend on the flavor structure of the anarchic matrices. This effect encapsulates the misalignment between the bulk mass parameters and the Yukawa matrices in flavor space. We quantify how these models realize this misalignment.
We also review tree-level lepton flavor bounds in these models and show that these are are in mild tension with the $\mu\to e\gamma$ bounds from typical models with a 3 TeV Kaluza-Klein scale.
Further, we illuminate the nature of the one-loop finiteness of these diagrams and show how to accurately determine the degree of divergence of a five-dimensional loop diagram using both the five-dimensional and KK formalism. This power counting can be obfuscated in the four-dimensional Kaluza-Klein formalism and we explicitly point out subtleties that ensure that the two formalisms agree. 
Finally, we remark on the existence of a perturbative regime in which these one-loop results give the dominant contribution.

\end{quote}}

\end{center}

\end{titlepage}

\begin{spacing}{.775}
\tableofcontents
\end{spacing}

\newpage

\section{Introduction}
\label{sec:intro}

The Randall-Sundrum (RS) set up for a warped extra dimension is a novel framework for models of electroweak symmetry breaking \cite{Randall:1999ee}.  When fermion and gauge fields are allowed to propagate in the bulk, these models can also explain the fermion mass spectrum through the split fermion proposal \cite{split,GN,GherghettaPomarol}. In these anarchic flavor models each element of the Yukawa matrices can take natural ${\cal O}(1)$ values because the hierarchy of the fermion masses is generated by the exponential localization of the fermion wave functions away from the Higgs field~\cite{APS1,Agashe:2004cp}.

The same small wavefunction overlap that yields the fermion mass spectrum also gives hierarchical mixing angles~\cite{APS1,Huber,Kitano:ij,Moreau:fv} and suppresses tree-level flavor-changing neutral currents (FCNCs) by the RS-GIM mechanism \cite{APS1,Agashe:2004cp}.
This built-in protection, however, may not always be sufficient to completely protect against the most dangerous types of experimental FCNC constraints. In the quark sector, for example, the exchange of Kaluza-Klein (KK) gluons induces left-right operators that contribute to CP violation in kaons and result in generic bounds of ${\cal O}(10-20$ TeV) for the KK gluon mass~\cite{CFW1,Buras,Neubert, Buras:2009ka, Albrecht:2009xr, Blanke:2008yr}. 
To reduce this bound one must either introduce additional structure (such as horizontal symmetries~\cite{Jose,CFW2} or flavor alignment~\cite{5DMFV,CPSW}) or alternately gain several ${\cal O}(1)$ factors~\cite{Kaustubh} by promoting the Higgs to a bulk field, inducing loop-level QCD matching, etc. This latter approach is limited by tension with loop-induced flavor-violating effects \cite{Gedalia:2009ws}. 

The leptonic sector of the anarchic model is similarly bounded by FCNCs. Agashe, Blechman and Petriello recently studied the two dominant constraints in the lepton sector: the loop-induced $\mu\to e\gamma$ photon penguin from Higgs exchange and the tree-level contribution to $\mu\to 3e$ and $\mu \to e$ conversion from the exchange of the $Z$ boson KK tower ~\cite{Agashe:2006iy}.
These processes set complementary bounds due to their complementary dependence on the overall magnitude of the anarchic Yukawa coupling, $Y_*$. While $\mu \to e\gamma$ is proportional to $Y_*^3$ due to two Yukawa couplings and a chirality-flipping mass insertion, the dominant contribution to $\mu \to 3e$ and $\mu\to e$ conversion comes from the nonuniversality of the $Z$ boson near the IR brane.
In order to maintain the observed mass spectrum, increasing the Yukawa coupling pushes the bulk fermion profiles away from the IR brane and hence away from the flavor-changing part of the $Z$. This reduces the effective four-dimensional (4D) FCNC coupling so that these processes are proportional to $Y_*^{-1}$. For a given KK gauge boson mass, these processes then set an upper and lower bound on the Yukawa coupling which are usually mutually exclusive.

A key feature of the lepton sector is that one expects large mixing angles rather than the hierarchical angles in the Cabbibo-Kobayashi-Maskawa (CKM) matrix. One way to obtain this is by using a global flavor symmetry for the lepton sector~\cite{A4} (see also~\cite{Perez:2008ee, Chen:2008qg}). 
Including these additional global symmetries can relax the tension between the two bounds. For example, imposing an A$_4$ symmetry on the leptonic sector completely removes the tree-level constraints~\cite{A4}. Another interesting possibility for obtaining large lepton mixing angles is to have the wavefunction overlap for the neutrino Yukawa peak near the UV brane~\cite{Raman}. For generic models with anarchic fermions, however, \cite{Agashe:2006iy} found that the tension between $\mu\to e\gamma$ and tree-level processes ($\mu \to 3e$ and $\mu \to e$ conversion) push the gauge boson KK scale to be on the order of 5--10 TeV. 

The main goal of this paper is to present a detailed one-loop calculation of the $\mu\to e\gamma$ penguin in the RS model with a brane-localized Higgs and to show that this amplitude is finite.

\begin{figure}[ht]
    \centering
        \includegraphics{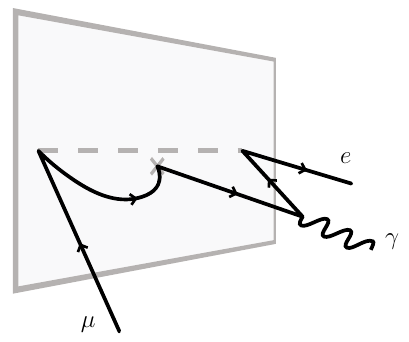}
    \caption{A contribution to $\meg$ from a brane-localized Higgs. The dashed line represents the Higgs while the cross represents a Yukawa coupling with a Higgs vev.}
    \label{fig:5Ddiagram}
\end{figure}

To perform the calculation and obtain a numerical result we choose to work in the five-dimensional (5D) mixed position/momentum space formalism \cite{Puchwein:2003jq, Carena:2004zn}. This setup is natural for calculating processes on an interval with brane-localized terms, as shown in Fig.~\ref{fig:5Ddiagram}.  In particular, there are no sums over KK modes, the chiral boundary conditions are fully incorporated in the 5D propagators, and the UV behavior is clear upon Wick rotation where the basis of Bessel functions becomes exponentials in the 4D loop momentum. The physical result is, of course, independent of whether the calculation was done in 5D or in 4D via a KK decomposition. We show explicit one-loop finiteness in the KK decomposed theory and remark upon the importance of taking into account the correct number of KK modes relative to the momentum cutoff when calculating finite 5D loops.

The paper is organized as follows: 
We begin in Sections~\ref{sec:anarchy} and \ref{sec:TreeOps} by reviewing the flavor structure of anarchic Randall-Sundrum models and summarizing tree-level constraints on the anarchic Yukawa scale.
We then proceed the analysis of $\mu\to e \gamma$.
The dipole operators involved in this process are discussed in Section~\ref{sec:OperatorAnalysis} and the relevant coefficient is calculated using 5D methods in Section~\ref{sec:warpedXD}.
In Section~\ref{sec:heuristic} we discuss the origin of finiteness in these operators in both the 5D and 4D frameworks. We remark on subtleties in counting the superficial degree of divergence, the matching of the number of KK modes with any effective 4D momentum cutoff, and remark on the expected two-loop degree of divergence.
We conclude with an outlook for further directions in Section~\ref{sec:conclusion}.
In Appendix~\ref{app:5D:EFT} we highlight the matching of local 4D effective operators to \textit{nonlocal} 5D amplitudes. Next in Appendices~\ref{app:estimates} and \ref{app:analytic:expressions} we give estimates for the size of each diagram and analytic expressions for the (next-to)leading $\mu\to e\gamma$ diagrams. 
Appendices~\ref{sec:5D:momentum:space},~\ref{app:bulk:Feynman:rules}, and~\ref{app:propagator:derivation} focus on the formalism of quantum field theory in mixed position/momentum space, respectively focusing on a discussion of power counting, a summary of RS Feynman rules, and details on the derivation of the bulk fermion propagators.
Finally, in Appendix~\ref{app:Finiteness} we explicitly demonstrate a subtle cancellation in the single-mass insertion neutral Higgs diagram that is referenced in Section~\ref{sec:heuristic}.

\section{Review of anarchic Randall-Sundrum models}
\label{sec:anarchy}

We now summarize the main results for anarchic RS models. For a review see, e.g. Refs~\cite{Csaki:2005vy}. We consider a 5D warped interval $z\in[R,R']$ with a UV~brane at $z=R$ and an IR~brane at $z=R'$. The metric is
\begin{align}\label{eq:metric}
    ds^ 2&=\left(\frac{R}{z}\right)^2 (dx_\mu dx_\nu \eta^{\mu\nu} -dz^2),
\end{align}
where we see that $R$ is also the AdS curvature scale so that $R/R' \sim \text{TeV}/M_{\text{Pl}}$. These conformal coordinates are natural in the context of the AdS/CFT correspondence but differ from the classical RS conventions $z = R\exp(ky)$ and $k=1/R$.
The relevant scales have magnitudes $R^{-1}\sim M_{\text{Pl}}$ and $R'^{-1} \sim$ TeV.
Fermions are bulk Dirac fields which propagate in the full 5D space and can be decomposed into left- and right-handed Weyl spinors $\chi$ and $\bar\psi$ via
\begin{align}
    \Psi(x,z) =
        \begin{pmatrix}
        \chi(x,z) \\ \bar\psi(x,z)
        \end{pmatrix}.
\end{align}
In order to obtain a chiral zero mode spectrum, these fields are subject to the chiral (orbifold) boundary conditions
\begin{align}
    \psi_L(x^\mu,R)=\psi_L(x^\mu,R') = 0 \quad\quad\quad\quad\quad\quad \chi_R(x^\mu,R) =\chi_R(x^\mu,R') = 0,
\end{align}
where the subscripts $L$ and $R$ denote the $SU(2)_L$ doublet ($L$) and singlet ($R$) representations, i.e.\ the chirality of the zero mode.
The fermion bulk masses are given by $c/R$ where $c$ is a dimensionless parameter controlling the localization of the normalized 5D zero mode profiles,
\begin{align}
    \chi^{(0)}_{c} (x,z) &=\frac{1}{\sqrt{R'}} \left(\frac{z}{R}\right)^2 \left(\frac{z}{R'}\right)^{-c} f_c \;\chi^{(0)}_c(x)
    \quad\quad\quad \text{and }\quad\quad\quad  \psi_c^{(0)} (x,z) = \chi_{-c}^{(0)} (x,z)    ,\label{eq:fermion:profile}
\end{align}
where we have defined the usual RS flavor function
\begin{align}
f_c &=\sqrt{\frac{1-2c}{1-(R/R')^{1-2c}}}.\label{eq:flavor:function}
\end{align}

We assume that the Higgs is localized on the IR brane.
The Yukawa coupling is
\begin{align}
 S_{\text{Yuk}} = \int d^4 x\; \left(\frac{R}{R'}\right)^4  \bar{E}_i\left(R\,Y_{ij}\right) L_j\cdot H  + \text{h.c.}
\end{align}
where $Y_{ij}$ is a dimensionless 3$\times$3  matrix such that $(Y_5)_{ij}=RY_{ij}$ is the dimensionful parameter appearing in the 5D Lagrangian. In the anarchic approach $Y$ is assumed to be a random matrix with average elements of order $Y_*$. After including all warp factors and rescaling to canonical fields the effective 4D Yukawa and mass matrices for the zero modes are 
\begin{align}
y^{\text{SM}}_{ij} &= f_{c_{L_i}} Y_{ij} f_{-c_{R_j}} \quad\quad\quad\quad\quad\quad m_{ij} = \frac{v}{\sqrt{2}}y^{\text{SM}}_{ij},
\label{eq:RS:anarchy:zero:mode:Yukawa}
\end{align}
so that the fermion mass hierarchy is set by the $f_1 \ll f_2 \ll f_3$ structure for both left- and right-handed zero modes. In other words, the choice of $c$ for each fermion family introduces additional flavor structure into the theory which generates the zero mode spectrum while allowing the fundamental Yukawa parameters to be anarchic.

In the Standard Model the diagonalization of the fermion masses transmits the flavor structure of the Yukawa sector to the kinetic terms via the CKM matrix where it is manifested in the flavor-changing charged current through the $W^\pm$ boson. We shall use the analogous mass basis in Section~\ref{sec:TreeOps} for our calculation of the Yukawa constraints from $\mu\to 3e$ and $\mu\to e$ conversion operators. The key point is that in the gauge basis the interaction of the neutral gauge bosons is flavor diagonal but not flavor universal. The different fermion wave functions cause the overlap integrals to depend on the bulk mass parameters. Once we rotate into the mass eigenbasis we obtain flavor changing couplings for the neutral KK gauge bosons.

In the lepton sector this does not occur for the zero mode photon since its wavefunction remains flat after electroweak symmetry breaking and hence $\mu\to e \gamma$ remains a loop-level process. Thus for the primary analysis of this paper we choose a basis where the 5D fields are diagonal with respect to the bulk masses while the Yukawas are completely general. In this basis all of the relevant flavor-changing effects occur due to the Yukawa structure of the theory with no contributions from $W$ loops.
In the Standard Model, this corresponds to the basis before diagonalizing the fermion masses so that all flavor-changing effects occur through off-diagonal elements in the Yukawa matrix manifested as mass insertions or Higgs interactions. This basis is particularly helpful in the 5D mixed position/momentum space framework since the Higgs is attached to the IR brane, which simplifies loop integrals.

\section{Tree-level constraints from $\mu\to 3e$ and $\mu \to e$ conversion}
\label{sec:TreeOps}

For a fixed KK gauge boson mass $M_{\text{KK}}$, limits on $\mu \to 3e$ and $\mu \to e$ conversion in nuclei provide the strongest \textit{lower} bounds on the anarchic Yukawa scale $Y_*$. These tree-level processes are parameterized by Fermi operators generated by $Z$ and $Z'$ exchange, where the prime indicates the KK mode in the mass basis.
The effective Lagrangian for these lepton flavor-violating Fermi operators are traditionally parameterized as \cite{Chang:2005ag}
\begin{eqnarray}
{\cal L}&=& \frac{4 G_F}{\sqrt{2}} \left[ g_3 (\bar{e}_R \gamma^\mu \mu_R)(\bar{e}_R \gamma_\mu e_R) +g_4 (\bar{e}_L \gamma^\mu \mu_L)(\bar{e}_L \gamma^\mu e_L)+g_5 (\bar{e}_R \gamma^\mu \mu_R)(\bar{e}_L\gamma_\mu e_L) \right.
\nonumber \\ && \left.+g_6 (\bar{e}_L \gamma^\mu \mu_L)(\bar{e}_R \gamma_\mu e_R) \right]
 +\frac{G_F}{\sqrt{2}} \bar{e} \gamma^\mu (v-a \gamma_5) \mu \sum_q \bar{q} \gamma_\mu (v^q-a^q \gamma_5) q,
 \label{eq:effLag}
\end{eqnarray}
where we have only introduced the terms that are non-vanishing in the RS set up, and use the normalization where $v^q=T_3^q-2 Q^q \sin^2\theta$. The axial coupling to quarks, $a^q$, vanishes in the dominant contribution coming from coherent scattering off the nucleus. The $g_{3,4,5,6}$ are responsible for $\mu\to 3e$ decay, while the $v,a$ are responsible for $\mu\to e$ conversion in nuclei. The rates are given by (with the conversion rate normalized to the muon capture rate):
\begin{align}
\text{Br}(\mu\to 3e)&= 2 (g_3^2+ g_4^2)+g_5^2+g_6^2\ ,  \label{eq:Br:mu3e}\\
\text{Br}(\mu\to e)&= \frac{ p_e E_e G_F^2 F_p^2 m_\mu^3 \alpha^3 Z_{eff}^4}{\pi^2 Z \Gamma_{\text{capt}}} Q_N^2  (v^2+a^2),\label{eq:Br:mue}
\end{align}
where the parameters for the conversion depend on the nucleus and are calculated in the Feinberg-Weinberg approximation \cite{Feinberg:1959ui} and we write the charge for a nucleus with atomic number $Z$ and neutron number $N$ as
\begin{align}
    Q_N = v^u (2Z+N)+v^d(2N+Z).
\end{align}.
The most sensitive experimental constraint comes from muon conversion in ${}_{22}^{48}\text{Ti}$, for which
\begin{align}
    E_e\sim p_e \sim m_\mu,
    \quad\quad\quad\quad
    F_p \sim 0.55,
    \quad\quad\quad\quad
    Z_{\text{eff}}\sim 17.61,
    \quad\quad\quad\quad
    \Gamma_{\text{capt}} \sim 2.6 \cdot \frac{10^6}{\text{s}}.
\end{align}
We now consider these constraints for a minimal model (where $f_{e_L}=f_{e_R}$, $f_{\mu_L}=f_{\mu_R}$) and for a model with custodial protection.

\subsection{Minimal RS model}

In order to calculate the coefficients in the effective Lagrangian (\ref{eq:effLag}), we need to estimate the flavor-violating couplings of the neutral gauge bosons in the theory. In the basis of physical KK states all lepton flavor-violating couplings are the consequence of the non-uniformity of the gauge boson wave functions. Let us first consider the effect of the ordinary $Z$ boson, whose wave function is approximately (we use the approximation (2.19) of \cite{Csaki:2002gy} with a prefactor for canonical normalization)
\begin{align}
h^{(0)}(z) =\frac{1}{\sqrt{R\log \frac{R'}{R}}} \left[1+ \frac{M_Z^2}{4}z^2\left(1-2 \log \frac{z}{R}\right)\right].
\end{align}
The coupling of the $Z$ to fermions can be calculated by performing the overlap integral with the fermion profiles in (\ref{eq:fermion:profile}) and is found to be
\begin{align}
g^{Zff}=g^{Z}_{\text{SM}}\left(1+ \frac{(M_Z R')^2 \log \frac{R'}{R}}{2(3-2c)} f_c^2\right).
\end{align}
After rotating the fields to the mass eigenbasis we find that the off-diagonal coupling of the $Z$ boson to charged leptons is given by the nonuniversal term and is approximately
\begin{align}
g_{L,R}^{Ze\mu} \approx \left(g^{Z}_{\text{SM}}\right)^{L,R} \Delta^{(0)}_{e\mu} \equiv \left(g_{\text{SM}}^{Z}\right)^{L,R} \frac{(M_Z R')^2 \log \frac{R'}{R}}{2(3-2c)} f_{e_{L,R}}f_{\mu_{L,R}}.\label{eq:Zcoupl:LR}
\end{align}

Using these couplings one can estimate the coefficients of the 4-Fermi operators in (\ref{eq:effLag}),
\begin{align}
    g_{3,4} = 2g_{L,R}^2\Delta^{(0)}_{e\mu}\quad\quad\quad\quad\quad\quad
    g_{5,6} = 2g_L g_R \Delta^{(0)}_{e\mu}\quad\quad\quad\quad\quad\quad
    (v\pm a) = 2 g_{L,R} \Delta^{(0)}_{e\mu},
    \label{eq:tree:Z:couplings}
\end{align}
where the $g_{L,R}$ are proportional to the left- and right-handed charged lepton couplings to the $Z$ in the Standard Model, $g_L =-\frac{1}{2}+s^2_W$ and $g_R= s^2_W$.
The $Z'$ exchange contribution to $\mu\to 3e$ ($\mu\to e$) is a 15\% (5\%) correction and the $\gamma'$ exchange diagram is an additional 5\% (1\%) correction; we shall ignore both here.
We make the simplifying assumption that $f_{e_L}=f_{e_R}$ and $f_{\mu_L}=f_{\mu_R}$ and then express these in terms of the Standard Model Yukawa couplings as  $f= \sqrt{\lambda/{Y_*}}$. The expressions for the lepton flavor-violating processes are then
\begin{align}
\text{Br}(\mu\to 3 e) &= \phantom{1\cdot} 10^{-13} \left( \frac{3\ {\rm TeV}}{M_{\text{KK}}}\right)^4 \left(\frac{2}{Y_*}\right)^2  \\
\text{Br}(\mu\to e)_{\text{Ti}} &=   2 \cdot 10^{-12} \left( \frac{3\ {\rm TeV}}{M_{\text{KK}}}\right)^4 \left(\frac{2}{Y_*}\right)^2\ .
\end{align}

The current experimental bounds are $\text{Br}(\mu \to 3e)< 10^{-12}$ \cite{Bellgardt:1987du} and $\text{Br}(\mu\to e)_{\text{Ti}}< 6.1 \cdot 10^{-13}$ \cite{Wintz:1996va} so that $\mu\to e$ conversion provides the most stringent constraint,
\begin{equation}
\left( \frac{3\ {\rm TeV}}{M_{\text{KK}}}\right)^2 \left(\frac{2}{Y_*}\right) < 0.5. \label{eq:upper:bound}
\end{equation}
For a 3 TeV $Z'$, the anarchic Yukawa scale must satisfy $Y_*\gtrsim 3.7$, which agrees with \cite{Agashe:2006iy}.

\subsection{Custodially protected model}
\label{sec:tree:custodial}

Since the bound in (\ref{eq:upper:bound}) is model dependent, one might consider weakening this constraint by having the leptons transform under the custodial group
\begin{align}
    \text{SU(2)}_L\times\text{SU(2)}_R\times\text{U(1)}_X\times \text{P}_{LR},
\end{align}
where P$_{LR}$ is a discrete $L\leftrightarrow R$ exchange symmetry. Such a custodial protection was introduced in \cite{Agashe:2006at} to eliminate large corrections to the $Zb\bar{b}$ vertex in the quark sector. It was later found that this symmetry also eliminates some of the FCNCs in the $Z$ sector \cite{Albrecht:2009xr} so that one might also expect it to alleviate the lepton flavor violation bounds.
We shall now estimate the extent to which custodial symmetry can relax the bound on $Y_*$.
Further discussion including neutrino mixing can be found in \cite{Agashe:2009hc}. 

To custodially protect the charged leptons one choses the
$(L,R)_X$ representation
$(\mathbf{2},\mathbf{2})_0$ for the left-handed leptons, $(\mathbf{3},\mathbf{1})_0\oplus(\mathbf{1},\mathbf{3})_0$ for the charged right-handed leptons, and $(\mathbf{1},\mathbf{1})_0$ for the right-handed neutrinos. There are two neutral zero mode gauge bosons, the Standard Model $Z$ and $\gamma$, and three neutral KK excitations, $\gamma', Z'$ and $Z_H$, where the latter two are linear combinations of the $Z$ and $Z_X$ boson modes.
The coupling of the left handed leptons to the ordinary $Z$ and the $Z'$ are protected since those couplings are exactly flavor universal in the limit where P$_{LR}$ is exact. The breaking of P$_{LR}$ on the UV brane leads to small residual contributions which we neglect. The remaining flavor-violating couplings for the left-handed leptons come from the exchange of $Z_H$ and the $\gamma'$, while the right-handed leptons are unprotected.

Since $(v-a)$ couples to right-handed leptons its coupling is unprotected and is the same as in (\ref{eq:tree:Z:couplings}). For $(v+a)$, on the other hand, the leading-order effect comes from the $Z^{(1)}$ component of the $Z_H$, whose composition in terms of gauge KK states is~\cite{Albrecht:2009xr}
\begin{equation}
Z_H=\cos \xi Z^{(1)}+ \sin\xi Z_X^{(1)} +\beta Z^{(0)},
\end{equation}
where $Z^{(0)}$ is the flat zero mode $Z$-boson which does not contribute to FCNCs, $\cos \xi \approx \sqrt{\frac{1}{2}-s_W^2}/c_W$, and $\beta$ is a small correction of order $\mathcal O(v^2/M_{\text{KK}}^2)$.
 The flavor-changing coupling of the KK gauge bosons is analogous to that of KK gluons in \cite{CFW1},
\begin{align}
g^{Z^{(1)}e\mu}_{L,R} \approx \left(g^{Z}_{\text{SM}}\right)^{L,R} \Delta^{L,R(1)}_{e\mu} \equiv \left(g_{SM}^{Z}\right)^{L,R}\sqrt{ \log \frac{R'}{R}}\,\gamma_c\, f_{e_{L,R}}f_{\mu_{L,R}},
\label{eq:Zpcoupl}
\end{align}
where
\begin{align}
    \gamma_c = \frac{\sqrt{2}}{J_1(x_1)}\int_0^1 dx\; x^{1-2c}J_1(x_1\, x) \approx \frac{\sqrt{2}}{J_1(x_1)} \frac{0.7 x_1}{2(3-2c)}
\end{align}
and $x_1 = M_{\text{KK}}R'$ is the first zero of $J_0(x)$.
The analogous $\gamma^{(1)}$ coupling is given by $g^Z_\text{SM} \to e$. Taking into account the $Z_H$ and $\gamma^{(1)}$, the $(v+a)$ effective coupling to left-handed leptons is
\begin{align}
(v + a)= 2 g_L\, g_\text{KK}\, \frac{M_Z^2}{M_{\text{KK}}^2} \left(\cos^2\xi + \frac{Q_N^{Z_X}}{Q_N} \cos\xi\sin\xi\right) \Delta^{L(1)}_{e\mu} + 2 s_W^2 c_W^2 \, g_{\text{KK}}\, \frac{M_Z^2}{M_{\text{KK}}^2} \frac{Q_N^\gamma}{Q_N}\Delta^{L(1)}_{e\mu}.
\end{align}
The $\cos\xi\sin\xi$ term in the parenthesis represents the $Z_{X}^{(1)}$ component of the $Z_H$ which couples to the quarks in the nucleus via
\begin{align}
    Q_N^{Z_X} = -\frac{1}{\sqrt{2}}c_W\cos\xi\left(5Z+7N\right) -  \frac{2\sqrt{2}}{\cos \xi} s_W \frac{g'}{g} (Z+N),
    \quad\quad\quad\quad\quad
    g_{\text{KK}} = \frac{1}{\sqrt{\log R'/R}}. \label{eq:QNZX}
\end{align}
The $g_{\text{KK}}$ factor gives the universal (flavor-conserving) coupling of KK gauge bosons to zero mode fermions. $Q^\gamma_N$ is the electric charge of the nucleus normalized according to (\ref{eq:Br:mue}), $Q^\gamma_N = 2Z$.

Minimizing over the flavor factors $f_{e_{L,R}}$ and $f_{\mu_{L,R}}$ subject to the zero mode fermion mass spectrum and comparing to the experimental bound listed above (\ref{eq:upper:bound}), we find that the conversion rate must satisfy
\begin{equation}
\left( \frac{3\ {\rm TeV}}{M_{\text{KK}}}\right)^2 \left(\frac{2}{Y_*}\right) < 1.6. \label{eq:upper:bound:custodial}
\end{equation}
lowering the bound to $Y_*\gtrsim 1$ for a $3$ TeV KK gauge boson scale.

\section{Operator analysis of $\mu\to e\gamma$}
\label{sec:OperatorAnalysis}

We work in 't Hooft--Feynman gauge ($\xi=1$) and a flavor basis where all bulk masses $c_i$ are diagonal. The 5D amplitude for $\mu \to e\gamma$ takes the form
\begin{align}
 \label{eq:5Doperator}
    C H \cdot \bar{L}_i \sigma^{MN}E_j F_{MN},
\end{align}
where it is understood that the 5D fields should be replaced by the appropriate external states which each carry an independent $z$ position in the mixed position/momentum space formalism. These positions must be separately integrated over when matching to an effective 4D operator so that (\ref{eq:5Doperator}) can be thought of as a dimension-8 5D scattering amplitude whose prefactor $C$ is a function of the external state positions, as explained in Appendix \ref{app:5D:EFT}.
When calculating this amplitude in the mixed position/momentum space formalism, the physical external state fields have definite KK number, which we take to be zero modes. The external field profiles and internal propagators depend on 4D momenta and $z$-positions so that vertex $z$-positions are integrated from $z=R$ to $z=R'$ while loop momenta are integrated as usual.

After plugging in the wave functions for the fermion and photon zero modes, including all warp factors, matching the gauge coupling, and expanding in Higgs-induced mass insertions, the leading order 4D operator and coefficients for $\mu\to e \gamma$ are
\begin{align}
 R'^2 \frac{e}{16\pi^2} \frac{v}{\sqrt{2}}  f_{L_i} \left( a_{k\ell} Y_{ik}Y^\dagger_{k\ell} Y_{\ell j}  + b_{ij}Y_{ij} \right) f_{-E_j} \bar{L}^{(0)}_i \sigma^{\mu\nu} E_j^{(0)} F_{\mu\nu}^{(0)} + \text{h.c.}
\label{eq:finalop}
\end{align}
The term proportional to three Yukawa matrices comes from the diagrams shown in Figs.~\ref{fig:a:e:diagrams} and \ref{fig:a:n:diagrams}, while the single-Yukawa term comes from those in Fig.~\ref{fig:b:diagrams}. 
In the limit where the bulk masses are universal, we may treat the Yukawas as spurions of the U(3)$^3$ lepton flavor symmetry and note that these are the products of Yukawas required for a chirality-flipping, flavor-changing operator.
%
%

In anarchic flavor models, however, the bulk masses for each fermion species is independent and introduce an additional flavor structure into the theory so that the U$(3)^3$ lepton flavor symmetry is not restored even in the limit $Y\to 0$.  The indices on the dimensionless $a_{k\ell}$ and $b_{ij}$ coefficients encode this flavor structure as carried by the internal fermions of each diagram. Because the lepton hierarchy does not require very different bulk masses, both $a_{k\ell}$ and $b_{ij}$ are nearly universal.

Next note that the zero-mode mass matrix (\ref{eq:RS:anarchy:zero:mode:Yukawa}) introduces a preferred direction in flavor space which defines the mass basis. In fact, up to the non-universality of $b_{ij}$, the single-Yukawa term in (\ref{eq:finalop}) is proportional to---or aligned---with (\ref{eq:RS:anarchy:zero:mode:Yukawa}).
Hence upon rotation to the mass basis, the off-diagonal elements of this term are typically much smaller than its value in the flavor basis \cite{Agashe:2009di, Azatov:2009na} and would be identically zero if the bulk masses were universal. Given a set of bulk mass parameters, the extent to which a specific off-diagonal element of the $b_{ij}$ term is suppressed depends on the particular structure of the anarchic 5D Yukawa matrix. This is a novel feature since the structure of the underlying anarchic Yukawa is usually washed out in observables by the hierarchies in the $f_c$ flavor functions.


On the other hand, a product of anarchic matrices typically indicates a very different direction in flavor space from the original matrix so that the $a_{ij}$ term is not aligned and we may simplify the product to 
\begin{align}
\sum_{k,\ell}a_{k\ell} Y_{ik}Y^\dagger_{k\ell} Y_{\ell j} &= a Y_*^3\label{eq:def:a:anarchic}
\end{align}
for each $i$ and $j$. Here we have \textit{defined} the prefactor $a$; different definitions can include an overall $\mathcal O(1)$ factor from the sum over anarchic matrix elements.
We have used the anarchic limit and the assumption that neither $a_{k\ell}$ nor $b_{ij}$ vary greatly over realistic bulk mass values. This assumption is justified in Section~\ref{sec:warpedXD} where we explicitly calculate these coefficients to leading order. Further, we have assumed that the scales of the anarchic electron and neutrino Yukawa matrices are the same so that $(Y_E)_{ij} \sim (Y_N)_{ij} \sim Y_*$.

To determine the physical $\mu\to e \gamma$ amplitude from this expression we must go to the standard 4D mass eigenbasis by performing a bi-unitary transformation to diagonalize the Standard Model Yukawa,
\begin{align}
\lambda^{\text{SM}} &=U_L \lambda^{(\text{diag})} U_R^\dagger,\label{eq:diagonalize:yukawa}
\end{align}
where the magnitudes of the elements of the unitary matrices $U_{L,R}$ are set, in the anarchic scenario, by the hierarchies in the flavor constants
\begin{align}
(U_L)_{ij} \sim \frac{f_{L_i}}{f_{L_j}} \ \ \text{for} \ \ f_{L_i}<f_{L_j}.
\label{eq:Umagnitudes}
\end{align}
For future simplicity, let us define the relevant part of the $b_{ij}Y_{ij}$ matrix after this rotation,
\begin{align}
	b Y_*=\sum_{k,\ell}(U_L)_{2k }b_{k\ell}Y_{k\ell}(U_R^\dag)_{\ell 1}.\label{eq:def:b}
\end{align}

The traditional parameterization for the $\mu\to e\gamma$ amplitude is written as \cite{Agashe:2006iy}
\begin{align}
\frac{-i C_{L,R}}{2m_\mu} \bar{u}_{L,R}\, \sigma^{\mu\nu}\, u_{R,L} F_{\mu\nu},
\label{eq:amplitude}
\end{align}
where $u_{L,R}$ are the left- and right-handed Dirac spinors for the leptons. Comparing (\ref{eq:finalop}) with (\ref{eq:amplitude}) and using the magnitudes of the off-diagonal terms in the $U_L$ rotation matrix in (\ref{eq:Umagnitudes}), we find that in the mass eigenbasis the coefficients are given by
\begin{align}
C_L&= \left(a Y_*^3 + bY_*\right) R'^2 \frac{e}{16\pi^2}  \frac{v}{\sqrt{2}} 2 m_\mu f_{L_2} f_{-E_1},  \\
C_R&= \left(a Y_*^3 + bY_*\right) R'^2 \frac{e}{16\pi^2} \frac{v}{\sqrt{2}} 2 m_\mu f_{L_1} f_{-E_2}.\label{eq:Cs}
\end{align}
The $\mu\to e \gamma$ branching fraction and its experimental bound are given by
\begin{align}
{\rm Br}(\mu\to e\gamma)_{\text{thy}}&=\frac{12\pi^2}{(G_F m_\mu^2)^2}(\left|C_L\right|^2 +\left|C_R\right|^2), \\
{\rm Br}(\mu\to e\gamma)_{\text{exp}}&< 1.2 \cdot 10^{-11}.
\end{align}
While the generic expression for Br$(\mu\to e\gamma )$ depends on the individual wave functions $f_{L,-E}$, the product $C_L C_R$ is fixed by the physical lepton masses and the relation $C_L^2+C_R^2 \geq 2 C_L C_R$ so that one can put a lower bound on the branching ratio
\begin{align}
	\text{Br} (\mu\to e\gamma ) \geq 6\, \left|aY_*^2+b\right|^2  \frac{\alpha}{4\pi} \left( \frac{R'^2}{G_F}\right)^2 \frac{m_e}{m_\mu} \approx 5.1 \cdot 10^{-8}\, \left|aY_*^2+b\right|^2 \left(\frac{3\ {\rm TeV}}{M_\text{KK}}\right)^4.\label{eq:bound}
\end{align}
Thus for a 3 TeV KK gauge boson scale we obtain an upper bound on $Y_*$
\begin{align}
|aY_*^2+b| \left(\frac{3\ {\rm TeV}}{M_\text{KK}}\right)^2 \leq 0.015. \label{eq:Ystar:bound:with:b}
\end{align}
Note that the $b$ coefficient is independent of $Y_*$ so that sufficiently large $b$ can rule out the assumption that the 5D Yukawa matrix can be completely anarchic---i.e.\ with no assumed underlying flavor structure---at a given KK scale no matter how small one picks $Y_*$. This is a new type of constraint on anarchic flavor models in a warped extra dimension. Conversely, if $b$ is of the same order as $a$ and has the opposite sign, then the bounds on the anarchic scale $Y_*$ are alleviated. We will show below that $b$ is typically suppressed relative to $a$ but can, in principle, take a range of values between $b=-0.5$ and $0.5$. For simplicity we may use the case $b=0$ as a representative and plausible example, in which case the bound on the anarchic Yukawa scale is
\begin{align}
Y_* \leq 0.12 \,|a|^{-\frac{1}{2}}. \label{eq:Ystar:bound}	
\end{align}
In Section~\ref{sec:constraints:and:tension} we quantify the extent to which the $b$ term may affect this bound.
Combined with the lower bounds on $Y_*$ from tree-level processes in Section \ref{sec:TreeOps}, this bound typically introduces a tension in the preferred value of $Y_*$ depending on the value of $a$. In other words, it can force one to either increase the KK scale or introduce additional symmetry structure into the 5D Yukawa matrices which can reduce $a$ in (\ref{eq:def:a:anarchic}) or force a cancellation in (\ref{eq:Ystar:bound:with:b}).


\section{Calculation of $\mu\to e \gamma$ in a warped extra dimension}
\label{sec:warpedXD}

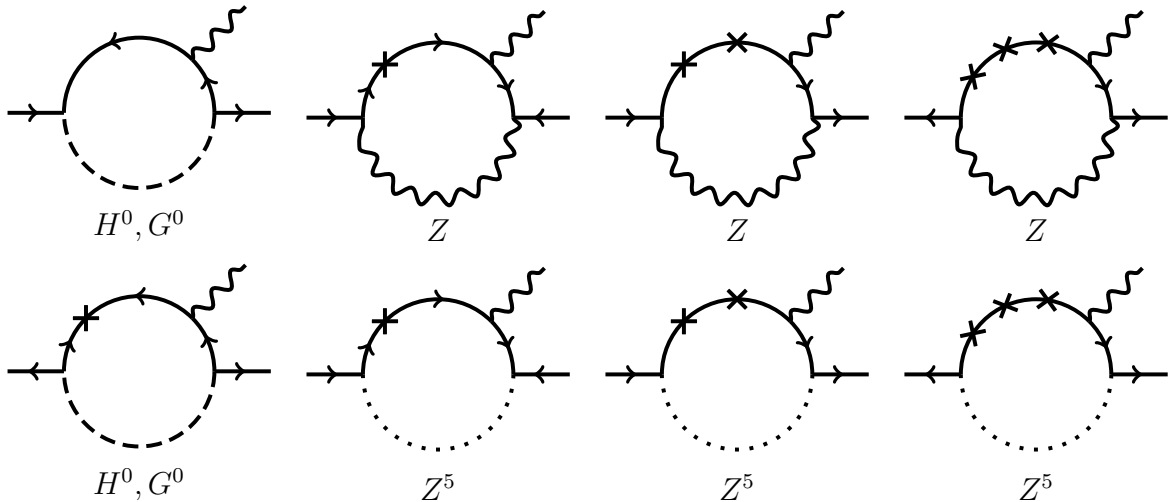
\begin{figure}[t!]
	\centering
		\begin{tabular}{cccc}

	\begin{tikzpicture}[line width=1.5 pt, scale=1]
		\draw[fermion] (-1.75,0) -- (-1,0);
		\draw[fermion] (1,0) -- (1.75,0);
		\draw[fermionbar] (-1,0) arc (180:45:1);
		\draw[fermionbar] (45:1) arc (45:0:1); 
		\draw[scalarnoarrow] (1,0) arc (0:-180:1);
		\draw[vector] (45:1) -- (45: 2);
	\node at (0,-1.5) {$H^0, G^0$};
	\end{tikzpicture}
	
	&

	\begin{tikzpicture}[line width=1.5 pt, scale=1]
		\draw[fermion] (-1.75,0) -- (-1,0);
		\draw[fermionbar] (1,0) -- (1.75,0);
		\draw[fermion] (-1,0) arc (180:135:1);
		\draw[fermion] (135:1) arc (135:45:1);
		\draw[fermion] (45:1) arc (45:0:1); 
		\draw[vector] (1,0) arc (0:-180:1);
		\draw[vector] (45:1) -- (45:2);
	%
	\begin{scope}[rotate=135]
	\begin{scope}[shift={(1,0)}] 
		\clip (0,0) circle (.175cm);
		\draw[fermionnoarrow] (-1,1) -- (1,-1);
		\draw[fermionnoarrow] (1,1) -- (-1,-1);
	\end{scope}	
	\end{scope}
	\node at (0,-1.5) {$Z$};
	\end{tikzpicture}

	&

	\begin{tikzpicture}[line width=1.5 pt, scale=1]
		\draw[fermion] (-1.75,0) -- (-1,0);
		\draw[fermion] (1,0) -- (1.75,0);
		\draw[fermionnoarrow] (-1,0) arc (180:45:1);
		\draw[fermion] (45:1) arc (45:0:1); 
		\draw[vector] (1,0) arc (0:-180:1);
		\draw[vector] (45:1) -- (45:2);
	%
	\begin{scope}[rotate=135]
	\begin{scope}[shift={(1,0)}] 
		\clip (0,0) circle (.175cm);
		\draw[fermionnoarrow] (-1,1) -- (1,-1);
		\draw[fermionnoarrow] (1,1) -- (-1,-1);
	\end{scope}	
	\end{scope}
	\begin{scope}[rotate=90]
	\begin{scope}[shift={(1,0)}] 
		\clip (0,0) circle (.175cm);
		\draw[fermionnoarrow] (-1,1) -- (1,-1);
		\draw[fermionnoarrow] (1,1) -- (-1,-1);
	\end{scope}	
	\end{scope}
	\node at (0,-1.5) {$Z$};
	\end{tikzpicture}
	
	&
	
	\begin{tikzpicture}[line width=1.5 pt, scale=1]
		\draw[fermionbar] (-1.75,0) -- (-1,0);
		\draw[fermion] (1,0) -- (1.75,0);
		\draw[fermionnoarrow] (-1,0) arc (180:45:1);
		\draw[fermion] (45:1) arc (45:0:1); 
		\draw[vector] (1,0) arc (0:-180:1);
		\draw[vector] (45:1) -- (45:2);
	%
	\begin{scope}[rotate=147]
	\begin{scope}[shift={(1,0)}] 
		\clip (0,0) circle (.175cm);
		\draw[fermionnoarrow] (-1,1) -- (1,-1);
		\draw[fermionnoarrow] (1,1) -- (-1,-1);
	\end{scope}	
	\end{scope}
	\begin{scope}[rotate=114]
	\begin{scope}[shift={(1,0)}] 
		\clip (0,0) circle (.175cm);
		\draw[fermionnoarrow] (-1,1) -- (1,-1);
		\draw[fermionnoarrow] (1,1) -- (-1,-1);
	\end{scope}	
	\end{scope}
	\begin{scope}[rotate=81]
	\begin{scope}[shift={(1,0)}] 
		\clip (0,0) circle (.175cm);
		\draw[fermionnoarrow] (-1,1) -- (1,-1);
		\draw[fermionnoarrow] (1,1) -- (-1,-1);
	\end{scope}	
	\end{scope}
	\node at (0,-1.5) {$Z$};
	\end{tikzpicture}

%
%

\\

\begin{tikzpicture}[line width=1.5 pt, scale=1]
	\draw[fermionbar] (-1.75,0) -- (-1,0);
	\draw[fermion] (1,0) -- (1.75,0);
	\draw[fermion] (-1,0) arc (180:135:1);
	\draw[fermionbar] (135:1) arc (135:45:1);
	\draw[fermionbar] (45:1) arc (45:0:1); 
	\draw[scalarnoarrow] (1,0) arc (0:-180:1);
	\draw[vector] (45:1) -- (45: 2);
%
\begin{scope}[rotate=135]
\begin{scope}[shift={(1,0)}] 
	\clip (0,0) circle (.175cm);
	\draw[fermionnoarrow] (-1,1) -- (1,-1);
	\draw[fermionnoarrow] (1,1) -- (-1,-1);
\end{scope}	
\end{scope}
\node at (0,-1.5) {$H^0, G^0$};
\end{tikzpicture}

&

\begin{tikzpicture}[line width=1.5 pt, scale=1]
	\draw[fermion] (-1.75,0) -- (-1,0);
	\draw[fermionbar] (1,0) -- (1.75,0);
	\draw[fermion] (-1,0) arc (180:135:1);
	\draw[fermion] (135:1) arc (135:45:1);
	\draw[fermion] (45:1) arc (45:0:1); 
	\draw[vectorscalar] (1,0) arc (0:-180:1);
	\draw[vector] (45:1) -- (45:2);
%
\begin{scope}[rotate=135]
\begin{scope}[shift={(1,0)}] 
	\clip (0,0) circle (.175cm);
	\draw[fermionnoarrow] (-1,1) -- (1,-1);
	\draw[fermionnoarrow] (1,1) -- (-1,-1);
\end{scope}	
\end{scope}
\node at (0,-1.5) {$Z^5$};
\end{tikzpicture}

	&

	\begin{tikzpicture}[line width=1.5 pt, scale=1]
		\draw[fermion] (-1.75,0) -- (-1,0);
		\draw[fermion] (1,0) -- (1.75,0);
		\draw[fermionnoarrow] (-1,0) arc (180:45:1);
		\draw[fermion] (45:1) arc (45:0:1); 
		\draw[vectorscalar] (1,0) arc (0:-180:1);
		\draw[vector] (45:1) -- (45:2);
	%
	\begin{scope}[rotate=135]
	\begin{scope}[shift={(1,0)}] 
		\clip (0,0) circle (.175cm);
		\draw[fermionnoarrow] (-1,1) -- (1,-1);
		\draw[fermionnoarrow] (1,1) -- (-1,-1);
	\end{scope}	
	\end{scope}
	\begin{scope}[rotate=90]
	\begin{scope}[shift={(1,0)}] 
		\clip (0,0) circle (.175cm);
		\draw[fermionnoarrow] (-1,1) -- (1,-1);
		\draw[fermionnoarrow] (1,1) -- (-1,-1);
	\end{scope}	
	\end{scope}
	\node at (0,-1.5) {$Z^5$};
	\end{tikzpicture}

	&

	\begin{tikzpicture}[line width=1.5 pt, scale=1]
		\draw[fermionbar] (-1.75,0) -- (-1,0);
		\draw[fermion] (1,0) -- (1.75,0);
		\draw[fermionnoarrow] (-1,0) arc (180:45:1);
		\draw[fermion] (45:1) arc (45:0:1); 
		\draw[vectorscalar] (1,0) arc (0:-180:1);
		\draw[vector] (45:1) -- (45:2);
	%
	\begin{scope}[rotate=147]
	\begin{scope}[shift={(1,0)}] 
		\clip (0,0) circle (.175cm);
		\draw[fermionnoarrow] (-1,1) -- (1,-1);
		\draw[fermionnoarrow] (1,1) -- (-1,-1);
	\end{scope}	
	\end{scope}
	\begin{scope}[rotate=114]
	\begin{scope}[shift={(1,0)}] 
		\clip (0,0) circle (.175cm);
		\draw[fermionnoarrow] (-1,1) -- (1,-1);
		\draw[fermionnoarrow] (1,1) -- (-1,-1);
	\end{scope}	
	\end{scope}
	\begin{scope}[rotate=81]
	\begin{scope}[shift={(1,0)}] 
		\clip (0,0) circle (.175cm);
		\draw[fermionnoarrow] (-1,1) -- (1,-1);
		\draw[fermionnoarrow] (1,1) -- (-1,-1);
	\end{scope}	
	\end{scope}
	\node at (0,-1.5) {$Z^5$};
	\end{tikzpicture}

%
%
	
\end{tabular}
	\caption{Neutral boson diagrams contributing to the $a$ coefficient defined in (\ref{eq:def:a:anarchic}). Fermion arrows denote the zero mode chirality, i.e.\ the SU(2) representation. External legs whose arrows do not point outward have an implicit external mass insertion. Dotted lines represent the fifth component of a bulk gauge field. Analytic forms for these diagrams are given in Appendix~\ref{app:analytic:expressions}.}
	\label{fig:a:e:diagrams}
\end{figure}

\begin{figure}[t!]
	\centering
		\begin{tabular}{cccc}

	\begin{tikzpicture}[line width=1.5 pt, scale=1]
		\draw[fermion] (-1.75,0) -- (-1,0);
		\draw[fermion] (1,0) -- (1.75,0);
		\draw[vector] (-1,0) arc (180:45:1);
		\draw[vector] (45:1) arc (45:0:1); 
		\draw[fermion] (1,0) arc (0:-180:1);
		\draw[vector] (45:1) -- (45:2);
	%
	\begin{scope}[rotate=-135]
	\begin{scope}[shift={(1,0)}] 
		\clip (0,0) circle (.175cm);
		\draw[fermionnoarrow] (-1,1) -- (1,-1);
		\draw[fermionnoarrow] (1,1) -- (-1,-1);
	\end{scope}	
	\end{scope}
	\begin{scope}[rotate=-45]
	\begin{scope}[shift={(1,0)}] 
		\clip (0,0) circle (.175cm);
		\draw[fermionnoarrow] (-1,1) -- (1,-1);
		\draw[fermionnoarrow] (1,1) -- (-1,-1);
	\end{scope}	
	\end{scope}
	\node at (0,-1.5) {$W$};
	\end{tikzpicture}

	&

	\begin{tikzpicture}[line width=1.5 pt, scale=1]
		\draw[fermion] (-1.75,0) -- (-1,0);
		\draw[fermion] (1,0) -- (1.75,0);
		\draw[vector] (-1,0) arc (180:45:1);
		\draw[vectorscalar] (45:1) arc (45:0:1); 
		\draw[fermion] (1,0) arc (0:-180:1);
		\draw[vector] (45:1) -- (45:2);
	%
	\begin{scope}[rotate=-135]
	\begin{scope}[shift={(1,0)}] 
		\clip (0,0) circle (.175cm);
		\draw[fermionnoarrow] (-1,1) -- (1,-1);
		\draw[fermionnoarrow] (1,1) -- (-1,-1);
	\end{scope}	
	\end{scope}
	\begin{scope}[rotate=-45]
	\begin{scope}[shift={(1,0)}] 
		\clip (0,0) circle (.175cm);
		\draw[fermionnoarrow] (-1,1) -- (1,-1);
		\draw[fermionnoarrow] (1,1) -- (-1,-1);
	\end{scope}	
	\end{scope}
	\node at (0,-1.5) {$W, W^5$};
	\end{tikzpicture}

	&

	\begin{tikzpicture}[line width=1.5 pt, scale=1]
		\draw[fermion] (-1.75,0) -- (-1,0);
		\draw[fermion] (1,0) -- (1.75,0);
		\draw[vectorscalar] (-1,0) arc (180:45:1);
		\draw[vector] (45:1) arc (45:0:1); 
		\draw[fermion] (1,0) arc (0:-180:1);
		\draw[vector] (45:1) -- (45:2);
	%
	\begin{scope}[rotate=-135]
	\begin{scope}[shift={(1,0)}] 
		\clip (0,0) circle (.175cm);
		\draw[fermionnoarrow] (-1,1) -- (1,-1);
		\draw[fermionnoarrow] (1,1) -- (-1,-1);
	\end{scope}	
	\end{scope}
	\begin{scope}[rotate=-45]
	\begin{scope}[shift={(1,0)}] 
		\clip (0,0) circle (.175cm);
		\draw[fermionnoarrow] (-1,1) -- (1,-1);
		\draw[fermionnoarrow] (1,1) -- (-1,-1);
	\end{scope}	
	\end{scope}
	\node at (0,-1.5) {$W^5, W$};
	\end{tikzpicture}
	
	&
	
	\begin{tikzpicture}[line width=1.5 pt, scale=1]
		\draw[fermion] (-1.75,0) -- (-1,0);
		\draw[fermion] (1,0) -- (1.75,0);
		\draw[vectorscalar] (-1,0) arc (180:45:1);
		\draw[vectorscalar] (45:1) arc (45:0:1); 
		\draw[fermion] (1,0) arc (0:-180:1);
		\draw[vector] (45:1) -- (45:2);
	%
	\begin{scope}[rotate=-135]
	\begin{scope}[shift={(1,0)}] 
		\clip (0,0) circle (.175cm);
		\draw[fermionnoarrow] (-1,1) -- (1,-1);
		\draw[fermionnoarrow] (1,1) -- (-1,-1);
	\end{scope}	
	\end{scope}
	\begin{scope}[rotate=-45]
	\begin{scope}[shift={(1,0)}] 
		\clip (0,0) circle (.175cm);
		\draw[fermionnoarrow] (-1,1) -- (1,-1);
		\draw[fermionnoarrow] (1,1) -- (-1,-1);
	\end{scope}	
	\end{scope}
	\node at (0,-1.5) {$W^5$};
	\end{tikzpicture}

\\

%

\begin{tikzpicture}[line width=1.5 pt, scale=1]
	\draw[fermion] (-1.75,0) -- (-1,0);
	\draw[fermion] (1,0) -- (1.75,0);
	\draw[scalarnoarrow] (-1,0) arc (180:45:1);
	\draw[scalarnoarrow] (45:1) arc (45:0:1); 
	\draw[fermion] (1,0) arc (0:-180:1);
	\draw[vector] (45:1) -- (45: 2);
\node at (0,-1.5) {$H^\pm$};
\end{tikzpicture}

&


\begin{tikzpicture}[line width=1.5 pt, scale=1]
	\draw[fermionbar] (-1.75,0) -- (-1,0);
	\draw[fermion] (1,0) -- (1.75,0);
	\draw[scalarnoarrow] (-1,0) arc (180:45:1);
	\draw[scalarnoarrow] (45:1) arc (45:0:1); 
	\draw[fermion] (1,0) arc (0:-90:1);
	\draw[fermionbar] (-90:1) arc (-90:-180:1);
	\draw[vector] (45:1) -- (45: 2);
%
\begin{scope}[rotate=-90]
\begin{scope}[shift={(1,0)}] 
	\clip (0,0) circle (.175cm);
	\draw[fermionnoarrow] (-1,1) -- (1,-1);
	\draw[fermionnoarrow] (1,1) -- (-1,-1);
\end{scope}	
\end{scope}
\node at (0,-1.5) {$H^\pm$};
\end{tikzpicture}

	&


	\begin{tikzpicture}[line width=1.5 pt, scale=1]
		\draw[fermionbar] (-1.75,0) -- (-1,0);
		\draw[fermion] (1,0) -- (1.75,0);
		\draw[scalarnoarrow] (-1,0) arc (180:45:1);
		\draw[vector] (45:1) arc (45:0:1); 
		\draw[fermion] (1,0) arc (0:-180:1);
		\draw[vector] (45:1) -- (45:2);
	%
	\begin{scope}[rotate=-135]
	\begin{scope}[shift={(1,0)}] 
		\clip (0,0) circle (.175cm);
		\draw[fermionnoarrow] (-1,1) -- (1,-1);
		\draw[fermionnoarrow] (1,1) -- (-1,-1);
	\end{scope}	
	\end{scope}
	\begin{scope}[rotate=-45]
	\begin{scope}[shift={(1,0)}] 
		\clip (0,0) circle (.175cm);
		\draw[fermionnoarrow] (-1,1) -- (1,-1);
		\draw[fermionnoarrow] (1,1) -- (-1,-1);
	\end{scope}	
	\end{scope}
	\node at (0,-1.5) {$H^\pm,W$};
	\end{tikzpicture}

	&
	

%
%
%
\end{tabular}
	\caption{Charged boson diagrams contributing to the $a$ coefficient following the conventions in Fig.~\ref{fig:a:e:diagrams}. Analytic forms for these diagrams are given in Appendix~\ref{app:analytic:expressions}.}
	\label{fig:a:n:diagrams}
\end{figure}
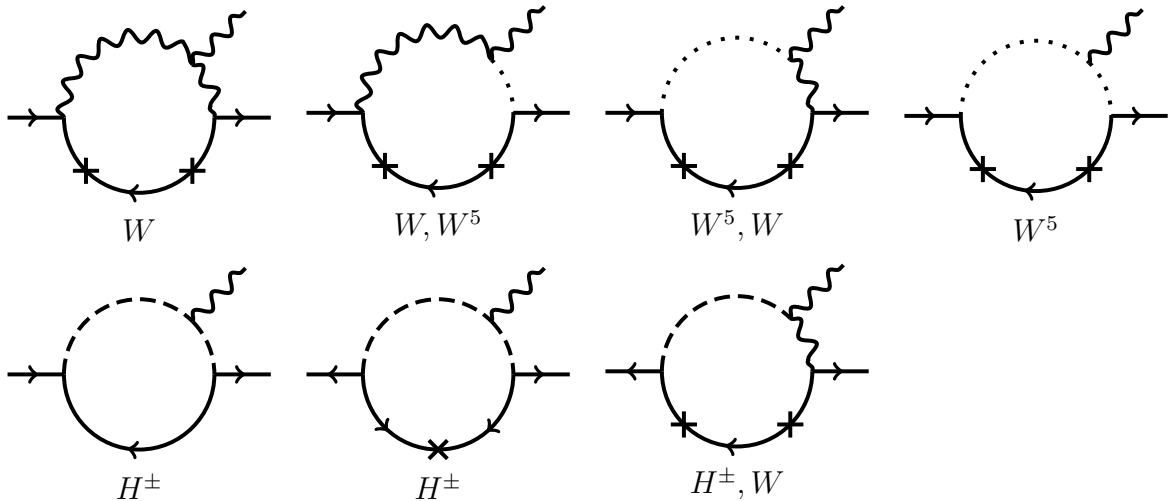

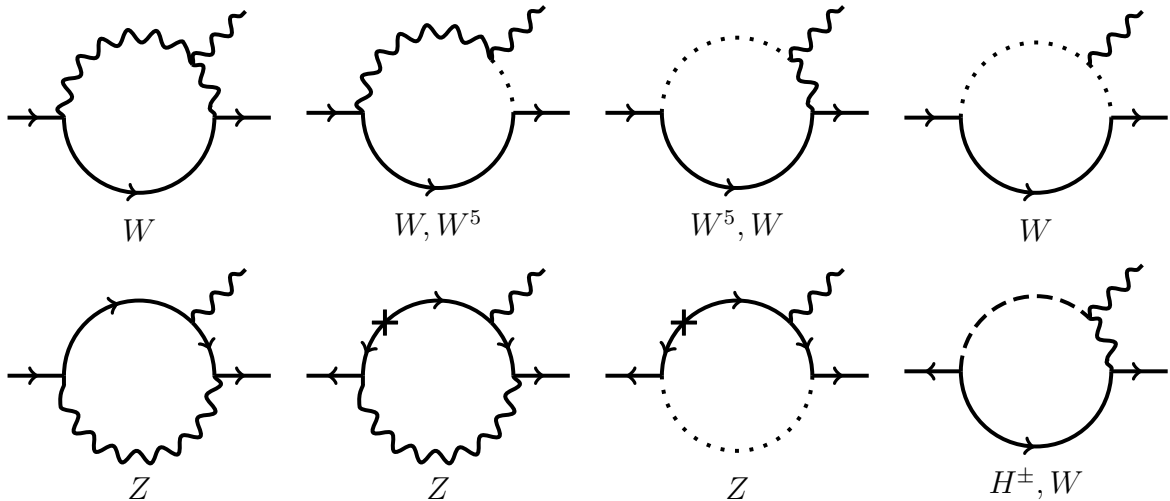
\begin{figure}[t!]
	\centering
		\begin{tabular}{cccc}

	\begin{tikzpicture}[line width=1.5 pt, scale=1]
		\draw[fermion] (-1.75,0) -- (-1,0);
		\draw[fermion] (1,0) -- (1.75,0);
		\draw[vector] (-1,0) arc (180:45:1);
		\draw[vector] (45:1) arc (45:0:1); 
		\draw[fermionbar] (1,0) arc (0:-180:1);
		\draw[vector] (45:1) -- (45: 2);
	\node at (0,-1.5) {$W$};
	\end{tikzpicture}

	&

	\begin{tikzpicture}[line width=1.5 pt, scale=1]
		\draw[fermion] (-1.75,0) -- (-1,0);
		\draw[fermion] (1,0) -- (1.75,0);
		\draw[vector] (-1,0) arc (180:45:1);
		\draw[vectorscalar] (45:1) arc (45:0:1); 
		\draw[fermionbar] (1,0) arc (0:-180:1);
		\draw[vector] (45:1) -- (45: 2);
	\node at (0,-1.5) {$W, W^5$};
	\end{tikzpicture}

	&

	\begin{tikzpicture}[line width=1.5 pt, scale=1]
		\draw[fermion] (-1.75,0) -- (-1,0);
		\draw[fermion] (1,0) -- (1.75,0);
		\draw[vectorscalar] (-1,0) arc (180:45:1);
		\draw[vector] (45:1) arc (45:0:1); 
		\draw[fermionbar] (1,0) arc (0:-180:1);
		\draw[vector] (45:1) -- (45: 2);
	\node at (0,-1.5) {$W^5, W$};
	\end{tikzpicture}

	&
	
	\begin{tikzpicture}[line width=1.5 pt, scale=1]
		\draw[fermion] (-1.75,0) -- (-1,0);
		\draw[fermion] (1,0) -- (1.75,0);
		\draw[vectorscalar] (-1,0) arc (180:45:1);
		\draw[vectorscalar] (45:1) arc (45:0:1); 
		\draw[fermionbar] (1,0) arc (0:-180:1);
		\draw[vector] (45:1) -- (45: 2);
	\node at (0,-1.5) {$W$};
	\end{tikzpicture}

\\

%
%

\begin{tikzpicture}[line width=1.5 pt, scale=1]
	\draw[fermion] (-1.75,0) -- (-1,0);
	\draw[fermion] (1,0) -- (1.75,0);
	\draw[fermion] (-1,0) arc (180:45:1);
	\draw[fermion] (45:1) arc (45:0:1); 
	\draw[vector] (1,0) arc (0:-180:1);
	\draw[vector] (45:1) -- (45: 2);
\node at (0,-1.5) {$Z$};
\end{tikzpicture}

&

\begin{tikzpicture}[line width=1.5 pt, scale=1]
	\draw[fermionbar] (-1.75,0) -- (-1,0);
	\draw[fermion] (1,0) -- (1.75,0);
	\draw[fermionbar] (-1,0) arc (180:135:1);
	\draw[fermion] (135:1) arc (135:45:1);
	\draw[fermion] (45:1) arc (45:0:1); 
	\draw[vector] (1,0) arc (0:-180:1);
	\draw[vector] (45:1) -- (45: 2);
%
\begin{scope}[rotate=135]
\begin{scope}[shift={(1,0)}] 
	\clip (0,0) circle (.175cm);
	\draw[fermionnoarrow] (-1,1) -- (1,-1);
	\draw[fermionnoarrow] (1,1) -- (-1,-1);
\end{scope}	
\end{scope}
\node at (0,-1.5) {$Z$};
\end{tikzpicture}

&

\begin{tikzpicture}[line width=1.5 pt, scale=1]
	\draw[fermionbar] (-1.75,0) -- (-1,0);
	\draw[fermion] (1,0) -- (1.75,0);
	\draw[fermionbar] (-1,0) arc (180:135:1);
	\draw[fermion] (135:1) arc (135:45:1);
	\draw[fermion] (45:1) arc (45:0:1); 
	\draw[vectorscalar] (1,0) arc (0:-180:1);
	\draw[vector] (45:1) -- (45: 2);
%
\begin{scope}[rotate=135]
\begin{scope}[shift={(1,0)}] 
	\clip (0,0) circle (.175cm);
	\draw[fermionnoarrow] (-1,1) -- (1,-1);
	\draw[fermionnoarrow] (1,1) -- (-1,-1);
\end{scope}	
\end{scope}
\node at (0,-1.5) {$Z$};
\end{tikzpicture}


&

\begin{tikzpicture}[line width=1.5 pt, scale=1]
	\draw[fermionbar] (-1.75,0) -- (-1,0);
	\draw[fermion] (1,0) -- (1.75,0);
	\draw[scalarnoarrow] (-1,0) arc (180:45:1);
	\draw[vector] (45:1) arc (45:0:1); 
	\draw[fermionbar] (1,0) arc (0:-180:1);
	\draw[vector] (45:1) -- (45:2);
\node at (0,-1.5) {$H^\pm,W$};
\end{tikzpicture}


%
%
	
\end{tabular}
	\caption{Diagrams contributing to the $b$ coefficient following the conventions in Fig.~\ref{fig:a:e:diagrams}. Not shown: zero mass-insertion $Z^5$ diagram. Analytic forms for these diagrams are given in Appendix~\ref{app:analytic:expressions}.}
	\label{fig:b:diagrams}
\end{figure}

In principle, there are a large number of diagrams contributing to the $a$ and $b$ coefficients even when only considering the leading terms in a mass insertion expansion. These are depicted in Figs.~\ref{fig:a:e:diagrams}--\ref{fig:b:diagrams}. Fortunately, many of these diagrams are naturally suppressed and the dominant contribution to each coefficient is given by the two diagrams shown in Fig.~\ref{fig:main:diagrams}. Analytic expressions for the leading and next-to-leading diagrams are given in Appendix \ref{app:analytic:expressions} along with an estimate of the size of each contribution.

The flavor structure of the diagrams contributing to the $b$ coefficient is aligned with the fermion zero-mode mass matrix \cite{Agashe:2004cp, Agashe:2006iy, Kaustubh}. The rotation of the external states to mass eigenstates thus suppresses these diagrams up to the bulk mass ($c$) dependence of internal propagators which point in a different direction in flavor space and are not aligned. Since KK modes do not carry very strong bulk mass dependence, the diagrams which typically give the largest contribution after alignment are those which permit zero mode fermions in the loop. We provide a precise definition of the term ``typically'' in Section \ref{sec:calc:b}.


The Ward identity requires that the physical amplitude for a muon of momentum $p$ to decay into a photon of polarization $\epsilon$ and an electron of momentum $p'$ takes the form
\begin{align}
	\mathcal M = \epsilon_\mu \mathcal M^\mu \sim \epsilon_\mu\bar u_{p'}\left[(p+p')^\mu - (m_\mu + m_e)\gamma^\mu\right] u_p.\label{eq:amplitude:general:form}
\end{align} 
This is the combination of masses and momenta that gives the correct chirality-flipping tensor amplitude in (\ref{eq:amplitude}). This simplifies the calculation of this process since one only has to identify the coefficient of the $\bar u_{p'}(p+p')^\mu u$ term to determine the entire amplitude; all other terms are redundant by gauge invariance \cite{Lavoura:2003xp}. 
The general strategy is to use the Clifford algebra and the equations of motion for the external spinors to determine this coefficient. This allows us to directly write the finite physical contribution to the amplitude without worrying about the regularization of potentially divergent terms which are not gauge invariant. 
In Section \ref{sec:heursitic:4D5D} we will further use this observation to explain the finiteness of this amplitude in 5D.
%
%

In addition to the diagrams in Figs.~\ref{fig:a:e:diagrams}--\ref{fig:b:diagrams}, there are higher-order diagrams with an even number of additional mass insertions and brane-to-brane propagators. Following the Feynman rules in Appendix~\ref{app:bulk:Feynman:rules}, each higher-order pair of mass insertions is suppressed by an additional factor of
\begin{align}
    \left(\frac{\slashed k}{k}\frac{R'^4}{R^4}\cdot (-i)\frac{R^3}{R'^3} RY_* \frac{v}{\sqrt{2}}\right)^2 \sim \frac{1}{2}\left(Y_* R' v\right)^2 \sim \mathcal O(10^{-2}),
\end{align}
since we assume anarchic Yukawa matrices, $Y_* \sim 2$. We are thus justified in considering only the leading-order terms in the mass insertion approximation.

We now present the leading contributions to the $a$ and $b$ coefficients.
Other diagrams give a correction on the order of 10\% of these results. We provide explicit formulas and numerical estimates for the next-to-leading order corrections in Appendix \ref{app:analytic:expressions}.


\subsection{Calculation of $a$}
\label{sec:calc:a}

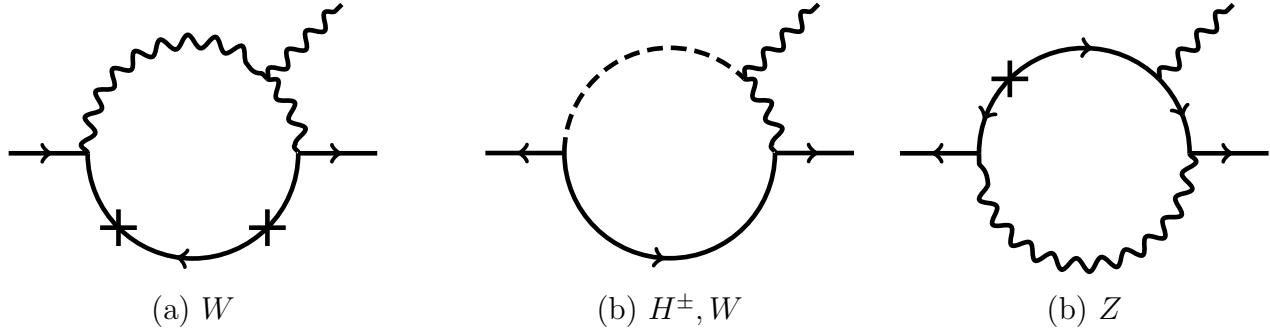
\begin{figure}[t!]
	\centering
			\begin{tikzpicture}[line width=1.75 pt, scale=1.4]
					%
					\draw[fermion] (-1.75,0) -- (-1,0);
					\draw[fermion] (1,0) -- (1.75,0);
					\draw[vector] (-1,0) arc (180:45:1);
					\draw[vector] (45:1) arc (45:0:1); 
					\draw[fermion] (1,0) arc (0:-180:1);
					\draw[vector] (45:1) -- (45:2);
				%
				\begin{scope}[rotate=-135]
				\begin{scope}[shift={(1,0)}] 
					\clip (0,0) circle (.175cm);
					\draw[fermionnoarrow] (-1,1) -- (1,-1);
					\draw[fermionnoarrow] (1,1) -- (-1,-1);
				\end{scope}	
				\end{scope}
				\begin{scope}[rotate=-45]
				\begin{scope}[shift={(1,0)}] 
					\clip (0,0) circle (.175cm);
					\draw[fermionnoarrow] (-1,1) -- (1,-1);
					\draw[fermionnoarrow] (1,1) -- (-1,-1);
				\end{scope}	
				\end{scope}
				\node at (0,-1.5) {(a) $W$};
				\end{tikzpicture}
			\qquad\quad
	\begin{tikzpicture}[line width=1.75 pt, scale=1.4]
		\draw[fermionbar] (-1.75,0) -- (-1,0);
		\draw[fermion] (1,0) -- (1.75,0);
		\draw[scalarnoarrow] (-1,0) arc (180:45:1);
		\draw[vector] (45:1) arc (45:0:1); 
		\draw[fermionbar] (1,0) arc (0:-180:1);
		\draw[vector] (45:1) -- (45:2);
	\node at (0,-1.5) {(b) $H^\pm,W$};
	\end{tikzpicture}
	\quad		
	\begin{tikzpicture}[line width=1.75 pt, scale=1.4]
		\draw[fermionbar] (-1.75,0) -- (-1,0);
		\draw[fermion] (1,0) -- (1.75,0);
		\draw[fermionbar] (-1,0) arc (180:135:1);
		\draw[fermion] (135:1) arc (135:45:1);
		\draw[fermion] (45:1) arc (45:0:1); 
		\draw[vector] (1,0) arc (0:-180:1);
		\draw[vector] (45:1) -- (45: 2);
	%
	\begin{scope}[rotate=135]
	\begin{scope}[shift={(1,0)}] 
		\clip (0,0) circle (.175cm);
		\draw[fermionnoarrow] (-1,1) -- (1,-1);
		\draw[fermionnoarrow] (1,1) -- (-1,-1);
	\end{scope}	
	\end{scope}
	\node at (0,-1.5) {(b) $Z$};
	\end{tikzpicture}
			
	\caption{The leading diagrams contributing to the $a$ and $b$ coefficients following the same conventions as Fig.~\ref{fig:a:e:diagrams}.}
	\label{fig:main:diagrams}
\end{figure}

We now calculate the leading-order contribution to the amplitude to determine the $a$ coefficient in (\ref{eq:def:a:anarchic}). As discussed above, it is sufficient to compute the coefficient of the $(p+p')^\mu$ term in the amplitude. 
The dominant contribution to $a$ comes from the $W$ boson diagrams in Fig.~\ref{fig:main:diagrams}a. This is because diagrams with 5D gauge bosons are enhanced relative to the Higgs diagrams by a factor of $\ln R'/R \sim 37$. Further, the $W$ diagrams are enhanced over the $Z$ diagrams due to the size of their respective Standard Model couplings to leptons.
Additional suppression factors can arise from the structure of each diagram and are discussed in Appendix~\ref{app:estimates}. Explicit calculation confirms that the $W$ loop with two internal mass insertions indeed gives the leading contribution to $a$. 

The charged and neutral boson diagrams have independent flavor structures, $(Y_E Y_N^\dag Y_N)_{\mu e}$ and $(Y_EY^\dag_E Y_E)_{\mu e}$ respectively. The anarchic Yukawa assumption implies that both of these terms should be of the same order, $Y_*^3$. However one must remember that there may be a relative sign between these contributions depending on the specific anarchic $Y_N$ and $Y_E$ matrices. In other words, $a = a_\text{charged} \pm a_\text{neutral}$ where the sign cannot be specified generically. However, because $a_\text{neutral} \ll a_\text{charged}$, we ignore the neutral boson loops, though these neutral boson diagrams may become appreciable if one allows a hierarchy between the overall scales of the $Y_N$ and $Y_E$ matrices.

The $W$ loop in Fig.~\ref{fig:main:diagrams}a contains an implicit mass insertion on the external muon leg. As explained in Appendix~\ref{app:estimates}, the 5D fermion propagator between this mass insertion and the loop vertex is dominated by the KK mode which changes fermion chirality. This is because the chirality-preserving piece of the propagator goes like $\slashed{p}$. Invoking the muon equation of motion gives a factor of $f_\mu^{(0)}(vR') f_\mu^{(0)}\sim (m_\mu R')$ for the external leg. This is much smaller than the $f_\mu^{(0)}(vR')f_\mu^{(\text{KK})}$ factor from the chirality-flipping part of the propagator. Compared to the mass insertion connecting the zero mode external muon to a KK intermediate state, the mass insertion connecting two zero mode fermions is smaller by a factor of the exponentially suppressed zero mode profile\footnote{We thank Martin Beneke, Paramita Dey, and J\"urgen Rohrwild for pointing this out.}.

Using the Feynman rules in Appendix~\ref{app:bulk:Feynman:rules}, the amplitude this diagram is
\begin{align}
	\left.\mathcal M^\mu\right|_{(p+p')} 
	=
	\frac{i}{16\pi^2}(R')^2 
	f_{c_{L_\mu}}Y_*^3 f_{-c_{E_e}} 
	\frac{ev}{\sqrt{2}} 
	\left(\frac{g^2}{2}\ln \frac{R'}{R}\right) 
	\left(\frac{R'v}{\sqrt{2}}\right)^2 
	I_{2\text{MI}W}\,\bar u_{p'}(p+p')^\mu u_p,
	\label{eq:a:amplitude}
\end{align}
where $I_{2\text{MI}W}=-0.31$ is a dimensionless loop integral. Taking $R'v/\sqrt{2} = .17$ and $g^2/2\, \ln(R'/R) = 7.3$, the $a$ coefficient in (\ref{eq:def:a:anarchic}) is 
\begin{align}
	a = -0.065.
\end{align}

\subsection{Calculation of $b$}
\label{sec:calc:b}

As discussed above, the diagrams contributing to $b$ are sensitive to the structure of the anarchic Yukawa matrix relative to that of the non-universal internal bulk fermion masses. For example, if the bulk mass parameters were universal, then the $b$ coefficient operator would be aligned and the off-diagonal element would vanish.
The sign of this off-diagonal term is a function of the initial anarchic matrix so that the $b$ term may interfere constructively or destructively with the $a$ term calculated above. We numerically generate anarchic matrices whose elements have random sign and random values between 0.5 and 2 to determine the distribution of probable Yukawa structures. Such a distribution is peaked about zero so that the choice $b=0$ is a reasonable simplifying assumption. For a more detailed description of the range of bounds accessible by the anarchic RS scenario, one may use the 1$\sigma$ value of $|b|$ as characteristic measure of how large an effect one should expect from generic anarchic Yukawas.

The dominant contributions to the $b$ coefficient are shown in Fig.~\ref{fig:main:diagrams}b. These are the diagram with a charged Goldstone and a $W$ in the loop and the diagram with a $Z$ and a single mass insertion in the loop. Following the analysis in in Appendix~\ref{sec:robustness:vs:alignment}, these diagrams can have zero mode fermions propagating in the loop and hence are sensitive to the bulk mass parameters of the internal fermions being summed in the loop. This, in turn, implies that the diagrams are more robust against alignment upon rotating to the zero mode mass basis. 
%

%
The amplitudes associated with this diagram are 
\begin{align}
         \left.\mathcal M(1\text{MI}Z)\right|_{(p+p')^\mu} &= \frac{i}{16\pi^2} 
	\left(R'\right)^2 f_{c_{L}} Y_E f_{-c_{E}} \frac{ev}{\sqrt{2}}
	\left(g_{Z_L}g_{Z_R}\,\ln\frac{R'}{R}\right)
	\times I_{1\text{MI}Z},
	\\
	\left.\mathcal M(0\text{MI}HW)\right|_{(p+p')^\mu} &= \frac{i}{16\pi^2} 
	\left(R'\right)^2  f_{c_{L}} Y_E  f_{-c_{E}}  \frac{ev}{\sqrt{2}}
	\left(\frac{g^2}{2}\,\ln\frac{R'}{R}\right)\times I_{0\text{MI}HW},
	%
\end{align}
where $g_{Z_{L,R}}$ is the Standard Model coupling of the $Z$ to left- and right-handed leptons respectively. 
The values for the dimensionless integrals are given in (\ref{eq:0MIHW:new}) and (\ref{eq:I:1MIZ}). 
%
%


After scanning over anarchic matrices as defined above, the $1\sigma$ value for the $b$ coefficient is
\begin{align}
	\left| b^{1\sigma}\right| = 0.03.
\end{align}
%
Here we take the $1\sigma$ value of the $b$ coefficient assuming the bulk masses of the minimal model $c_L = c_R$ as a representative benchmark for a plausible general estimate of the generically allowed range of $b$.

\subsection{Modifications in custodial modes}

In Section~\ref{sec:tree:custodial} it was shown that custodial symmetry weakens the bounds from tree-level FCNCs. Since we would like to assess the tension between tree- and loop-level bounds, we should also examine the effect of the additional custodial modes on $\mu\to e\gamma$. These additional diagrams are described by the same topologies as those in Figs.~\ref{fig:a:e:diagrams}--\ref{fig:b:diagrams} but differ by replacing internal lines with custodial bosons and fermions. The expression for the amplitude differs by coupling constants and the use of propagators with different boundary conditions, but not in the overall structure of each amplitude and so are straightforward to extract from the minimal model expressions. 
The leading topologies are unchanged so that it is sufficient to consider the custodial versions of the diagrams in Fig.~\ref{fig:main:diagrams}.

For the two-mass-insertion $W$ diagram, there are two additional diagrams with custodial fermions: one with a $W_L$ and the other with a $W_R$ in the loop. The $P_\text{LR}$ symmetry enforces that the couplings are identical while the different boundary conditions modify the definitions of the internal propagators so that the only difference comes from the value of the dimensionless integral in (\ref{eq:a:amplitude}). The each diagram contributes a dimensionless integral $I=-0.2$, so that the $a$ coefficient is modified to
\begin{align}
	a_\text{cust.} = -0.15.
\end{align}

Custodial diagrams do not contribute to the $b$ coefficient at leading order. For example, one might consider the diagram with a $Z$ loop where the $Z$ is replaced by a $Z_X$, the orthogonal mixture of the custodial $X$ and $W^3_R$ bosons. However, leptons carry no $X$ charge so that the effective coupling is only to right chiral modes. For $\mu_R \to e_L\gamma$, such a diagram would not be allowed. The leading custodial $b$ coefficient diagrams are an order of magnitude smaller than the minimal model diagrams and we shall ignore them in this paper.



\subsection{Constraints and tension}
\label{sec:constraints:and:tension}

\begin{figure}[th]
  \centering
  \subfloat[Minimal model]{\label{fig:gull}\includegraphics[width=0.45\textwidth]{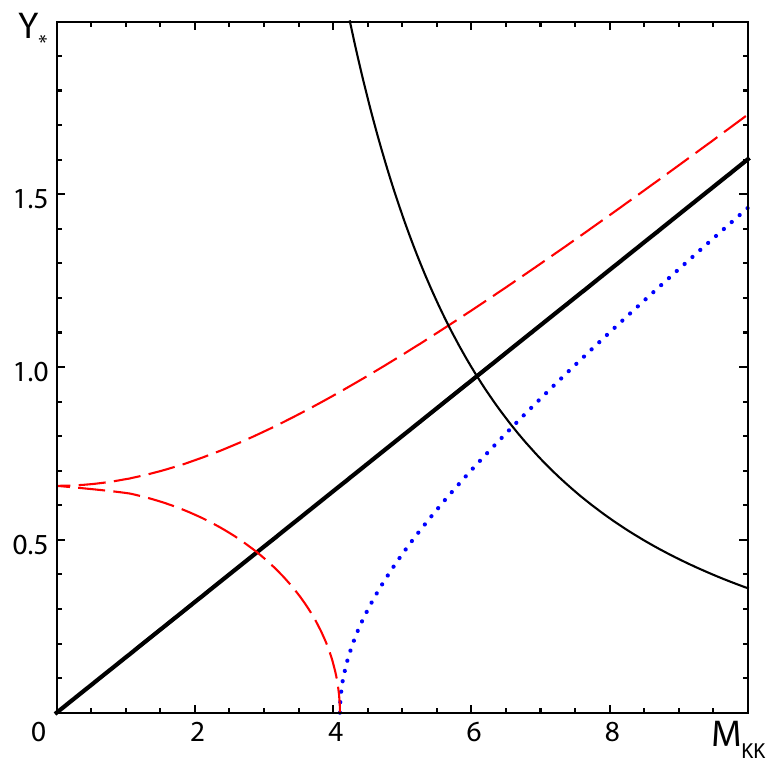}}
                \qquad
  \subfloat[Custodial model]{\label{fig:tiger}\includegraphics[width=0.45\textwidth]{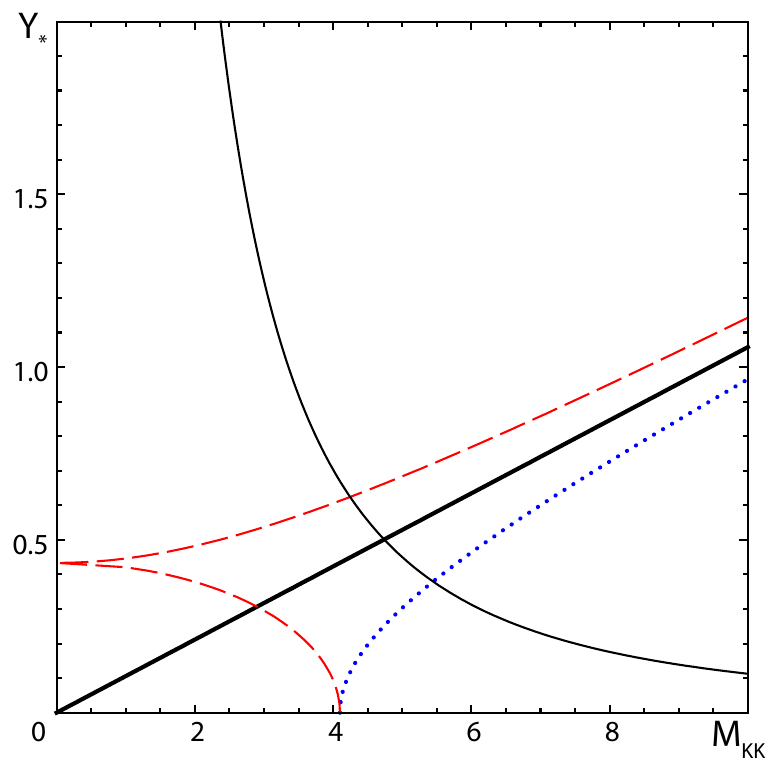}}
  \caption{Bounds on the anarchic Yukawa and KK scales in the minimal (a) and custodial (b) models from tree- and loop-level constraints, (\ref{eq:upper:bound}), (\ref{eq:upper:bound:custodial}), and (\ref{eq:Ystar:bound:with:b}). Each curve rules out the region to its left. The solid hyperbola is the appropriate tree-level bound. The thick solid straight line is the $b=0$ loop-level bound. The red dashed (blue dotted) curve is the loop-level bounds in the case where $b$ has the same (opposite) sign as $a$ and takes its $1\sigma$ magnitude $|b|= |b|_{1\sigma}=0.03$.}
  \label{fig:results} 
\end{figure}

We can now estimate the upper bound on the anarchic Yukawa scale $Y_*$ in (\ref{eq:Ystar:bound:with:b}),
\begin{align}
	\left|aY_*^2+b\right| \left(\frac{3\ {\rm TeV}}{M_\text{KK}}\right)^2 \leq 0.015.\tag{\ref{eq:Ystar:bound:with:b}}
\end{align}
First let us consider the scenario where the $b$ coefficient takes its statistical mean value, $b=0$, and $M_\text{KK}=3\text{ TeV}$. In this case the minimal model suffers a $\mathcal O(10)$ tension between the tree-level lower bound on $Y_*$ and the loop-level upper bound,
\begin{align}
	Y_*>4 
	\hspace{3 cm}
	Y_* < 0.5.
\end{align}
The custodial model slightly alleviates this tension,
\begin{align}
	Y_*>1.25
	\hspace{3 cm}
	Y_* < 0.3.
\end{align}
These discrepancies should be interpreted as an assessment on the extent to which the 5D Yukawa matrices may be generically anarchic. The tension in the bounds above imply that for $M_\text{KK}=3\text{ TeV}$, one must accept some mild tuning in the relative sizes of the 5D Yukawa matrix. This is shown by the hyperbola and solid line in Fig.~\ref{fig:results}.


Alternately, one may ask that assuming totally anarchic Yukawas, what is the minimum value of $M_\text{KK}$ for which the tension is alleviated? In the minimal model the tree- and loop-level bounds allow mutually consistent Yukawa scales for $M_\text{KK}>6$ starting at $Y=1$. Similarly, for the custodial model the tree- and loop-level bounds allow consistent values for $M_\text{KK}> 4.75$ starting at $Y=0.5$. 

Next one may consider the effect of the $b$ coefficient which is sensitive to the particular flavor structure of the anarchic 5D Yukawa matrix relative to the choice of fermion bulk mass parameters. The $1\sigma$ range of $b$ values for randomly generated anarchic matrices is $b \in (-0.03,0.03)$.
%
Because this term is independent of $Y_*$, the value of $b$ can directly constraint the KK scale.
For the $1\sigma$ value this sets $M_\text{KK}\gtrsim 4$ TeV, as can be seen from the intersection of the red dashed lines and blue dotted lines with the horizontal axes in Fig.~\ref{fig:results}.

The most interesting range for $b$, however, is the regime where it can cancel the $a$ term in term in (\ref{eq:Ystar:bound:with:b}). In such a regime the loop level bounds can deviate significantly from the prediction with only the $a$ coefficient, allowing one to relax the constraints on $Y_*$ and $M_\text{KK}$. However, because the $1\sigma$ value of $b$ is an order of magnitude smaller than $a$ in the lepton sector, this region is disfavored by tree-level bounds.
For broad model-building purposes, the key point is that the effect of the $b$ coefficient lines in Fig.~\ref{fig:results} represent the freedom to reduce (or enhance) the loop-level constraints through the misalignment of the anarchic Yukawas relative to the bulk masses. This misalignment comes from the choice of two independent spurions in flavor space and is not a tuning in the hierarchies of the Yukawa matrices.

%

%
In Fig.~\ref{fig:results} the red dashed line shows the bound when $b$ takes its $1\sigma$ magnitude and has an opposite sign from $a$; the cusp at $M_\text{KK}=0$ represents the case where the $a$ and $b$ terms cancel. The blue dotted line shows the case where $b$ takes its $1\sigma$ magnitude and has the same sign as $a$.
What is important to note is that as one takes $|b|$ less than $|b|_{1\sigma}$, these lines continuously converge upon the straight line corresponding to $b=0$ so that any combination of $Y_*$ and $M_\text{KK}$ between the upper red dashed line and the blue dotted line can be plausibly achieved within the anarchic paradigm.
Let us make the caveat that the above values are estimates at $\mathcal O(10\%)$ accuracy. Specific results depend on model-dependent factors such as the extent to which the matrices are anarchic, the relative scale of the charged lepton and neutrino anarchic values, or extreme values for bulk masses. For completeness we provide analytic formulas for the leading and next-to-leading order diagrams in Appendix \ref{app:analytic:expressions}.

\section{Power counting and finiteness}
\label{sec:heuristic}

We now develop an intuitive understanding of the finiteness of this 5D process, highlight some subtleties associated with the KK versus 5D calculation of the loop diagrams\footnote{The finiteness of dipole operators has been investigated in gauge-higgs unified models where a higher-dimensional gauge invariance can render these terms finite \cite{Adachi:2009tz}. Here we do not assume the presence of such additional symmetries.}, and estimate the degree of divergence of the two-loop result.
Our primary tool is na\"ive dimensional analysis, from which we may determine the superficial degree of divergence for a given 5D diagram. Special care is given to the treatment of brane-localized fields and the translation between the manifestly 5D and KK descriptions. 

\subsection{4D and 5D theories of bulk fields}
\label{sec:heursitic:4D5D}

It is instructive to review key properties of $\mu\to e \gamma$ in the Standard Model. This amplitude was calculated by several authors \cite{Lavoura:2003xp, Cheng:1976uq, Petcov:1976ff, Marciano:1977wx, Lee:1977tib}.
Two key features are relevant for finiteness:
    \begin{enumerate}
        \item \textbf{Gauge invariance} cancels the leading order divergences. The Ward identity requires $q_\mu\mathcal M^\mu=0$, where $\mathcal M^\mu$ is the amplitude with the photon polarization peeled off and $q_\mu$ is the photon momentum. This imposes a nontrivial $q$-dependence on $\mathcal M$ and reduces the superficial degree of divergence by one.
        \item \textbf{Lorentz invariance} prohibits divergences which are odd in the loop momentum, $k$. In other words, $\int d^4k\, \slashed k/k^{2n}=0$. After accounting for the Ward identity, the leading contribution to the dipole operator is odd in $k$ and thus must vanish. Specifically, one of the $\slashed{k}$ terms in a fermion propagator must be replaced by the fermion mass $m$.
    \end{enumerate}
Recall that the \textbf{chiral structure} of this magnetic operator requires an explicit internal mass insertion. In the Standard Model this is related to both gauge and Lorentz invariance so that it does not give an additional reduction in the superficial degree of divergence.
Before accounting for these two features, na\"ive power counting in the loop integrals appears to suggest that the Standard Model amplitude is logarithmically divergent from diagrams with two internal fermions and a single internal boson. Instead, one finds that these protection mechanisms force the amplitude to go as $M^{-2}$ where $M$ is the characteristic loop momentum scale.

We can now extrapolate to the case of a 5D theory. First suppose that the theory is modified to include a \textit{noncompact} fifth dimension: then we could trivially carry our results from 4D momentum space to 5D except that there is an additional loop integral. By the previous analysis, this would give us an amplitude that goes as $M^{-1}$ and is thus finite. Such a theory is not phenomenologically feasible but accurately reproduces the UV behavior of a bulk process in a compact extra dimension so long as we consider the UV limit where the loop momentum is much larger than the compactification and curvature scales. This is because the UV limit of the loop probes very small length scales that are insensitive to the compactification and any warping. This confirms the observation that $\mu\to e \gamma$ in Randall-Sundrum models with all fields (including the Higgs) in the bulk is UV--finite~\cite{Agashe:2006iy}. 
%
In the case where there are brane-localized fields, this heuristic picture is complicated since the $\mu\to e\gamma$ loop is intrinsically localized near the brane and is sensitive to its physics; we address this issue below.

\subsection{Bulk fields in the 5D formalism}
\label{sec:bulk:field:power:counting}

We may formalize this power counting in the mixed position/momentum space formalism. This also generalizes the above argument to theories on a compact interval. Each loop carries an integral $d^4k$ and so contributes $+4$ to the superficial degree of divergence. We can now consider how various features of particular diagrams can render this finite.

\begin{enumerate}
\item \textbf{Gauge invariance ($p+p'$)}. 
As argued above and shown explicitly in (\ref{eq:amplitude:general:form}), the Ward identity identifies the gauge invariant contribution to this process to be proportional to $(p+p')^\mu$, which reduces the overall degree of divergence by one. 
%
%
\item \textbf{Bulk Propagators}.
The bulk fermion propagators in the mixed position/momentum space formalism have a momentum dependence of the form $\slashed{k}/k \sim 1$ while the bulk boson propagators go like $1/k$. This matches the power counting from summing a tower of KK modes. Note that this depends on $k=\sqrt{k^2}$ so that the Lorentz invariance in Section~\ref{sec:heursitic:4D5D} for a noncompact extra dimension is no longer valid. 
%
%
\item \textbf{Bulk vertices ($dz$), overall $z$-momentum conservation}.
Each bulk vertex carries an integral over the vertex position which brings down 
an inverse power of the momentum flowing through it. This can be seen from the form of the bulk propagators, which depend on $z$ in the dimensionless combination $kz$ up to overall warp factors. In the Wick-rotated UV limit, the integrands reduce to exponentials so that their integrals go like $1/k$. In momentum space this suppression is manifested as the momentum-conserving $\delta$ function in the far UV limit where the loop momentum is much greater than the curvature scale. 

An alternate and practical way to see the $1/k$ scaling of an individual $dz$ integral comes from the Jacobian as one shifts to dimensionless integration variables,
\begin{align}
	y  = k_ER'  \qquad\qquad\qquad x = k_Ez \label{eq:dimensionless:vars}
\end{align}
so that $y \in [0,\infty]$ plays the role of the loop integrand and $x\in \left[yR/R',y\right]$ plays the role of the integral over the interval extra dimension. These are the natural objects that appear as arguments in the Bessel functions contained in the bulk field propagators, as demonstrated in Appendix \ref{app:warpedGreensFunc:Euc}. In these variables each $dx$ brings down a factor of $1/y$ from the Jacobian of the integration measure. These variables are natural choices because they relate distance intervals in the extra dimension to the scales that are being probed by the loop process. The \textit{physically} relevant distance scales are precisely these ratios.
\item\textbf{Overall $z$-momentum conservation}. We must make one correction to the bulk vertex suppression due to overall $z$-momentum conservation. This is most easily seen in momentum space where  one $\delta$-function from the bulk vertices conserves overall external momentum in the extra dimension and hence does not affect the loop momentum. In mixed position/momentum space this is manifested as one $dz$ integral bringing down an inverse power of only external momenta without any dependence on the loop momentum. We review this in Appendix~\ref{sec:5D:momentum:space}, where we discuss the passage between position and momentum space. The overall $z$-momentum conserving $\delta$-function thus adds one unit to the superficial degree of divergence to account for the previous overcounting of $dz\sim 1/k$ suppressions.

%
%
\item\textbf{Derivative coupling}.\label{item:bulk:suppression:rules}
The photon couples to charged bosons through a derivative coupling which is proportional to the momentum flowing through the vertex. This gives a contribution that is linear in the loop momentum, $k^\mu$. 
%
%
\item\textbf{Chirality: mass insertion, equation of motion}.\label{item:bulk:suppression:chirality} To obtain the correct chiral structure for a dipole operator, each diagram must either have an explicit fermion mass insertion or must make use of the external fermion equation of motion (EOM).
For a bulk Higgs field, each fermion mass insertion carries a $dz$ integral which goes like $1/k$. As described in Section~\ref{sec:warpedXD}, the use of the EOM corresponds to an explicit external mass insertion. Thus fermion chirality reduces the degree of divergence by one unit.

%
%
\end{enumerate}

We may now straightforwardly count the powers of the loop momentum to determine the superficial degree of divergence for the case where the photon is emitted from a fermion (one boson and two fermions in the loop) or a boson (two bosons and one fermion in the loop). The latter case differs from the former in the number of boson propagators and the factor of $k^\mu$ in the photon Feynman rule.
	\begin{center}
	\begin{tabular}{rll}
											 & Neutral 					 & Charged \\
											 & Boson 					 & Boson \\
		Loop integral ($d^4k$) 			\quad& $+4$					\quad& $+4$\\
		Gauge invariance ($p+p'$)		\quad& $-1$                 \quad& $-1$\\
		Bulk fermion propagators		\quad& $\phantom{+}0$       \quad& $\phantom{+}0$\\
		Bulk boson propagator			\quad& $-1$                 \quad& $-2$\\
		Bulk vertices ($dz$)			\quad& $-3$                 \quad& $-3$\\
		Overall $z$-momentum			\quad& $+1$                 \quad& $+1$\\
		Derivative coupling				\quad& $\phantom{+}0$       \quad& $+1$\\
		Mass insertion/EOM				\quad& $-1$                 \quad& $-1$\\
		\hline
		\textit{Total degree of divergence} 	\quad& $-1$			\quad& $-1$
	\end{tabular}
	\end{center}
The $WH^\pm$ diagram in Fig.~\ref{fig:b:diagrams} is a special case since it has neither a derivative coupling nor an additional chirality flip, but these combine to make no net change to the superficial degree of divergence.
We confirm our counting in Section~\ref{sec:heursitic:4D5D} that the superficial degree of divergence for universal extra dimension where all fields propagate in the bulk is $-1$ so that the flavor-changing penguin is manifestly finite.

Before moving on to the case of a brane-localized boson, let us remark that this bulk counting may straightforwardly be generalized to the case of a bulk boson with brane-localized mass insertions. To do this, we note that the brane-localized mass insertion breaks momentum conservation in the $z$ direction and this no longer contributes $+1$ to the degree of divergence. On the other hand, each mass insertion no longer contributes $-1$ from the $dz$ integral so that the changes in the ``overall $z$-momentum'' and ``mass insertion/EOM'' counting cancel out. 
We find that diagrams with a bulk gauge boson and brane-localized mass insertions have the same superficial degree of divergence as the lowest order diagrams in a \textit{bulk} mass insertion expansion.

\subsection{Bulk fields in the KK formalism}
\label{sec:heuristic:KK:bulk}

All of the power counting from the 5D position/momentum space formalism carries over directly to the KK formalism with powers of $m_{\text{KK}}$ treated as powers of $k$. The position/momentum space propagators already carry the information about the entire KK tower as well as the profiles of each KK mode. Explicitly converting from a 5D propagator to a KK reduction,
\begin{align}
	\Delta_{5D}(k, z, z') = \sum_{n}f^{(n)}(z)\Delta_{\text{KK}}^{(n)}(k)f^{(n)}(z'),
\end{align}
where $f^{(n)}$ is the profile of the $n^{\text{th}}$ KK mode. The sum over KK modes is already accounted for in the 5D propagator; for example, for a boson $\Delta^{(n)}_{\text{KK}}\sim 1/k^2$ while $\Delta_{5D} \sim 1/k$. The vertices between KK modes are given by the $dz$ integral over each profile, which reproduces the same counting since each profile depends on $z$ as a function of $m^{(n)}_{\text{KK}}z$. Conservation of $z$-momentum is replaced by conservation of KK number in the UV limit of large KK number.

Indeed, it is almost tautological that the KK and position/momentum space formalisms should match for bulk fields since the process of KK reducing a 5D theory implicitly passes through the position/momentum space construction. This will become slightly more nontrivial in the case of brane-localized fields. We shall postpone a discussion of mixing between KK states until Section~\ref{sec:heuristic:KK:brane}. 


\subsection{Brane fields in the 5D formalism}
\label{sec:heuristic:brane:5D}

The power counting above appears to fail for loops containing a brane-localized Higgs field. 
The brane-localized Higgs propagator goes like $1/k^2$ rather than $1/k$ for the bulk propagator, but this comes at the cost of two vertices that must also be brane-localized, thus negating the suppression from the $dz$ integrals. The charged Higgs has two brane-localized Higgs propagators, but loses a third $dz$ integral from the brane-localized photon emission. Finally, there are no additional contributions from the brane-localized fermion mass insertions nor are there any corrections from the conservation of overall $z$-momentum since it is manifestly violated by the brane-localized vertices (see Appendix~\ref{sec:5D:momentum:space} for a detailed discussion). In the absence of any additional brane effects, both types of loops would be logarithmically divergent, as discussed in \cite{Agashe:2006iy}.

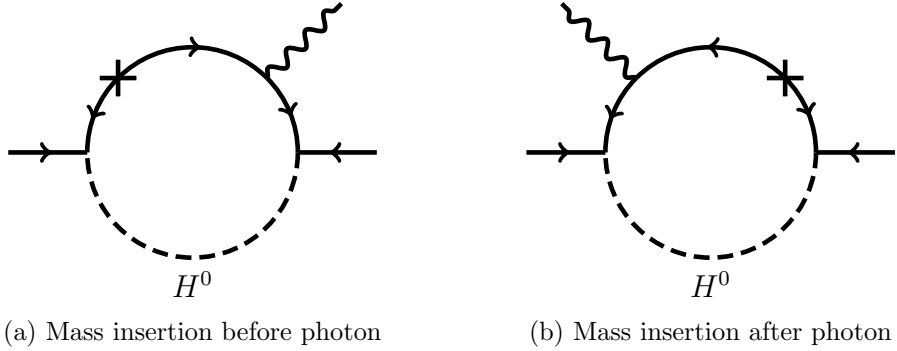
\begin{figure}[t]
  \centering
  \subfloat[Mass insertion before photon]
		{\label{fig:1MIH0:a}
		\begin{tikzpicture}[line width=1.75 pt, scale=1.4] 
			\draw[fermion] (-1.75,0) -- (-1,0);
			\draw[fermionbar] (1,0) -- (1.75,0);
			\draw[fermionbar] (-1,0) arc (180:135:1);
			\draw[fermion] (135:1) arc (135:45:1);
			\draw[fermion] (45:1) arc (45:0:1); 
			\draw[scalarnoarrow] (1,0) arc (0:-180:1);
			\draw[vector] (45:1) -- (45: 2);
			%
			\begin{scope}[rotate=135]
				\begin{scope}[shift={(1,0)}] 
					\clip (0,0) circle (.175cm);
					\draw[fermionnoarrow] (-1,1) -- (1,-1);
					\draw[fermionnoarrow] (1,1) -- (-1,-1);
				\end{scope}	
			\end{scope}
			\node at (0,-1.25) {$H^0$};
		\end{tikzpicture}
	}                
  \qquad \qquad
  \subfloat[Mass insertion after photon]
		{\label{fig:1MIH0:b}
		\begin{tikzpicture}[line width=1.75 pt, scale=1.4] 
			\draw[fermion] (-1.75,0) -- (-1,0);
			\draw[fermionbar] (1,0) -- (1.75,0);
			\draw[fermionbar] (-1,0) arc (180:135:1);
			\draw[fermionbar] (135:1) arc (135:45:1);
			\draw[fermion] (45:1) arc (45:0:1); 
			\draw[scalarnoarrow] (1,0) arc (0:-180:1);
			\draw[vector] (135:1) -- (135: 2);
			%
			\begin{scope}[rotate=45]
				\begin{scope}[shift={(1,0)}] 
					\clip (0,0) circle (.175cm);
					\draw[fermionnoarrow] (-1,1) -- (1,-1);
					\draw[fermionnoarrow] (1,1) -- (-1,-1);
				\end{scope}	
			\end{scope}
			\node at (0,-1.25) {$H^0$};
		\end{tikzpicture}
	}
  \caption{One-mass-insertion neutral scalar diagrams. The leading order $k$-dependence of each diagram cancels when the two are summed together.}
  \label{fig:1MIH0}
\end{figure}

Fortunately, two such brane effects appear. First consider the two neutral Higgs diagrams in Fig.~\ref{fig:a:e:diagrams}. The diagram with no mass insertion requires the use of an external fermion equation of motion which still reduces the superficial degree of divergence by one so that it is finite. The diagram with a single mass insertion is finite in the Standard Model due to a cancellation between the Higgs and neutral Goldstone diagrams, as discussed in Section~\ref{sec:warpedXD}. More generally, even for a single type of brane-localized field, there is a cancellation between diagrams in Fig.~\ref{fig:1MIH0} where the photon is emitted before and after the mass insertion.
This can be seen by writing down the Dirac structure coming from the fermion propagators to leading order in the loop momentum,
\begin{align}
	\mathcal M_{a} & \quad\sim \quad
	\slashed{k}\gamma^\mu \slashed{k}\slashed{k} - k\gamma^\mu k \slashed{k}
	\quad = \quad
	k^2 \left( 
	\slashed{k}\gamma^\mu  - \gamma^\mu \slashed{k}
	\right) \label{eq:brane:higgs:chiral:cancellation:a}
	\\
	\mathcal M_{b} & \quad\sim \quad
	\slashed{k} \slashed{k}\gamma^\mu\slashed{k} -  \slashed{k} k \gamma^\mu k
	\quad=\quad
	k^2 \left( 
	\gamma^\mu\slashed{k}  -  \slashed{k}\gamma^\mu
	\right) \label{eq:brane:higgs:chiral:cancellation:b}
\end{align}
The terms with three factors of $\slashed{k}$ are contributions where ``correct-chirality'' fermions propagate into the bulk, while the terms with only one $\slashed{k}$ are contributions where ``wrong-chirality'' fermions propagate into the bulk. The structure of the latter terms comes from the $\gamma^5\partial_z$ term in the Dirac operator. The structures above multiply scalar functions which, to leading order in $k$, are identical for each term. From the Clifford algebra it is clear that (\ref{eq:brane:higgs:chiral:cancellation:a}) and (\ref{eq:brane:higgs:chiral:cancellation:b}) cancel so that the contribution that is nonvanishing in the UV must be next-to-leading order in the loop momentum. In Appendix~\ref{app:Finiteness} this cancellation is connected to the chiral boundary conditions on the brane and is demonstrated with explicit flat-space fermion propagators. We thus find that the brane-localized neutral Higgs diagrams have an additional $-1$ contribution to the superficial degree of divergence. 

Next we consider the charged Goldstone diagrams. These diagrams have an additional momentum suppression coming from a positive power of the charged Goldstone mass $M_W^2$ appearing in the numerator due to a cancellation within each diagram. In fact, we have already seen in Section~\ref{sec:calc:a} how such a cancellation appears. 
For the single-mass-insertion charged Goldstone diagram in Fig.~\ref{fig:a:n:diagrams}, we saw in (\ref{eq:1MIHpm:cancellation}) that the form of the 4D scalar propagators and the photon-scalar vertex cancels the leading-order loop momentum term multiplying the required $(p+p')^\mu$. The cancellation introduces an additional factor of $M_W^2/(k^2-M_W^2)$ so that the superficial degree of divergence is reduced by two. Note that the position/momentum space propagators for a \textit{bulk} Higgs have a different form than that of the 4D brane-localized Higgs and do not display the same cancellation. In the KK picture this is the observation that the cancellation in (\ref{eq:1MIHpm:cancellation}) takes the form $M_\text{KK}^2/(k^2-M_\text{KK}^2)$, which does not provide any suppression for heavy KK Higgs modes.

Finally, the diagrams where the photon emission vertex mixes the $W$ and brane-localized charged Goldstone are special cases. The photon vertex carries neither a $dz$ integral nor a $k^\mu$ Feynman rule and hence makes no net contribution to the degree of divergence. A straightforward counting including the brane-localized Goldstone, bulk $W$, and the single bulk vertex thus gives a degree of divergence of $-1$. 

We summarize the power counting for a brane-localized Higgs as follows:
\begin{center}
\begin{tabular}{rlll}
										 & Neutral 					 & Charged 					 & $W$--$H^\pm$ \\
										 & boson 					 & boson              		 & mixing \\
	Loop integral ($d^4k$) 			\quad& $+4$					\quad& $+4$                \quad& $+4$\\
	Gauge invariance ($p+p'$)		\quad& $-1$                 \quad& $-1$                \quad& $-1$\\
	Brane boson propagators 		\quad& $-2$                 \quad& $-4$                \quad& $-2$\\
	Bulk boson propagator			\quad& $\phantom{+}0$       \quad& $\phantom{+}0$      \quad& $-1$\\
	Bulk vertices ($dz$)			\quad& $-1$                 \quad& $\phantom{+}0$      \quad& $-1$\\
	Photon Feynman rule				\quad& $\phantom{+}0$       \quad& $+1$                \quad& $\phantom{+}0$\\
	Brane chiral cancellation		\quad& $-1$ 	    		\quad& $\phantom{+}0$      \quad& $\phantom{+}0$\\
	Brane $M_W^2$ cancellation		\quad& $\phantom{+}0$		\quad& $-2$  			   \quad& $\phantom{+}0$\\
	\hline
	\textit{Total degree of divergence} 	\quad& $-1$			\quad& $-2$					\quad& $-1$
\end{tabular}
\end{center}
It may seem odd that the brane-localized charged Higgs loop has a different superficial degree of divergence than the other 5D cases, which heretofore have all been $-1$. This, however, should not be surprising since the case of a brane-localized Higgs is manifestly different from the universal extra dimension scenario. It is useful to think of the brane-localized Higgs as a limiting form of a KK reduction where the zero mode profile is sharply peaked on the IR brane. The difference between the bulk and brane-localized scenarios corresponds to whether or not one includes the rest of the KK tower. 

\subsection{Brane fields in the KK formalism}
\label{sec:heuristic:KK:brane}

Let us now see how the above power counting for the brane-localized Higgs manifests itself in the Kaluza-Klein picture \cite{Agashe:2006iy}. Observe that this power counting for both the $W$--$H^\pm$ and the charged boson loops are trivially identical to the 5D case due to the arguments in Section~\ref{sec:heuristic:KK:bulk}. For example, the $M_W^2$ cancellation is independent of how one treats the bulk fields. The neutral Higgs loop, however, is somewhat subtle since the ``chiral cancellation'' is not immediately obvious in the KK picture.


We work in the mass basis where the fermion line only carries a single KK sum (not independent sums for each mass insertion) and the zero mode photon coupling preserves KK number due to the flat $A^{(0)}$ profile. In this basis the internal fermion line carries one KK sum and it is sufficient to show that for a single arbitrarily large KK mode the process scales like $1/M_{\text{KK}}^2$. The four-dimensional power counting in Section~\ref{sec:heursitic:4D5D} appears to give precisely this, except that Lorentz invariance no longer removes a degree of divergence. This is because this suppression came from the replacement of a loop momentum $\slashed{k}$ by the fermion mass $m$. For an arbitrarily large KK mode, the fermion mass itself is the loop momentum scale and so does not reduce the degree of divergence. 
%
In the absence of any additional suppression coming from the mixing of KK modes, it would appear that the KK power counting only goes like $1/M_{\text{KK}}$ so that the sum over KK modes should be logarithmically divergent, in contradiction with the power counting for the same process in the 5D formalism.

We shall now show that the pair of Yukawa couplings for the neutral Higgs also carries the expected $1/k$ factor that renders these diagrams finite and allows the superficial degrees of divergence to match between the KK and 5D counting.
It is instructive to begin by defining a basis for the zero and first KK modes in the weak (chiral) basis. We denote left (right) chiral fields of KK number $a$ by $\chi^{(a)}_{L,R}$ $(\psi^{(a)}_{L,R})$ where the $L,R$ refers to SU(2)$_L$ doublets and singlets respectively. We can arrange these into vectors
\begin{align}
    \chi = \left(\chi_{L_i}^{(0)},\chi_{R_i}^{(1)},\chi_{L_i}^{(1)}\right)
    \quad\quad\quad\quad\quad\quad
    \psi = \left(\psi_{R_i}^{(0)},\psi_{R_i}^{(1)},\psi_{L_i}^{(1)}\right),\label{eq:KK:fields}
\end{align}
where $i$ runs over flavors. It is helpful to introduce a single index $J=3a+i$ where $i=1,2,3$ according to flavor and $a=0,1,2$ according to KK mode (writing $a=2$ to mean the \textit{first} KK mode with opposite chirality as the zero mode). Thus the external muon and electron are $\chi_2$ and $\psi_1$ respectively, while an internal KK mode takes the form $\chi_J$ or $\psi_J$ with $J>3$.
This convention in (\ref{eq:KK:fields}) differs from that typically used in the literature (e.g.\ \cite{Agashe:2006iy}) in the order of the last two elements of $\psi$. This basis is useful because the KK terms are already diagonal in the mass matrix ($\psi M \chi + \text{h.c.}$),
\begin{align}
    M =
    \begin{pmatrix}
        m^{11} & 0 & m^{13}\\
        m^{21} & M_{\text{KK},1}  & m^{23}\\
        0 & 0 & M_{\text{KK},2}
    \end{pmatrix}
    \label{eq:KK:mass}
\end{align}
where each element is a $3\times 3$ block in flavor space and we have written
\begin{align}
    m = \frac{v}{\sqrt{2}}  f^{(a)}_{R_i} Y_* f^{(b)}_{L_j} \ll M_{\text{KK}},
\end{align}
with indices as appropriate and $M_{\text{KK}}$ diagonal. 
Let us define $\epsilon = v/M_{\text{KK}}$ to parameterize the hierarchies in the mass matrix. For a bulk Higgs, these terms are replaced by overlap integrals and the $M_{32}$ block is nonzero, though this does not affect our argument. Note that $M_{\text{KK},1}$ and $M_{\text{KK},2}$ are typically not degenerate due to $\mathcal{O}(m)$ differences in the doublet and singlet bulk masses. In the gauge eigenbasis the Yukawa matrix is given by
\begin{align}
    y = \left.\frac{\sqrt{2}}{v} M\right|_{M_\text{KK}=0}
    \sim
    \begin{pmatrix}
        1 &  0 &  1\\
        1 & 0 & 1\\
        0 & 0 & 0
    \end{pmatrix},
\end{align}
where we have assumed $f_L, f_R, Y_* \sim \mathcal O(1)$ for simplicity since the hierarchies in the $f^{(0)}$s do not affect our argument. The $1$ elements thus refer to blocks of the same order of magnitude that are \textit{not} generically diagonal. The 0 blocks must vanish by gauge invariance and chirality.

\begin{figure}[t]
  \centering
	\begin{tikzpicture}[line width=1.75 pt, scale=1.7] 
	\draw[fermion] (-2,0) -- (-1,0);
	\draw[fermionbar] (-1,0) -- (0,0);
	\draw[fermion] (0,0) -- (1,0);
	\draw[fermionbar] (1,0) -- (2,0);
	\draw[scalarnoarrow] (-1,0) arc (180:150:1);
	\draw[scalarnoarrow] (1,0) arc (0:30:1);
	\begin{scope} 
		\clip (0,0) circle (.1cm);
		\draw[fermionnoarrow] (-1,1) -- (1,-1);
		\draw[fermionnoarrow] (1,1) -- (-1,-1);
	\end{scope}
	\node[left] at (-2,0.1) {$\displaystyle{ \hat\chi_2 }$};
	\node[right] at (2,0.1) {$\displaystyle{ \hat\psi_1 }$};
	\node[above] at (-.5,0.1) {$\displaystyle{ \hat\psi_J }$};
	\node[above] at (.5,0.1) {$\displaystyle{ \hat\chi_J }$};
	\node[below] at (-1,-0) {$\displaystyle{ \hat y_{2J} } $};
	\node[below] at (0,-0.1) {$\displaystyle{ M_{JJ} } $};
	\node[below] at (1,-0) {$\displaystyle{ \hat y_{J1} } $};
	\end{tikzpicture}
  \caption{The fermion line in the mass basis for diagrams with an internal KK mode ($J>3$). For simplicity we do not show the internal photon insertion.}
  \label{fig:KK:yukawa}
\end{figure}
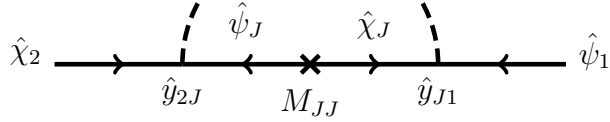

We now rotate the fields in (\ref{eq:KK:fields}) to diagonalize the mass matrix (\ref{eq:KK:mass}); we indicate this by a caret, e.g.\ $\hat \chi$. In this basis the Yukawa matrix is also rotated $y\to \hat y$. The fermion line for this process is shown in Fig.~\ref{fig:KK:yukawa}; the Yukawa dependence of the amplitude is
\begin{align}
	\mathcal M \sim \hat y_{1J} \hat y_{J2}.\label{eq:KK:H0:y:dependence:of:M}
\end{align}
First let us note that in the unrealistic case where $\hat y = y$, one of the Yukawa factors in (\ref{eq:KK:H0:y:dependence:of:M}) is identically zero for all internal KK modes, $J>3$.
%
One might then expect that the mass rotation would induce a mixing of the zero modes with the KK modes that induces $\mathcal O(\epsilon)$ blocks into the Yukawa matrix,
\begin{align}
    \hat{y} \stackrel{?}{\sim}
    \begin{pmatrix}
        1 &  \epsilon &  1\\
        1 & \cdots & \cdots\\
        \epsilon & \cdots & \cdots
    \end{pmatrix}. \label{eq:KK:yukawa:mass:basis:wrong}
\end{align}
If this were the case then the product $\hat y_{1J} \hat y_{J2}$ would not vanish, but would be proportional to $\epsilon \sim 1/M_{\text{KK}}$, which is precisely the KK dependence that we wanted to show. While this intuition is correct and captures the correct physics, the actual Yukawa matrix in the mass basis has the structure (c.f.\ (67) in \cite{Agashe:2006iy})
\begin{align}
    \hat{y} \sim
    \begin{pmatrix}
        1 &  1+\epsilon &  -1+\epsilon\\
        1+\epsilon & \cdots & \cdots\\
        1-\epsilon & \cdots & \cdots
    \end{pmatrix}. \label{eq:KK:yukawa:mass:basis}
\end{align}
The new $\mathcal O(1)$ elements come from the large rotations induced by the $m^{21}$ and $m^{13}$ blocks. These factors cancel out so that we still have the desired $\hat y_{1J} \hat y_{J2} \sim \epsilon$ relation. Physically this is because these $\mathcal O(1)$ factors come from the ``large'' rotation from chiral zero modes to light Dirac SM fermions. Thus they represent the ``wrong-chirality'' coupling of the external states induced by the usual mixing of Weyl states from a Dirac mass. This does \textit{not} include the mixing with the heavy KK modes, which indeed carries the above $\epsilon$ factors so that the final result is 
\begin{align}
	\hat y_{1J} \hat y_{J2} \sim \epsilon\sim \frac{1}{M_{\text{KK}}},
\end{align}
giving the correct $-1$ contribution to the superficial degree of divergence for the neutral Higgs diagrams to render them manifestly finite. 

%
%

A few remarks are in order. 
First let us emphasize again that promoting the Higgs to a bulk field makes the 3--2 block of the $y$ matrix nonzero. This does not affect the above argument so that the KK decomposition confirms the observation that the amplitude with a bulk Higgs is also finite \cite{Agashe:2006iy}. Of course, for a bulk Higgs the power counting in Section~\ref{sec:bulk:field:power:counting} gives a more direct check of finiteness.
Next, note that without arguing the nature of the zeros in the gauge basis Yukawa matrix or the physical nature of the $\epsilon$ mixing with KK modes, it may appear that the $1/M_{\text{KK}}$ dependence of $\hat y_{1J}\hat y_{J 2}$ requires a ``miraculous'' fine tuning between the matrix elements of (\ref{eq:KK:yukawa:mass:basis}). Our discussion highlights the physical nature of this cancellation as the mixing with heavy states that is unaffected by the $\mathcal O(1)$ mixing of light chiral states.

Finally, let us point out that the above arguments are valid for the neutral Higgs diagram where $y=y_E$, the charged lepton Yukawa matrix. The analogous charged Higgs diagram contains neutrino Yukawa matrices $y_N$ so that there is no additional $1/k$ from mixing.
%
%

\subsection{Matching KK and loop cutoffs}

There is one particularly delicate point in the single-mass-insertion neutral Higgs loop in the KK reduction that is worth pointing out because it highlights the relation between the KK scales $M^{(n)}_{\text{KK}}$ and the 5D loop momentum. 
To go from the 5D to the 4D formalism we replace our position/momentum space propagators with a sum of Kaluza-Klein propagators,
\begin{align}
    \Delta_{5\text{D}}(k,z,z') = \sum_{n=0}^{N}f^{(n)}(z) \frac{\slashed{k}+M_n}{k^2-M_n^2}f^{(n)}(z').
\end{align}
The full 5D propagator is exactly reproduced by summing the infinite tower of states, $N\to \infty$. More practically, the 5D propagator with characteristic momentum scale $k$ is well-approximated by at least summing up to modes with mass $M_n\approx k$. Modes that are much heavier than this decouple and do not give an appreciable contribution.
Thus, when calculating low-energy, tree-level observables in 5D theories, it is sufficient to consider only the effect of the first few KK modes. On the other hand, this means that one must be careful in loop diagrams where internal lines probe the UV structure of the theory. In particular, significant contributions from internal propagators near the threshold $M_n\approx k$ would be missed if one sums only to a finite KK number while taking the loop integral to infinity. This is again a concrete manifestation of the remarks below (\ref{eq:dimensionless:vars}) that the length scales probed by a process depend on the characteristic momentum scale of the process. 

Indeed, a Kaluza-Klein decomposition for a single neutral Higgs yields
\begin{align}
    \left|\mathcal M\right|_{(p+p')^\mu} = \frac{gv}{16\pi^2} f_\mu f_{-e} \bar u_e (p+p')^\mu u_\mu \times \frac{1}{M^2}
    \left[c_0 + c_1\left(\frac{v}{M}\right)^2 + \mathcal O\left(\frac{v}{M}\right)^3\right]
\end{align}
for some characteristic KK scale $M\approx M_{\text{KK}}$ and dimensionless coefficients $c_i$ that include a loop integral and KK sums. In order to match the 5D calculation detailed above, we shall work in the mass insertion approximation so that there are now two KK sums in each coefficient. The leading $c_0$ term is especially sensitive to the internal loop momentum cutoff $\Lambda$ relative to the internal KK masses,
\begin{align}
    c_0=  -\lambda^2\sum_{n=1}^{N} \sum_{m=1}^{N}
    \frac{
        \lambda^2\left(n^2+m^2\right) + 2  n^2 m^2
    }{
        4\left(n^2+\lambda^2\right)^2 \left(m^2+\lambda^2\right)^2
    }
    \equiv -\frac{1}{\lambda^2} \sum_{n=1}^{N} \sum_{m=1}^{N} \hat c_0(n,m),
    \label{eq:c0:matching:KKsum:loopintegral}
\end{align}
where we have written mass scales in terms of dimensionless numbers with respect to the mass of the first KK mode: $M_n \sim n M_\text{KK}$ and $\Lambda \sim \lambda M_{\text{KK}}$. It is instructive to consider the limiting behavior of each term $\hat c(n,m)$ for different ratios of the KK scale (assume $n=m$) to the cutoff scale $\lambda$:
\begin{align}
    \hat c_0(n,n) \longrightarrow \left(\frac{n}{\lambda}\right)^2
    & \quad\quad\quad \text{for } n \ll \lambda\\
    \hat c_0(n,n) \longrightarrow \left(\frac{n}{\lambda}\right)^0
    & \quad\quad\quad \text{for } n \approx \lambda\\
    \hat c_0(n,n) \longrightarrow \left(\frac{\lambda}{n}\right)^{4}
    & \quad\quad\quad \text{for } n \gg \lambda.
\end{align}
We see that the dominant contribution comes from modes whose KK scale is near the loop momentum cutoff while the other modes are suppressed by powers of the ratio of scales. In particular, if one calculates the loop for any internal mode of \textit{finite} KK number while taking the loop cutoff to infinity, then the $c_0$ contribution vanishes because the $n\approx \lambda$ contributions are dropped. From this one would incorrectly conclude that the leading order term is $c_1$ and that the amplitude is orders of magnitude smaller than our 5D calculation. Thus one cannot consistently take the 4D momentum to infinity without simultaneously taking the 5D momentum (i.e.\ KK number) to infinity. Or, in other words, one must always be careful to include the nonzero contribution from modes with $n\approx \lambda$.  One can see from power counting on the right-hand side of (\ref{eq:c0:matching:KKsum:loopintegral}) that so long as the highest KK number $N$ and the dimensionless loop cutoff $\lambda$ are matched, $c_0$ gives a nonzero contribution even in the $\lambda \to \infty$ limit.

This might seem to suggest UV sensitivity or a nondecoupling effect\footnote{Further discussion of these points can be found in the appendix of \cite{Blanke:2012tv}.}. However, we have already shown that $\mu\to e \gamma$ is UV-finite in 5D. Indeed, our previous arguments about UV finiteness tell us that the overall contribution to the amplitude from large loop momenta (and hence high KK numbers) must become negligible; we see this explicitly in the UV limit of (\ref{eq:c0:matching:KKsum:loopintegral}). The key statement is that the KK scale and the UV cutoff of the loop integral must be \textit{matched}, $N \gtrsim \lambda$. This can be understood as maintaining momentum-space rotational invariance in the microscopic limit of the effective theory (much smaller than the curvature scale). Further, the prescription that one must match our KK and loop cutoffs $N\gtrsim \lambda$ is simply the statement that we must include all the available modes of our effective theory. It does \textit{not} mean that one must sum a large number of modes in an effective KK theory. In particular, one is free to perform the loop integrals with a low cutoff $\Lambda \sim M_{\text{KK}}$ so that only a single KK mode runs in the loop. This result gives a nonzero value for $c_0$ which matches the order of magnitude of the full 5D calculation and hence confirms the decoupling of heavy modes. 



\subsection{Two-loop structure}
\label{sec:heuristic:twoloop}

As with any 5D effective theory, the RS framework is not UV complete. This nonrenormalizability means that it is possible for processes to be cutoff-sensitive. Since an effective $\mu\to e\gamma$ operator (in the sense of Appendix~\ref{app:5D:EFT}) cannot be written at tree level, there can be no tree-level counter term and so we expect the process to be finite at one-loop order, as we have indeed confirmed above. In principle, however, higher loops need not be finite.

The one-loop analysis presented thus far assumes that we may work in a regime where the relevant couplings are perturbative. In other words, we have assumed that higher-loop diagrams are negligible due to an additional $g^2/16\pi^2$ suppression, where $g$ is a generic internal coupling. This naturally depends on the divergence structure of the higher-loop diagrams. If such diagrams are power-law divergent then it is possible to lose this window of perturbativity even for relatively low UV cutoff $\Lambda\sim M_{\text{KK}}$. We have shown that even though na\"ive dimensional analysis suggests that the $\mu\to e \gamma$ amplitude should be linearly divergent in 5D, the one-loop amplitudes are manifestly finite. 

\begin{figure}[t]
  \centering
		\begin{tikzpicture}[line width=1.75 pt, scale=1.4] 
			\draw[fermion] (-2,0)--(-1,0);
			\draw[fermion] (-1,0) arc (180:270:1);
			\draw[fermion] (0,-1) arc (-90:90:.5) arc (270:90:.5);
			\draw[fermion] (0,1) arc (90:0:1);
			\draw[fermion] (1,0) --(2,0);
			\draw[dotted] (-1,0) arc (180:90:1);
			\draw[dotted] (1,0) arc (0:-90:1);
		\end{tikzpicture}
  \qquad \qquad
		\begin{tikzpicture}[line width=1.75 pt, scale=1.4] 
			\draw[fermion] (-2.5,0) -- (-1.5,0);
			\draw[fermion] (-1.5,0) -- (-.5,0);
			\draw[fermion] (-.5,0) -- (.5,0);
			\draw[fermion] (.5,0) -- (1.5,0);
			\draw[fermion] (1.5,0) -- (2.5,0);
			\draw[dotted] (-1.5,0) arc (180:0:1.5);
			\draw[dotted] (-.5,0) arc (180:0:.5);
		\end{tikzpicture}
  \caption{Yin-Yang and double rainbow topologies of two-loop diagrams. The dotted line represents either a gauge or Higgs boson. We have omitted the photon emission and an odd number of mass insertions.}
%
  \label{fig:two:loop:topology}
\end{figure}
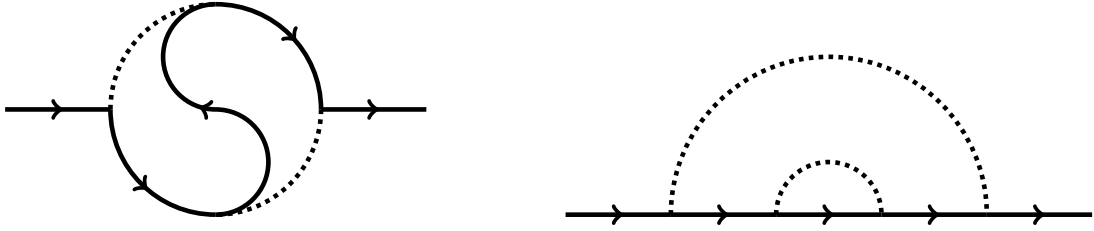

Here we argue that the two-loop diagrams should be no more than logarithmically divergent for bulk bosons
so that there is an appreciable region of parameter space where the process is indeed perturbative and the one-loop analysis can be trusted. This case is also addressed in \cite{Agashe:2006iy}. The relevant topologies are shown in Fig.~\ref{fig:two:loop:topology}.
In this case, the power counting arguments that we have developed in this section carry over directly to the two-loop diagrams:
\begin{center}
\begin{tabular}{rll}
	Loop integrals ($d^4k$) 		\quad& $+8$			   \\
	Gauge invariance ($p+p'$)		\quad& $-1$            \\
	Bulk boson propagators			\quad& $-2$            \\
	Bulk vertices ($dz$)			\quad& $-5$            \\
	\hline                                                 
	\textit{Total degree of divergence} 	\quad& $\phantom{+}0$	   
\end{tabular}
\end{center}
We find that the superficial degree of divergence is zero so that the process is, at worst, logarithmically divergent. 

The power counting for the brane-localized fields is more subtle, as we saw above. Na\"ive power counting suggests that the two-loop, brane-localized diagrams are no more than quadratically divergent. However, just as additional cancellations manifested themselves in the one-loop, brane-localized case, it may not be unreasonable to expect that those  cancellations might carry over to the two-loop diagrams. 
%
Checking the existence of such cancellations requires much more work we leave this to a full two-loop calculation.



\section{Outlook and Conclusion}
\label{sec:conclusion}

We have presented a detailed calculation of the $\mu\to e\gamma$ amplitude in a warped RS model using the mixed position/momentum representation of 5D propagators and the mass insertion approximation, where we have assumed that the localized Higgs VEV is much smaller than the KK masses in the theory. 
Our calculation reveals potential sensitivity to the specific flavor structure of the anarchic Yukawa matrices since this affects the relative signs of coefficients that may interfere constructively or destructively. We thus find that while generic flavor bounds can be placed on the lepton sector of RS models, one can systematically adjust the structure of the $Y_E$ and $Y_N$ matrices to alleviate the bounds while simultaneously maintaining anarchy.
In other words, there are regions of parameter space which can improve agreement with experimental constraints without fine tuning.
Conversely, one may generate anarchic flavor structures which---for a given KK scale---cannot satisfy the $\mu\to e\gamma$ constraints for \textit{any} value of the anarchic scale $Y_*$. Over a range of randomly generated anarchic matrices, the parameter controlling this $Y_*$-independent structure has a mean value of zero and a $1\sigma$ value which can push the KK scale to 4 TeV. 

It is interesting to consider the case where $M_{\text{KK}}=3$ TeV where KK excitations are accessible to the LHC. When the $b$ coefficient takes its statistical mean value, $b=0$, the minimal model suffers a $\mathcal O(10)$ tension between the tree-level lower bound on $Y_*$ and the loop-level upper bound,
\begin{align}
	Y_*>4 
	\hspace{3 cm}
	Y_* < 0.5.
\end{align}
This tension is slightly alleviated in the custodial model,
\begin{align}
	Y_*>1.25
	\hspace{3 cm}
	Y_* < 0.3.
\end{align}
Thus for $M_\text{KK}=3$ TeV one must one must accept some mild tuning in the relative sizes of the 5D Yukawa matrix. Fig.~\ref{sec:constraints:and:tension} summarizes the bounds including the effect of the $b$ coefficient.

On the other hand, we know that anarchic models generically lead to small mixing angles (see however~\cite{Raman}). These fit the observed quark mixing angles well but are in stark contrast with the lepton sector where neutrino mixing angles are large, ${\cal O}(1)$, and point to additional flavor structure in the lepton sector. 
For example in~\cite{A4} a bulk $A_4$ non-Abelian discrete symmetry is imposed on the lepton sector. This leads to a successful explanation of both the lepton mass hierarchy and the neutrino mixing angles (see also~\cite{delAguila:2010vg}) while all tree-level lepton number-violating couplings are absent, so the only bound comes from the $\mu\to e\gamma$ amplitude.

We have also provided different arguments for the one-loop finiteness of this amplitude which we verified explicitly through calculations. We have illuminated how to correctly perform the power counting to determine the degree of divergence from both the 5D and 4D formalisms. The transition between these two pictures is instructive and we have demonstrated the importance of matching the number of KK modes in a 4D EFT to any 4D momentum cutoff in loop diagrams. The power-counting analysis can be particularly subtle for the case of brane-localized fields and we have shown how one-loop finiteness can be made manifest. Finally, we have addressed the existence of a perturbative regime in which these one-loop results give the leading result by arguing that the bulk field two-loop diagrams should be at most logarithmically divergent and that it is at least feasible that the brane-localized two-loop diagrams may follow this power counting.

In addition to $\mu\to e\gamma$, there is an analogous flavor-changing dipole-mediated process in the quark sector, $b\to s \gamma$ with additional gluon diagrams with the same topology as the $Z$ diagrams described here.
%
Because of operator mixing, connecting the $b\to s\gamma$ amplitude to QCD observables requires the Wilson coefficients for both the photon penguin $C_{7\gamma}$ and the gluon penguin $C_{8g}$. A discussion can be found in~\cite{Agashe:2004cp}, though there it was expected that these penguins would be logarithmically divergent. Further, it would be interesting to note whether the experimental bounds on this process admits the small-$Y_*$ region of parameter space where the $b$ term may be of the same order as the $a$ term. We leave the explicit evaluation of the $b\to s\gamma$ amplitude in warped space to future work~\cite{Blanke:2012tv}.

\section*{Acknowledgements}

We thank Kaustubh Agashe, Monika Blanke, and Bibhushan Shakya for many extended discussions and useful comments on this manuscript. We also thank Kaustubh Agashe, Aleksandr Azatov, Monika Blanke, Yuko Hori, Takemichi Okui, Minho Son, and Lijun Zhu for discussions and, in particular, for pointing out the cancellation between the physical Higgs and neutral Goldstone loops. We thank Martin Beneke, Paramita Dey, and Ju\"rgen Rohwild for pointing out the importance diagrams with an external mass insertion. We would further like to acknowledge helpful conversations with Andrzej Buras, David Curtin, Gilad Perez, Michael Peskin, and Martin Schmaltz.
This research is supported in part by the NSF Grant No.~PHY-0757868. C.C. and Y.G. were also supported in part by  U.S./Israeli BSF grants. P.T. was also supported in part by an NSF graduate research fellowship and a Paul \& Daisy Soros Fellowship for New Americans. P.T. and Y.T. would like to thank the 2009 Theoretical Advanced Studies Institute and the Ku Cha House of Tea in Boulder, Colorado for their hospitality during part of this work.


\appendix

\section{Matching 5D amplitudes to 4D EFTs}
\label{app:5D:EFT}

The standard procedure for comparing the loop-level effects of new physics on low-energy observables is to work with a low-energy effective field theory in which the UV physics contributes to the Wilson coefficient of an appropriate local effective operator by matching the amplitudes of full and effective theories. 
In this appendix we briefly remark on the matching of 5D mixed position/momentum space amplitudes to 4D effective field theories, where some subtleties arise from notions of locality in the extra dimension.

The only requirement on the 5D amplitudes that must match to the 4D effective operator is that they are local in the four Minkowski directions. There is no requirement that the operators should be local in the fifth dimension since this dimension is integrated over to obtain the 4D operator. Thus the 5D amplitude should be calculated with independent external field positions in the extra dimension. Heuristically, one can write this amplitude as a \textit{nonlocal} 5D operator
\begin{align}
	\mathcal O_5(x,z_H,z_L,z_E,z_A) = H_5(x,z_H)\cdot \bar L_5(x,z_L)\,\sigma^{MN}\,E_5(x,z_E)F_{MN}(x,z_A).
\end{align}
Note that this object has mass dimension 8.
In the 5D amplitude the fields are replaced by external state wavefunctions and this is multiplied by a ``nonlocal coefficient'' $c_5(z_H,z_L,z_E,z_A)$ which includes integrals over internal vertices and loop momenta as well as the mixed position/momentum space propagators to the external legs. To match with the low-energy 4D operator we impose that the external states are zero modes and decompose them into 4D zero-mode fields multiplied by a 5D profile $f(z)$ of mass dimension 1/2,
\begin{align}
	\Phi_5(x,z)\to \Phi^{(0)}(x)f^{(0)}(z).
\end{align}
Further, we must integrate over each external field's $z$-position. Thus the 4D Wilson coefficient and operator are given by
\begin{align}
	c_4 \mathcal O_4(x) = \int \left[\prod_{i}dz_i\right] \, c_5(z_H,z_E,z_L,z_A) f_H^{(0)}(z_H)f_E^{(0)}(z_E)f_L^{(0)}(z_L)f_A^{(0)}(z_A)\;  
	H\cdot \bar L\,\sigma^{\mu\nu}\,E F_{\mu\nu},
\end{align}
where the fields on the right-hand side are all zero modes evaluated at the local 4D point $x$. Note that these indeed have the correct 4D mass dimensions, $[\mathcal O_4]=6$ and $[c]=-2$.

Finally, let us remark that we have treated the 5D profiles completely generally. In particular, there are no ambiguities associated with whether the Higgs field propagates in the bulk or is confined to the brane. One can take the Higgs profile to be brane-localized,
\begin{align}
	f_H(z_H) \sim \sqrt{R'}\delta(z-R'),
\end{align}
where the prefactor is required by the dimension of the profiles. With such a profile (or any limiting form thereof) the passage from 5D to 4D according to the procedure above gives the correct matching for brane-localized fields.

\section{Estimating the size of each diagram}
\label{app:estimates}

As depicted in Figs.~\ref{fig:a:e:diagrams}--\ref{fig:b:diagrams}, there are a large number of diagrams contributing to the $a$ and $b$ coefficients even when only considering the leading terms in a mass-insertion expansion. Fortunately, many of these diagrams are naturally suppressed and the dominant contribution to each coefficient is given by the two diagrams shown in Fig.~\ref{fig:main:diagrams}. This can be verified explicitly by using the analytic expressions for the leading and next-to-leading diagrams are given in Appendix~\ref{app:analytic:expressions}. In this appendix we provide some heuristic guidelines for estimating the relative sizes of these diagrams. 

\subsection{Relative sizes of couplings}

First note that after factoring out terms in the effective operator in (\ref{eq:finalop}), Yukawa couplings give order one contributions while gauge couplings give an enhancement of $g_\text{SM}^2 \ln R'/R$, where $g_\text{SM}$ is the appropriate Standard Model coupling. This gives a factor of $\sim 5$ ($7$) enhancement in diagrams with a $W$ over those with a $Z$ ($H$).

\subsection{Suppression mechanisms in diagrams}

Next one can count estimate suppressions to each diagram coming from the following factors

\begin{itemize}
	\item[A.] \textbf{Mass insertion}, $\sim10^{-1}$/insertion. Each fermion mass insertion on an internal line introduces a factor of $\mathcal O(vR')$. This comes from the combination of dimensionful factors in the Yukawa interaction and the additional fermion propagator. 
	\item[B$_1$.] \textbf{Equation of motion}, $\sim10^{-4}$. Higgs diagrams without an explicit chirality-flipping internal mass insertion must swap chirality using the muon equation of motion $\bar u(p) \slashed{p}=m_\mu u(p)$. This gives a factor of $\mathcal O(m_\mu R')$ and is equivalent to external mass insertion that picks up the zero-mode mass.
	\item[B$_2$.] \textbf{External mass insertion}, $\sim10^{-1}$. Alternately, when a loop vertex is in the bulk, an external mass insertion can pick up the diagonal piece of the propagator---see (\ref{eq:flat:5D:2x2:propagator})---representing the propagation of a zero mode into a `wrong-chirality' KK mode. Unlike the off-diagonal piece which imposes the equation of motion, this is only suppressed by the $\mathcal O(vR')$ mentioned above\footnote{We thank Martin Beneke, Paramita Dey, and J\"urgen Rohrwild for pointing this out.}. One can equivalently think of this as an insertion of the KK mass which mixes the physical zero and KK modes.
	
	\item[C.] \textbf{Higgs/Goldstone cancellation}, $\sim 10^{-3}$. The $H^0$ and $G^0$ one-mass-insertion loops cancel up to $\mathcal O\left((m_H^2-m_Z^2)/m_\text{KK}^2\right)$ because the two Goldstone couplings appear with factors of $i$ relative to the neutral Higgs couplings\footnote{We thank Yuko Hori and Takemichi Okui for pointing this out.}.
	\item[D.] \textbf{Proportional to charged scalar mass}, $\sim 10^{-2}$. The leading loop-momentum term in the one-mass-insertion brane-localized $H^\pm$ loop cancels due to the form of the photon coupling relative to the propagators. The gauge-invariant contribution from such a diagram is proportional to $(M_WR')^2$. This is shown explicitly in (\ref{eq:1MIHpm:cancellation}) below.
\end{itemize}

To demonstrate the charged scalar mass proportionality, we note that the amplitude for the one mass insertion charged Higgs diagram in Fig.~\ref{fig:a:n:diagrams} is
\begin{align}
	\mathcal M^\mu &=  - R^2 \left(\frac{R}{R'}\right)^6  \frac{ev}{\sqrt{2}}f_{c_{L_\mu}} Y_*^3 f_{-c_{E_e}} \int \frac{d^4k}{(2\pi)^4}\, \bar u_{p'}\Delta^R_k \Delta^L_k u_p \, \frac{(2k-p-p')^\mu}{[(k-p')^2-M_W^2][(k-p)^2-M_W^2]}.
\end{align}
Remembering that the 5D fermion propagators go like $\Delta \sim \slashed{k}/k$, this amplitude na\"ively appears to be logarithmically divergent. However, the Ward identity forces the form of the photon coupling to the charged Higgs to be such that the leading order term in $k^2$ cancels. This can be made manifest by expanding the charged Higgs terms in $p$ and $p'$,
\begin{align}
	\frac{(2k-p-p')^\mu}{[(k-p')^2-M_W^2][(k-p)^2-M_W^2]} = \frac{(p+p')^\mu}{(k^2-M_W^2)^2}\left[\frac{k^2}{k^2-M_W^2}-1\right] = \frac{M_W^2(p+p')^\mu}{(k^2-M_W^2)^3},\label{eq:1MIHpm:cancellation}
\end{align}
where we have dropped terms of order $\mathcal O(m_\mu^2/M_W^2)$.
Thus see that the coefficient of the gauge-invariant contribution is finite by power counting.
After Wick rotation, this amplitude takes the form
\begin{align}
	\left.\mathcal M^\mu(\text{1MI}H^\pm)\right|_{(p+p')} = \frac{2i}{16\pi^2}(R')^2 f_{c_{L_\mu}}Y_*^3 f_{-c_{E_e}} \frac{ev}{\sqrt{2}} (R'M_W)^2 I_{\text{1MI}H^\pm}\,\bar u_{p'}(p+p')u_p,
\end{align}
where $I_{\text{1MI}H^\pm}$ is a dimensionless integral given in (\ref{app:analytic:expressions}). We see that the amplitude indeed carries a factor of $(M_WR')^2$.

\subsection{Dimensionless integrals}

Estimating the size of dimensionless integrals over the loop momentum and bulk field propagators (such as $I_{\text{1MI}H^\pm}$) is more subtle and is best checked through explicit calculation. However, one may develop an intuition for the relative size of these integrals.

Note that the fifth component of a bulk gauge field naturally has boundary conditions opposite that of the four-vector \cite{Csaki:2005vy} so that the fifth components of Standard Model gauge fields have Dirichlet boundary conditions. This means that diagrams with a $W^5 H^\pm A$ vertex vanish since the brane-localized Higgs and bulk $W^5$ do not have overlapping profiles. Further, loops with fifth components of Standard Model gauge fields and internal mass insertions tend to be suppressed since the mass insertions attach the loop to the IR brane. In the UV limit the loop shrinks towards the brane and has reduced overlap with the fifth component gauge field.

Otherwise the loop integrals are typically $\mathcal O(0.1)$. The particular value depends on the propagators and couplings in the integrand.

\subsection{Robustness against alignment}
\label{sec:robustness:vs:alignment}

As discussed in Section~\ref{sec:calc:b}, the flavor structure of the diagrams contributing to the $b$ coefficient is aligned with the fermion zero-mode mass matrix \cite{Agashe:2004cp, Agashe:2006iy, Kaustubh}. Contributions to this coefficient vanish in the zero mode mass basis in the absence of additional flavor structure from the bulk mass ($c$) dependence of the internal fermion propagators. The diagrams which generally give the largest contribution after passing to the zero mode mass basis are those with with the strongest dependence on the fermion bulk masses. Since zero mode fermion profiles are exponentially dependent on the bulk mass parameter, a simple way to identify potential leading diagrams is to identify those which may have zero mode fermions propagating in the loop.

\begin{figure}[t]
  \centering
	\begin{tikzpicture}[line width=1.75 pt, scale=1.7] 
	\draw[fermionbar] (-3.5,0) -- (-2.5,0);
	\draw[fermion] (-2.5,0) -- (-1,0);
	\draw[fermion] (-1,0) -- (1,0);
	\draw[fermion] (1,0) -- (2,0);
	\draw[vector] (-1,0) arc (180:150:1);
	\draw[vector] (1,0) arc (0:30:1);
	\begin{scope}[shift={(-2.5,0)}] 
		\clip (0,0) circle (.1cm);
		\draw[fermionnoarrow] (-1,1) -- (1,-1);
		\draw[fermionnoarrow] (1,1) -- (-1,-1);
	\end{scope}
	\node[above] at (-3,0.0) {$\displaystyle{ \psi^{(0)} }$};
	\node[above] at (-2.15,0.0) {$\displaystyle{ \chi^{(n)} }$};
	\node[above] at (-1.3,0.0) {$\displaystyle{ \bar\psi^{(n)} }$};
	\node[above] at (-.6,0.0) {$\displaystyle{ \bar\psi^{(m)} }$};
	\node[above] at (.6,0.0) {$\displaystyle{ \chi^{(m)} }$};
	\node[above] at (1.5,0.0) {$\displaystyle{ \chi^{(0)} }$};
	\end{tikzpicture}
  \caption{
Alignment of the external mass insertion diagrams with Standard Model gauge bosons. $\chi$ and $\psi$ are left- and right-chiral Weyl spinors respectively.  The gauge boson vertices don't change fermion chirality so that the internal fermion must be a chirality-flipping KK mode. We have neglected the contribution where the external mass insertion connects two zero mode fermions since this is suppressed by $m_\mu R'$.
}
  \label{fig:external:MI:alignment}
\end{figure}
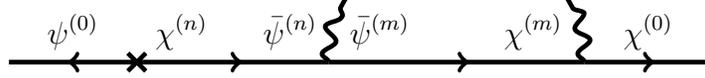

This allows us to neglect diagrams with an external mass insertion and a 4D vector boson in the loop. As shown in Fig.~\ref{fig:external:MI:alignment}, such diagrams do not permit intermediate zero modes to leading order.  Note, however, that diagrams with an external mass insertion and the fifth component of gauge boson are allowed to have zero mode fermions in the loop. Indeed, a diagram with a $W^5$ and $W^\mu$ in the loop would permit zero mode fermions but is numerically small due to the size of the $W^5 A W^\mu$ coupling. The dominant diagrams for the $b$ coefficient are the $H^\pm W^\pm$ loop and the $Z$ loop with an internal mass insertion. In the KK reduction, the misalignment comes from diagrams with zero mode fermions and KK gauge bosons.

%
%

%



\section{Analytic expressions}
\label{app:analytic:expressions}

We present analytic expressions for the leading and next-to-leading diagrams contributing to $\mu\to e\gamma$. We label the diagrams in Figs.~\ref{fig:a:e:diagrams}--\ref{fig:b:diagrams} according to the number of Higgs-induced mass insertions and the internal boson(s). For example, the two-mass-insertion $W$ diagram in Fig.~\ref{fig:main:diagrams}a is referred to as 2MI$W$.
Estimates for the size of each contribution are given in Appendix~\ref{app:estimates}. 
We shall only write the coefficient of the $\bar u_{p'} (p+p')^\mu u_p$ term since this completely determines the gauge-invariant contribution.

\subsection{Dominant diagrams}

As discussed in Section~\ref{sec:warpedXD}, the leading diagrams contributing to the $a$ and $b$ coefficients are
\begin{align}
	\mathcal M(2\text{MI}W)
	& =
	\frac{i}{16\pi^2}(R')^2 
	f_{c_{L_\mu}}Y_E Y_N^\dag Y_N f_{-c_{E_e}} 
	\frac{ev}{\sqrt{2}} 
	\left(\frac{g^2}{2}\ln \frac{R'}{R}\right) 
	\left(\frac{R'v}{\sqrt{2}}\right)^2 
	I_{2\text{MI}W}\\
	\mathcal M(0\text{MI}HW) &= \frac{i}{16\pi^2} 
	\left(R'\right)^2  f_{c_{L}} Y_E  f_{-c_{E}}  \frac{ev}{\sqrt{2}}
	\left(\frac{g^2}{2}\ln \frac{R'}{R}\right)  I_{0\text{MI}HW},
	\label{eq:M:0MIHW}\\
	\mathcal M(1\text{MI}Z) &= \frac{i}{16\pi^2} 
	\left(R'\right)^2 f_{c_{L}} Y_E f_{-c_{E}} \frac{ev}{\sqrt{2}}
	\left(g_{Z_L}g_{Z_R} \ln\frac{R'}{R}\right)
	I_{1\text{MI}Z},
	\label{eq:M:1MIZZ5}
\end{align}%
We have explicitly labeled the 4D (dimensionless) anarchic Yukawa matrices whose elements assumed to take values of order $(Y_E)_{ij} \sim (Y_N)_{ij} \sim Y_*$, but have independent flavor structure. Note that we have suppressed the flavor indices of the Yukawas and the dimensionless integrals.
Diagrams with a neutral boson and a Yukawa structure $Y_E Y_E^\dag Y_E$ also contribute to the $a$ coefficient, but these contributions are suppressed relative to the dominant charged boson diagrams above. These diagrams may become appreciable if one permits a hierarchy in the relative $Y_E$ and $Y_N$ anarchic scales, in which case one should also consider the $Z$ boson diagrams whose analytic forms are given below.
The dimensionless integrals are
\begin{align}
	I_{2\text{MI}W} =& 
	-
	\frac 32
	\int dy\, dx_1 dx_2 dx_3 \, 
	y^3
	\left(\frac{y}{x_1}\right)^{c_L+2}
	\left(\frac{y}{x_2}\right)^{4}
	\left(\frac{y}{x_3}\right)\nonumber\\
	& \phantom{-\frac 32\int } 
	\tF{+}{L}{y}{1}{y} 
	\tF{-}{R}{y}{y}{y} 
	\tDF{-}{L}{y}{y}{2} 
	\tF{+}{L}{y_\mu}{2}{y_\mu} 
	\frac{\partial}{\partial k_E}\left(G_{y}^{13}G_{y}^{32}\right)\label{eq:I:2MIW}\\
	I_{0\text{MIHW}} =& \phantom{+} \int dy\, dx\, 
			\left(\frac{y}{x}\right)^{2+c_L}\,			\Big(\frac{1}{2\sqrt{2}}\,\frac{y^2}{y^2+m_H^2R'^2}\,\tF{+}{L}{y}{1}{y}\,y\,\partial_{k_E}\,G^{xy}_y\Big)\label{eq:0MIHW:new}\\
	I_{1\text{MI}Z} =& -\int dy\, dx_1 dx_2 dx_3 \, 
	\left(\frac{y}{x_1}\right)^{2+c_L}
	\left(\frac{y}{x_2}\right)^{2-c_E}
	\left(\frac{y}{x_3}\right)^{4}
	\left(y\,\partial_{k_E}  G^{12}\right) y^2\times \nonumber\label{eq:I:1MIZ}\\
	&
	\phantom{-\int}
	\Big(
	-\tDF{+}{R}{y}{2}{3} \,
	\tDF{-}{R}{y}{3}{y}\,
	\tF{+}{L}{y}{y}{1} 
	+ 
	\tF{-}{R}{y}{2}{x_3}\,
	\tF{-}{R}{y}{3}{y}\,
	\tF{+}{L}{y}{y}{1} \nonumber
	\\
	&
	\phantom{-\int\Big(}- \tF{-}{R}{y}{2}{y}\,
	\tDF{-}{L}{y}{y}{3}\,
	\tDF{+}{L}{y}{3}{1}
	+ 
	\tF{-}{R}{y}{2}{y}\,
	\tF{+}{L}{y}{y}{3}\,
	\tF{+}{L}{y}{3}{1}
	\Big).
\end{align}
where $x=k_Ez$, $y=k_ER'$, and $y_\mu= m_\mu R'$. The significance of these dimensionless variables is discussed below (\ref{eq:dimensionless:vars}).
The dimensionless Euclidean-space propagator functions $\tilde F$ are defined in 
(\ref{eq:tilde:F:1} -- \ref{eq:tilde:F:2}),
where the upper indices of the $F$ functions define the propagation positions. For example, $F^{R3y}$ represents a propagator from $z=R'$ to $z=z_3$. Similarly, $G_y$ and $\bar G_y$ are defined in (\ref{eq:G:def}) and (\ref{eq:Gbar:def}). 

\subsection{Subdominant $a$ coefficient diagrams}
The diagrams containing a brane-localized Higgs loop are
\begin{align}
	\mathcal M(n\text{MI}H^\pm) &= \frac{i}{16\pi^2} 
	\left(R'\right)^2 f_{c_{L}} Y_E Y_N^\dag Y_N f_{-c_{E}} \frac{ev}{\sqrt{2}}
	I_{n\text{MI}H^\pm},
	\label{eq:M:1MIHpm}\\
	\mathcal M(n\text{MI}H^0) &= \frac{i}{16\pi^2} 
	\left(R'\right)^2 f_{c_L} Y_E Y_E Y_E^\dag f_{-c_{E}} \frac{ev}{\sqrt{2}}
	I_{0\text{MI}H^0}\label{eq:M:nMIH0}.
	\end{align}
Here $n=0,1$ counts the number of internal mass insertions in the diagram. The gauge boson loops are
\begin{align}
    \mathcal M(n\text{MI}Z^{(5)}) &= \frac{i}{16\pi^2} 
	\left(R'\right)^2  f_{c_{L}} Y_E Y_E^\dag Y_E f_{-c_{E}}  \frac{ev}{\sqrt{2}}
	\left(g_{Z_L}g_{Z_R}\ln\frac{R'}{R}\right)\,\left(\frac{v}{\sqrt{2}}R'\right)^2
	I_{n\text{MI}Z^{(5)}},
	\label{eq:M:nMIZ:a}\\
    \mathcal M(2\text{MI}ww) &= \frac{i}{16\pi^2} 
	\left(R'\right)^2  f_{c_{L}} Y_E Y_N^\dag Y_N f_{-c_{E}}  \frac{ev}{\sqrt{2}}
	\left(\frac{g^2}{2}\ln\frac{R'}{R}\right)\,\left(\frac{v}{\sqrt{2}}R'\right)^2\times I_{2\text{MI}ww}.
	\label{eq:M:2MIW5}
\end{align}
Where $n=2,(1+2),3$ with $(1+2)$ referring to a single internal mass insertion and two external mass insertions. 2MI$ww$ represents $2\text{MI}W^5W^5$, $2\text{MI}WW^5$ and $2\text{MI}W^5W$.
%
The dimensionless integrals are  
	\begin{align}
	I_{1\text{MI}H^0} =&\int dy \, dx\, y^2 \left(\frac yx\right)^4\Big[ 
	-2 \tF{+}{L}{y}{y}{x}\,
	\tF{+}{L}{y}{x}{y}\,
	\tF{-}{R}{y}{y}{y}
	\frac{y^2}{y^2+(M_HR')^2}
	\nonumber\\
	&
	+\tF{+}{L}{y}{y}{x}\,
	\tF+Lyxy \,
	\tF-Ryyy
	\frac{y^4}{(y^2+(m_HR')^2)^2}
	-
	\frac 12 \left(y\,\partial_{k_E}\tF+L{y}yx\right)
	\tF+Lyxy \,
	\tF-Ryyy
	\frac{y^2}{y^2+(M_HR')^2}
	\nonumber\\
	&-
	\frac 12 \left(y\,\partial_{k_E} \tDF-L{y}yx\right)
	\tDF+Lyxy \,
	\tF-Ryyy\,
	\frac{1}{y^2+(M_HR')^2}
	+2
	\tF+Lyyy\,
	\tDF+Ryyx\,
	\tDF-Ryxy
	\frac{1}{y^2+(M_HR')^2}
	\nonumber\\
	&
	-\tF+Lyyy \,
	\tDF+Ryyx \,
	\tDF-Ryxy
	\frac{y^2}{(y^2+(M_HR')^2)^2}
	+
	\frac 12 \left(y\,\partial_{k_E} \tF+L{y}yy\right)
	\tDF+Ryyx\,
	\tDF-Ryxy
	\frac{1}{y^2+(M_HR')^2}
	\nonumber \\
	&+
	\tF+Lyyy\,
	\tF-Ryyx\,
	\tF-Ryxy
	\frac{y^2}{y^2+(M_HR')^2}
	+
	\frac 12 \left(y\,\partial_{k_E} \tF+L{y}yy\right)
	\tF-Ryyx \,
	\tF-Ryxy
	\frac{y^2}{y^2+(M_HR')^2}
	\nonumber \\
	&
	+ \frac 12 \tF+Lyyy\,
	\left(y\,\partial_{k_E} \tF-R{y}yy\right)
	\tF-Ryxy
	\frac{y^2}{y^2+(M_HR')^2}
	+
	\frac 12 \tF+Lyyy\,
	\left(y\,\partial_{k_E} \tDF+R{y}yx\right)
	\tDF-R{y}xy 
	\frac{1}{y^2+(M_HR')^2},
	\Big].\label{eq:I:1MIH0}
	\end{align}
\begin{align}
	I_{1\text{MI}H^\pm} =& \int dy \, 
	\tF{+}{L}{y}{y}{y}
	\tF{+}{R}{y}{y}{y}
	\frac{2\,y^5}{(y^2+(M_WR')^2)^3}\label{eq:I:1MIHpm}\\
	I_{0\text{MI}H^\pm} =& \int dy\, 
	\tF-Ryyy 
	\frac{y^5}{(y^2+(M_HR')^2)^3}
	\label{eq:I:0MIHpm}
	\\
	I_{0\text{MI}H^0} =& \int dy \, dx\, y^2 \left(\frac{y}{x}\right)^4
	\tF+Lyyx \, \tF+Lyxy 
	\frac{y^2}{(y^2+(M_HR')^2)^2}
	\label{eq:I:0MIH0}\\
I_{2\text{MIZ}} =& \int dy\, dx_1 dx_2 dx_3 \, 
		\left(\frac{y}{x_1}\right)^{2+c_L}\,
		\left(\frac{y}{x_2}\right)^{4}\,
		\left(\frac{y}{x_3}\right)^{4}\,
		\times \nonumber\\
		&\Big\{y\,\partial_{k_E}\,G^{13}_y\,\tDF{+}{L}{y_{\mu}}{3}{y_{\mu}}\Big[y^2\Big(\tF{+}{L}{y}{12}\,\tF{+}{L}{y}{2}{y}\,\tF{-}{R}{y}{y}{y}\,\tDF{-}{L}{y}{y}{3}+\tF{+}{L}{y}{1}{y}\,\tF{-}{R}{y}{y}{2}\,\tF{-}{R}{y}{2}{y}\,\tDF{-}{L}{y}{y}{3}\nonumber
		\\
		&+\tF{+}{L}{y}{1}{y}\,\tF{-}{R}{y}{y}{y}\,\tDF{-}{R}{y}{2}{y}\,\tF{-}{L}{y}{2}{3}+\tF{+}{L}{y}{1}{y}\,\tF{-}{R}{y}{y}{y}\,\tF{+}{L}{y}{y}{2}\,\tDF{-}{L}{y}{2}{3}\Big)\nonumber
		\\
		&-\Big(\tDF{-}{L}{y}{1}{2}\,\tDF{+}{L}{y}{2}{y}\,\tF{-}{R}{y}{y}{y}\,\tDF{-}{L}{y}{y}{3}+\tF{+}{L}{y}{1}{y}\,\tDF{+}{R}{y}{y}{2}\,\tDF{-}{R}{y}{2}{y}\,\tDF{+}{L}{y}{y}{3}\Big)\Big]\Big\},\label{eq:I:2MIZ}\\	
I_{2\text{MIZ}^5} =&-\int dy\, dx_1 dx_2 dx_3 \, 
		\left(\frac{y}{x_1}\right)^{2+c_L}\,
		\left(\frac{y}{x_2}\right)^{4}\,
		\left(\frac{y}{x_3}\right)^{4}\times\nonumber\\
		&\frac{1}{2} \left[y\,\partial_{k_E}\,\bar{G}^{13}_y\,\tDF{+}{L}{y_{\mu}}{3}{y_{\mu}}\Big(y^2\,\tF{-}{L}{y}{1}{2}\,\tDF{+}{L}{y}{2}{y}\,\tF{+}{R}{y}{y}{y}\Big)\right],\label{eq:I2MIZ5:new}\\
I_{(1+2)\text{MIZ}} =& -\int dy\, dx_1 dx_2 dx_3 \, 
		\left(\frac{y}{x_1}\right)^{4}\,
		\left(\frac{y}{x_2}\right)^{4}\,
		\left(\frac{y}{x_3}\right)^{4}\,
		\times \nonumber\\
		&\Big[\tDF{+}{R}{y_e}{y_e}{1}\,\tDF{+}{L}{y_{\mu}}{2}{y_{\mu}}\,G^{21}_y-\left(4+y\,\partial_{k_E}\right)\Big(\tDF{-}{R}{y}{1}{y}\,\tF{-}{R}{y}{3}{y}\tDF{-}{L}{y}{3}{2}+\tF{+}{R}{y}{1}{3}\,\tDF{-}{R}{y}{3}{y}\,\tDF{-}{L}{y}{3}{2}\nonumber
		\\
		&+\tDF{-}{R}{y}{1}{y}\,\tDF{-}{L}{y}{y}{3}\,\tF{-}{L}{y}{3}{2}+\tDF{-}{R}{y}{1}{y}\,\tF{+}{L}{y}{y}{3}\,\tDF{-}{L}{y}{3}{2}\Big)
		\Big],\label{eq:I:1p2MIZ:new}\\		
I_{(1+2)\text{MIZ}^5} =&- \int dy\, dx_1 dx_2 dx_3 \, 
		\left(\frac{y}{x_1}\right)^{4}\,
		\left(\frac{y}{x_2}\right)^{4}\,
		\left(\frac{y}{x_3}\right)^4\,
		\tDF{+}{R}{y_e}{y_e}{1}\,\tDF{+}{L}{y_{\mu}}{3}{y_{\mu}}\,\bar{G}^{13}_y \nonumber\\
		&\frac{1}{2}\Big[\tF{-}{R}{y}{1}{2}\,y\,\partial_{k_E}\Big(\tF{-}{R}{y}{2}{y}\,\tF{+}{L}{y}{y}{3}\Big)+y\,\partial_{k_E}\Big(\tDF{+}{R}{y}{1}{2}\Big)\tDF{-}{R}{y}{2}{y}\,\tF{+}{L}{y}{y}{3}\nonumber
		\\
		&\tF{-}{R}{y}{1}{y}\,\tDF{-}{L}{y}{y}{2}\,y\,\partial_{k_E}\Big(\tDF{+}{L}{y}{2}{3}\Big)+y\,\partial_{k_E}\left(y^2\,\tF{-}{R}{y}{1}{y}\,\tF{+}{L}{y}{y}{2}\right)\tF{+}{L}{y}{2}{3}\Big],\label{eq:I:1p2MIZ5:new}
\end{align}
The integral for $3\text{MIZ}$ and $3\text{MIZ}^5$ can be written as
\begin{align}
I_{3\text{MIZ/}Z^5} =& \frac{1}{2}\int dy\, dx_1 dx_2 dx_3 \, 
		\left(\frac{y}{x_1}\right)^{2+c_L}\,
		\left(\frac{y}{x_2}\right)^{2-c_E}\,
		\left(\frac{y}{x_3}\right)^4G^{13}_y\,\displaystyle{\sum_{i=1}^8}\,M_i\,y\,\partial_{k_E}\,N_i.\label{eq:I:3MIZZ5}	
\end{align}
For $3\text{MIZ}$, the $(M,N)$ pairs are 
\begin{align}
		\Big(M_1\;,\;N_1\Big)&=\left(\tF{+}{L}{y}{1}{2}\;,\;y^4\,\tF{+}{L}{y}{2}{y}\,\tF{-}{R}{y}{y}{y}\,\tF{+}{L}{y}{y}{y}\,\tF{-}{R}{y}{y}{3}\right),
		\\
		\Big(M_2\;,\;N_2\Big)&=\left(-y^2\,\tDF{+}{L}{y}{2}{y}\,\tF{-}{R}{y}{y}{y}\,\tF{+}{L}{y}{y}{y}\,\tF{-}{R}{y}{y}{3}\;,\;\tDF{-}{L}{y}{1}{2}\right),
		\\
		\Big(M_3\;,\;N_3\Big)&=\left(-y^2\,\tF{-}{R}{y}{2}{y}\,\tF{+}{L}{y}{y}{y}\,\tF{-}{R}{y}{y}{3}\;,\;-y^2\,\tF{+}{L}{y}{1}{y}\,\tF{-}{R}{y}{y}{2}\right),
	         \\   
		\Big(M_4\;,\;N_4\Big)&=\left(\tF{+}{L}{y}{1}{y}\,\tDF{+}{R}{y}{y}{2}\;,\;-y^2\,\tDF{-}{R}{y}{2}{y}\,\tF{+}{L}{y}{y}{y}\,\tF{-}{R}{y}{y}{3}\right),
		\\
		\Big(M_5\;,\;N_5\Big)&=\left(-y^2\,\tF{+}{L}{y}{1}{y}\,\tF{-}{R}{y}{y}{y}\,\tF{+}{L}{y}{y}{2}\;,\;-y^2\,\tF{+}{L}{y}{2}{y}\,\tF{-}{R}{y}{y}{3}\right),
		\\
		\Big(M_6\;,\;N_6\Big)&=\left(\tDF{+}{L}{y}{2}{y}\,\tF{-}{R}{y}{y}{3}\;,\;-y^2\,\tF{+}{L}{y}{1}{y}\,\tF{-}{R}{y}{y}{y}\,\tDF{-}{L}{y}{y}{2}\right),
		 \\   
		\Big(M_7\;,\;N_7\Big)&=\left(\tF{-}{R}{y}{2}{3}\;,\;y^4\,\tF{+}{L}{y}{1}{y}\,\tF{-}{R}{y}{y}{y}\,\tF{+}{L}{y}{y}{y}\,\tF{-}{R}{y}{y}{2}\right),
		 \\   
		\Big(M_8\;,\;N_8\Big)&=\left(-y^2\,\tF{+}{L}{y}{1}{y}\,\tF{-}{R}{y}{y}{y}\,\tF{+}{L}{y}{y}{y}\,\tDF{+}{R}{y}{y}{2}\;,\;\tDF{-}{R}{y}{2}{3}\right).
\end{align}
For $3\text{MIZ}^5$, the $(M,N)$ pairs are 
\begin{align}
		\Big(M_1\;,\;N_1\Big)&=\left(-y^2\,\tDF{+}{L}{y}{1}{y}\,\tF{-}{R}{y}{y}{y}\,\tF{+}{L}{y}{y}{y}\,\tF{-}{R}{y}{y}{2}\;,\;\tDF{+}{R}{y}{2}{3}\right),
		\\
		\Big(M_2\;,\;N_2\Big)&=\left(\tF{+}{R}{y}{2}{3}\;,\;-y^2\,\tDF{+}{L}{y}{1}{y}\,\tF{-}{R}{y}{y}{y}\,\tF{+}{L}{y}{y}{y}\,\tDF{+}{R}{y}{y}{2}\right),
		\\
		\Big(M_3\;,\;N_3\Big)&=\left(\tDF{+}{L}{y}{1}{y}\,\tF{-}{R}{y}{y}{y}\,\tDF{-}{L}{y}{y}{2}\;,\;\tDF{+}{L}{y}{2}{y}\,\tDF{+}{R}{y}{y}{3}\right),
	         \\   
		\Big(M_4\;,\;N_4\Big)&=\left(\tF{+}{L}{y}{2}{y}\,\tDF{+}{R}{y}{y}{3}\;,\;-y^2\,\tDF{+}{L}{y}{1}{y}\,\tF{+}{R}{y}{y}{y}\,\tF{+}{L}{y}{y}{2}\right),
		\\
		\Big(M_5\;,\;N_5\Big)&=\left(\tDF{+}{L}{y}{1}{y}\,\tF{-}{R}{y}{y}{2}\;,\;-y^2\,\tF{-}{R}{y}{2}{y}\,\tF{+}{L}{y}{y}{y}\,\tDF{+}{R}{y}{y}{3}\right),
		\\
		\Big(M_6\;,\;N_6\Big)&=\left(\tDF{-}{R}{y}{2}{y}\,\tF{+}{L}{y}{y}{y}\,\tDF{+}{R}{y}{y}{3}\;,\;\tDF{+}{L}{y}{1}{y}\,\tDF{+}{R}{y}{y}{2}\right),
		 \\   
		\Big(M_7\;,\;N_7\Big)&=\left(\tF{-}{L}{y}{1}{2}\;,\;-y^2\,\tDF{+}{L}{y}{2}{y}\,\tF{-}{R}{y}{y}{y}\,\tF{+}{L}{y}{y}{y}\,\tDF{+}{R}{y}{y}{3}\right),
		 \\   
		\Big(M_8\;,\;N_8\Big)&=\left(-y^2\,\tF{+}{L}{y}{2}{y}\,\tF{-}{R}{y}{y}{y}\,\tF{+}{L}{y}{y}{y}\,\tDF{+}{R}{y}{y}{3}\;,\;\tDF{+}{L}{y}{1}{2}\right).
\end{align}
The integrals for the $W^5$ loops are
\begin{align}
I_{2\text{MIW}^5\text{W}^5} =& -\int dy\, dx_1 dx_2 dx_3 \, 
		\left(\frac{y}{x_1}\right)^{2+c_L}\,
		\left(\frac{y}{x_2}\right)^{4}\,
		\left(\frac{y}{x_3}\right)\,
		\times \nonumber\\
		&\Big\{\frac{1}{2}\,y^2\,\tDF{+}{L}{y}{1}{y}\,\tF{-}{R}{y}{y}{y}\,\tF{+}{L}{y}{y}{2}\,\tDF{+}{L}{y_{\mu}}{2}{y_{\mu}}\Big[4\,\bar{G}^{13}_y\,\bar{G}^{23}_y+y\,\partial_{k_E}\Big(\bar{G}^{13}_y\,\bar{G}^{23}_y\Big)\Big]\Big\},\label{eq:I:2MIW5W5}\\		
I_{2\text{MIW}^5\text{W}} =& -\int dy\, dx_1 dx_2 dx_3 \, 
\left(\frac{y}{x_1}\right)^{2+c_L}\,
\left(\frac{y}{x_2}\right)^{4}\,
\left(\frac{y}{x_3}\right)\,
\times \nonumber\\
		&\Big[\frac{1}{2}\,y^2\,\tF{+}{L}{y}{1}{y}\,\tF{-}{R}{y}{y}{y}\,\tF{+}{L}{y}{y}{2}\,\tDF{+}{L}{y_{\mu}}{2}{y_{\mu}}\Big(y\,\partial_{k_E}\,G^{13}_y\,\partial_z\,\bar{G}^{23}_y-y\,\partial_{k_E}\partial_z\,G^{13}_y\,\bar{G}^{23}_y\Big)\Big],\label{eq:I:2MIW5W},
\end{align}
\begin{align}		
		I_{2\text{MIW}\text{W}^5} =& -\int dy\, dx_1 dx_2 dx_3 \, 
		\left(\frac{y}{x_1}\right)^{2+c_L}\,
		\left(\frac{y}{x_2}\right)^{4}\,
		\left(\frac{y}{x_3}\right)\,
		\times \nonumber\\
		&\Big[\frac{1}{2}\,\tDF{+}{L}{y}{1}{y}\,\tF{-}{R}{y}{y}{y}\,\tDF{-}{L}{y}{y}{2}\,\tDF{+}{L}{y_{\mu}}{2}{y_{\mu}}\Big(y\,\partial_{k_E}\,G^{23}_y\,\partial_z\,\bar{G}^{13}_y-y\,\partial_{k_E}\partial_z\,G^{23}_y\,\bar{G}^{13}_y\Big)\Big].\label{eq:I:2MIWW5}	
\end{align}
\subsection{Subdominant $b$ coefficient diagrams}
\begin{align}
    \mathcal M(n\text{MI}Z\,/\,Z^5) &= \frac{i}{16\pi^2} 
	\left(R'\right)^2 f_{c_{L}} Y_E f_{-c_{E}} \frac{ev}{\sqrt{2}}
	\left(g_{Z_L}g_{Z_R}\ln\frac{R'}{R}\right)
	I_{n\text{MI}Z\,/\,Z^5},
	\label{eq:M:01MIZZ5}\\
	\mathcal M(0\text{MI}W) &= \frac{i}{16\pi^2} 
	\left(R'\right)^2  f_{c_{L}} Y_E  f_{-c_{E}}  \frac{ev}{\sqrt{2}}
	\left(\frac{g^2}{2}\ln\frac{R'}{R}\right) I_{0\text{MI}W},
	\label{eq:M:nMIZ:b}\\
	%
	%
	\mathcal M(0\text{MI}W^5)
	&=
	\frac{i}{16\pi^2}(R')^2 
	f_{c_{L_\mu}}Y_E f_{-c_{E_e}} 
	\frac{ev}{\sqrt{2}} 
	\left(\frac{g^2}{2}\ln \frac{R'}{R}\right) 
	I_{0\text{MI}W^5}
\end{align}
where $n=0,1$ counts the number of internal mass insertions.
\begin{align}
%
%
I_{1\text{MIZ}^5} =& \int dy\, dx_1 dx_2 dx_3 \, 
		\left(\frac{y}{x_1}\right)^{2+c_L}\,
		\left(\frac{y}{x_2}\right)^{2-c_E}\,
		\left(\frac{y}{x_3}\right)^4\,
		\times \nonumber\\
		&\frac{1}{2}\Big[\tF{-}{L}{y}{1}{3}\,y\,\partial_{k_E}\Big(\tDF{+}{L}{y}{3}{y}\,\tDF{+}{R}{y}{y}{2}\Big)\,\bar{G}^{12}_y-\tDF{+}{L}{y}{1}{3}\,y\,\partial_{k_E}\Big(\tF{+}{L}{y}{3}{y}\,\tDF{+}{R}{y}{y}{2}\,\bar{G}^{12}_y\Big)\nonumber
		\\
		&-4\,\tDF{+}{L}{y}{1}{3}\,\tF{+}{L}{y}{3}{y}\,\tDF{+}{R}{y}{y}{2}\,\bar{G}^{12}_y+\tDF{+}{L}{y}{1}{y}\,\tF{-}{R}{y}{y}{3}\Big(y\,\partial_{k_E}\,\tDF{+}{R}{y}{3}{2}\Big)\bar{G}^{12}_y\nonumber
		\\
		&-\tDF{+}{L}{y}{1}{y}\,\tDF{+}{R}{y}{y}{3}\,y\,\partial_{k_E}\Big(\tF{+}{R}{y}{32}\,\bar{G}^{12}_y\Big)-4\,\tDF{+}{L}{y}{1}{y}\,\tDF{+}{R}{y}{y}{3}\,\tF{+}{R}{y}{3}{2}\,\bar{G}^{12}_y\Big].\label{eq:I:1MIZ5}\\	
I_{0\text{MIZ}} =& \int dy\, dx_1 dx_2 dx_3 \, 
		\left(\frac{y}{x_1}\right)^{2+c_L}\,
		\left(\frac{y}{x_2}\right)^{4}\,
		\left(\frac{y}{x_3}\right)^{4}\,
		\times \nonumber\\
		&y\,\partial_{k_E}\,G^{13}_y\,\tDF{+}{L}{y_{\mu}}{3}{y_{\mu}}\Big(\tDF{-}{L}{y}{1}{2}\,\tF{-}{L}{y}{2}{3}+\tF{+}{L}{y}{1}{2}\,\tDF{-}{L}{y}{2}{3}\Big),\label{eq:I:0MIZ:new}\\
I_{0\text{MIZ}^5} =& -\int dy\, dx_1 dx_2 dx_3 \, 
		\left(\frac{y}{x_1}\right)^{2+c_L}\,
		\left(\frac{y}{x_2}\right)^{4}\,
		\left(\frac{y}{x_3}\right)^{4}\,
		\times \nonumber\\
		&\Big\{\frac{1}{4}\,\tDF{+}{L}{y}{2}{3}\,\tDF{+}{L}{y_{\mu}}{3}{y_{\mu}}\Big[\tF{-}{L}{y}{1}{2}\Big(4\bar{G}^{13}_y+y\,\partial_{k_E}\,\bar{G}^{13}_y\Big)+y\,\partial_{k_E}\,\tF{-}{L}{y}{1}{2}\,\bar{G}^{13}_y\Big]\Big\},\label{eq:I:0MIZ5:new}\\
		I_{0\text{MIW}} =& -\int dy\, dx_1 dx_2 dx_3 \, 
		\left(\frac{y}{x_1}\right)^{2+c_L}\,
		\left(\frac{y}{x_2}\right)\,
		\left(\frac{y}{x_3}\right)^{4}\,
		\times \nonumber\\
	&\frac{3}{2}\,y\,\partial_{k_E}\Big(G^{13}_y\,G^{32}_y\Big)\tDF{-}{L}{y}{1}{2}\,\tDF{L}{+}{y_{\mu}}{3}{y_{\mu}}\label{eq:I:0MIW:new}\\
	%
	I_{0\text{MI}W^5} =& 
	\int dy\, dx_1 dx_2 dx_3 \, 
	\left(\frac{y}{x_1}\right)^{c_L+2}
	\left(\frac{y}{x_2}\right)^{4}
	\left(\frac{y}{x_3}\right)\nonumber\\
	& \phantom{\int } 
	\bigg\{ \,
	\frac y2 
	\tF{+}{L}{y}{1}{y} 
	\tDF{+}{L}{y_\mu}{2}{y_\mu} 
	\left(\frac{\partial}{\partial k_E}\frac{\partial}{\partial x_3} G^{13}_y \right)\bar G^{32}_y
	+ 
	\frac y2
	\tF{+}{L}{y}{1}{2}
	\tDF{+}{L}{y_\mu}{2}{y_\mu}
	\left( 
	\frac{\partial}{\partial k_E}
	\frac{\partial}{\partial x_3} G^{32}_y
	\right) \bar G^{13}_y
	\nonumber\\
	& 
	\phantom{\int\bigg\{\, } 
	-
	\tDF{+}{L}{y}{1}{2}
	\tDF{+}{L}{y_\mu}{2}{y_\mu}
		\left[
		2\bar G^{13}_y \bar G^{23}_y
		+ \frac y2 \frac{\partial}{\partial k_E} 
			\left( \bar G^{13}_y \bar G^{32}_y
			\right)
		\right]
	\bigg\}.
	\label{eq:I:0MIW5}
	\end{align}

\subsection{Custodial Models}
For custodially protected models, one must include loops with the custodial partners of fermions and gauge bosons. See, e.g., \cite{Albrecht:2009xr} for details of the additional field content of such models. The new particles have mixed boundary conditions, $(-+)$ or $(+-)$. For the chirality flipping process $\mu\to e\,\gamma$, Yukawa insertions on the IR brane only allow fermions carrying either $(++)$ or $(-+)$ boundary conditions running in the loop. This limits the number of the new diagrams to be considered. 
The new fermion propagators can be obtained by making the replacement $\tilde F \to \tilde E$. Writing the boundary condition in terms of the Weyl components of the Dirac spinor, $\tilde{E}^L$ corresponds to the boundary condition $\left(\psi_{(+-)},\bar{\chi}_{(-+)}\right)$, while $\tilde{E}^R$ corresponds to $\left(\psi_{(-+)},\bar{\chi}_{(+-)}\right)$. For $x>x'$, the $\tilde E$-functions can be written as follows:

%
\begin{align}
\tilde E^L_{-} &=\phantom{+} \frac{(xx')^{5/2}}{y^5} \frac{S_{c}(x_-,y_-) T_{c}(x'_-,wy_+)}{T_{c}(y_-,wy_+)}
&
\tilde E^L_{+} &=- \frac{(xx')^{5/2}}{y^5} \frac{T_{c}(x_+,y_-) S_{c}(x'_+,wy_+)}{T_{c}(y_-,wy_+)}
\\
\tilde E^{R}_{-} &=- \frac{(xx')^{5/2}}{y^5} \frac{T_{c}(x_-,y_+) S_{c}(x'_-,wy_-)}{T_{c}(y_+,wy_-)}
&
\tilde E^R_{+} &=\phantom{+} \frac{(xx')^{5/2}}{y^5} \frac{S_{c}(x_+,y_+) T_{c}(x'_+,wy_-)}{T_{c}(y_+,wy_-)}.
\end{align}
The $x<x'$ expressions are obtained by replacing $x\leftrightarrow x'$. 
Gauge bosons with $(-+)$ boundary conditions can also appear in custodial loops. The corresponding propagator for $x>x'$ is $G\to H$ with
\begin{align}
H_k(x,x')\,=\,\frac{\left(R'\right)^2}{R}\frac{xx'}{y}\frac{T_{10}(x,y)\,S_{11}(x',wy)}{T_{10}(wy,y)}.
\end{align}
The $T$ and $S$ are defined in Appendix.~(\ref{app:bulk:Feynman:rules}), and the $x<x'$ case can be obtained by $x\leftrightarrow x'$.

\section{Position, momentum, and position/momentum space}
\label{sec:5D:momentum:space}

In order to elucidate the power counting in Section~\ref{sec:heuristic} and to provide some motivation for the structure of the propagators in Appendix~\ref{app:flatGreensFunc}, we review the passage between Feynman rules in position, momentum, and mixed position/momentum space. For simplicity we shall work with massless scalar fields on a flat (Minkowski) $d$-dimensional background, but the generalization of the salient features to higher spins is straightforward. In position space, the two-point Green's function for a particle propagating from $x'$ to $x$ is
\begin{align}
	D(x,x') = \int \dbar^dk \frac{i}{k^2}e^{-ik\cdot(x-x')}\label{eq:momentum:cons:2point},
\end{align}
a momentum-space integral over a power-law in $k$ times a product of exponentials in $k\cdot x$ and $k\cdot x'$. Each vertex carries a $d^dx$ integral representing each spacetime point at which the interaction may occur. When some dimensions are compact, the associated integrals are reverted to discrete sums and the particular linear combination of exponentials is shifted to maintain boundary conditions. Further, when dimensions are warped the exponentials become Bessel functions. In this Appendix we will neglect these differences and focus on general features since the UV behavior of each of the aforementioned scenarios (i.e.\ for momenta much larger than any mass, compactification, or warping scales) reduces to the flat noncompact case presented here.

In 4D it is conventional to work in full momentum space where the Feynman rules are derived by performing the $d^dx$ integrals at each vertex over the exponential functions from each propagator attached to the vertex and amputating the external propagators. This generates a momentum-conserving $\delta$-function at each vertex which can be used to simplify the $\dbar^dk$ integrals in each propagator. For each diagram one such $\delta$-function imposes overall conservation of the external momenta and hence has no dependence on any internal momenta. For a loop diagram this means that there is a leftover $\dbar^d k $ which corresponds to the integration over the loop momentum. Thus the momentum space formalism involves separating the exponentials in $k\cdot x$ from the rest of the Green's function and performing the $d^dx$ integral to obtain $\delta$-functions.

To go to the mixed position/momentum space formalism we pick one direction, $z$, and leave the dependence on that position in the propagator while integrating over the $z$-component of the momentum, $k^z$ in (\ref{eq:momentum:cons:2point}). We shall write the Minkowski scalar product of the $(d-1)$ momentum-space directions as $k^2$ so that the full $d$-dimensional scalar product is $k^2 - k_z^2$.
The Feynman rule for each vertex now includes an explicit $dz$ integral which must be performed \textit{after} including each of the position/momentum space propagators, which take the form
\begin{align}
	\Delta(k,z,z') = \int \dbar k_z \frac{i}{k^2-k_z^2}e^{ik_z(z-z')}.\label{eq:momentum:cons:prop:momentum}
\end{align}
The $(d-1)$ other exponentials and momentum integrals are accounted in the usual momentum-space formalism. This object goes like $\Delta \sim 1/k$, which indeed has the correct dimensionality for the sum over a KK tower of scalar propagators. Similarly, the massless bulk fermion propagator is
\begin{align}
	\Delta(k,z,z') = \int \dbar k_z \frac{i(\slashed{k}-k_z\gamma^5)}{k^2-k_z^2}e^{ik_z(z-z')}\label{eq:momentum:cons:prop:mixed},
\end{align}
where we may now identify the scalar functions $F\sim dk_z e^{ik_z(z-z')}/(k^2-k_z^2)$ in (\ref{eq:flat:G:fromF}) and (\ref{eq:warped:KG:greens}).

It is thus apparent that the mixed formalism contains all of the same integrals and factors as the momentum-space formalism, but that these are packaged differently between vertex and propagator Feynman rules. By identifying features between the two pictures one may glean physical intuition in one picture that is not manifest in the other. For example, the observation in the mixed formalism that each bulk vertex on a loop brings down a power of $1/k$ is straightforwardly understood to be a manifestation of momentum conservation in the momentum space picture.

On the other hand, the mixed formalism is much more intuitive for brane-localized effects. Interactions with fields on the brane at $z=L$ carry $\delta(z-L)$ factors in the vertex Feynman rules. Such interactions violate momentum conservation in the $z$-direction. In the KK formalism this manifests itself as the question of when it is appropriate to sum over an independent tower of KK modes. This is easily quantified in the mixed formalism since the $dz$ integrals are not yet performed in the Feynman rules and we may directly insert $\delta(z-L)$ terms in the expression for the amplitude. 

\begin{figure}[t]
  	\begin{center}
		\begin{tikzpicture}[scale=1.25, line width=1.75]
			\draw (0,0) circle (1);
			\draw (180:1) -- (180:2);
			\draw (60:1) -- (60:2);
			\draw (-60:1) -- (-60:2);
			\draw[shift={(0,.17)}, line width=.8,<-] (180:1.2) -- (180:1.8);
			\draw[shift={(.2,0)}, line width=.8,<-] (60:1.2) -- (60:1.8);
			\draw[shift={(-.2,0)}, line width=.8,<-] (-60:1.2) -- (-60:1.8);
			\draw[line width=.8, ->] (150:1.2) arc (150:90:1.2);
			\draw[line width=.8, ->] (30:1.2) arc (30:-30:1.2);
			\draw[line width=.8, ->] (270:1.2) arc (270:210:1.2);
			\node at (165:1.5) {$p_1$};
			\node at (45:1.6) {$p_2$};
			\node at (-75:1.5) {$p_3$};
			\node at (120:1.5) {$k_2$};
			\node at (0:1.5) {$k_3$};
			\node at (-120:1.4) {$k_1$};
			\node at (180:.7) {$z_1$};
			\node at (60:.7) {$z_2$};
			\node at (-60:.7) {$z_3$};
		\end{tikzpicture}	
	\end{center}
  \caption{A simple loop diagram to demonstrate the power counting principles presented. The lines labeled $p_i$ represent the net external momentum flowing into each vertex so that $p_i^z$ corresponds to the KK mass of the $i^\text{th}$ external particle.}
  \label{fig:momentum:conservation}
\end{figure}
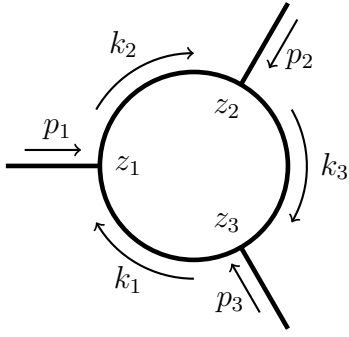

As a concrete example, consider the loop diagram with three vertices shown in Fig.~\ref{fig:momentum:conservation}. 
It is instructive to explicitly work out loop $z$-momentum structure of this diagram in the case where all vertices are in the bulk and observe how this changes as vertices are localized on the brane. 
To simplify the structure, let us define the product of momentum-space propagators
\begin{align}
	f(k_1, k_2, k_3) \equiv \prod_{i=1}^3\frac{i}{k_i^2 - (k_i^z)^2}.
\end{align}
Using $\int dz\,\exp(izk)=\delta(k)$, the bulk amplitude is proportional to
\begin{align}
	\mathcal M \sim& \int dz_1\,dz_2\,dz_3\; dk_1^z\,dk_2^z\,dk_3^z\;f(k_1,k_2,k_3)\,
	 e^{iz_1(k_1+p_1-k_2)^z}\,e^{iz_2(k_2+p_3-k_3)^z}\,e^{iz_3(k_3+p_3-k_1)^z}
	\label{eq:momentum:1}\\
	\sim&\int dz_2\,dz_3\;dk_2^z\,dk_3^z\;f(k_2-p_1,k_2,k_3)\,
	 e^{iz_2(k_2+p_3-k_3)^z}\,e^{iz_3(k_3+p_3-k_2+p_1)^z}
	\label{eq:momentum:2}\\
	\sim&\int dz_3\;dk_3^z\;f(k_3-p_2-p_1,k_3-p_2,k_3)\,
	 e^{iz_3(p_1+p_2+p_3)^z}.
	\label{eq:momentum:3}
\end{align}
We have implicitly performed the associated $d^{(d-1)}x$ integrals at each step.
The final $dz_3$ integral gives the required $\delta$-function of external momenta while leaving an unconstrained $dk_3^z$ loop integral. Each $dk^z/(k^2-k_z^2) \sim 1/k$ represents the entire KK tower associated with an internal line. The removal of two $dk^z$ integrals by $\delta$-functions is a manifestation of the $1/k$ suppression coming from each $dz$ integral with the caveat that the ``last'' $dz$ integral only brings down powers of external momenta and hence does not change the power of loop momenta. This explains the ``overall $z$-momentum'' contribution to the superficial degree of divergence in Section~\ref{sec:bulk:field:power:counting}.

Next consider the case when the $z_3$ vertex is brane localized so that its Feynman rule is proportional to $\delta(z_3-L)$. This only affects the last line of the simplification by removing the $dz_3$ integral. Physically this means that $z$-momentum (KK number) needn not be conserved for this process. Since the $z_3$ exponential is independent of any loop momenta, this does not affect the superficial degree of divergence. 

On the other hand, if $z_2$ is also brane localized, then the $\delta(z_2-L)$ from the vertex prevents the $dz_2$ integral in the second line from giving the $\delta(k_2+p_2-k_3)$ that cancels the $dk_2^z$ integral. Thus the process has an additional $dk_2^z$ integral which now increases the degree of divergence. In the 4D formalism this is manifested as an additional independent sum over KK states. It is now also clear that setting $z_1$ to be brane localized prevents the $dk_1^z$ from being cancelled and hence adds another unit to the degree of divergence. This counting is trivially generalized to an arbitrary number of vertices and different types of internal propagators. For a loop with $V$ vertices, $V_B$ of which are in the bulk, the key points are:
\begin{enumerate}
	\item If $V=V_B$, then the $dz$ integrals reduce the superficial degree of divergence by $(V_B-1)$.
	\item If, on the other hand, $V>V_B$ so that there is at least one brane-localized vertex, then the $dz$ integrals reduce the superficial degree of divergence by $V_B$.
\end{enumerate}

Intuitively the $z$-momentum nonconservation coming from brane-localized interactions can be understood as the particle picking up an arbitrary amount of momentum as it bounces off the brane (a similar picture can be drawn for the orbifold \cite{Georgi:2000ks}). Alternately, it reflects the uniform spread in momentum associated with complete localization in $z$-position. While this may seem to imply sensitivity to arbitrarily high scale physics on the brane, a negative degree of divergence will prevent the loop from being sensitive to UV physics. In other words, we are free to treat brane-localized fields as having $\delta$-function profiles independent of the physics that generates the brane.

Finally, note that we have assumed that each fermion mass insertion is brane localized. In 5D this means that higher-order diagrams in the fermion mass-insertion approximation are not suppressed by momentum since each additional brane-to-brane propagator goes like $\sim\slashed{k}/k$ after accounting for the $dk^z$ integrals. Instead, these mass insertions are suppressed only by the relative sizes of the Higgs vev and compactification scale, $(vR')^2 \sim .01$. It is perhaps interesting to note that our analysis further suggests that in 6D with a Higgs localized on a 4D subspace, there are two additional momentum integrals coming from a mass insertion so that each vev-to-vev propagator goes like a positive power of the momentum $\sim\slashed{k}$ causing the mass-insertion approximation to break down.

\section{Bulk Feynman Rules}
\label{app:bulk:Feynman:rules}

Here we summarize the 5D position/momentum space Feynman rules used to derive the amplitudes in this paper. All couplings are written in terms of 5D quantities. The brane-localized Higgs field is drawn as a dashed line and the fifth component of a bulk gauge boson is drawn as a dotted line.

\vspace{1em}

\begin{minipage}[!t]{0.4\linewidth}
		\begin{tikzpicture}[line width=1.5 pt, scale=1]
			\draw[vector] (0,0) -- (90:1);
			\draw[fermionbar] (0,0) -- (210:1);
			\draw[fermion] (0,0) -- (-30:1);
			\node[right] at (1.25,.25) {$\displaystyle{=ig_5\left(\frac{R}{z}\right)^4\gamma^\mu}$};
		\end{tikzpicture}

	\vspace{1em}

		\begin{tikzpicture}[line width=1.5 pt, scale=1]
			\draw[vector] (0,0) -- (90:1);
			\draw[scalarbar] (0,0) -- (210:1);
			\draw[scalarbar] (0,0) -- (-30:1);
			\node[right] at (1.25,.25) {$\displaystyle{=ie_5(p_+-p_-)_\mu}$};
		\end{tikzpicture}

	\vspace{1em}

		\begin{tikzpicture}[line width=1.5 pt, scale=1]
			\draw[vector] (0,0) -- (90:1);
			\draw[vector] (0,0) -- (210:1);
			\draw[scalar] (0,0) -- (-30:1);
			\node[right] at (1.25,.25) {$\displaystyle{=\frac{i}{2}e_5g_5\,v\,\eta^{\mu\nu}}$};
		\end{tikzpicture}

	\vspace{1em}

		\begin{tikzpicture}[line width=1.5 pt, scale=1]
			\draw[scalarnoarrow] (0,0) -- (90:1);
			\draw[fermionbar] (0,0) -- (210:1);
			\draw[fermionbar] (0,0) -- (-30:1);
			\node[right] at (1.25,.25) {$\displaystyle{=i\left(\frac{R}{R'}\right)^3 Y_{5}}$};
		\end{tikzpicture}

	%
\end{minipage}
\hspace{0.5cm}
\begin{minipage}[!t]{0.6\linewidth}

		\begin{tikzpicture}[line width=1.5 pt, scale=1]
			\draw[fermion] (-.5,0) -- (1.25,0);
			\node[right] at (1.5,0) {$\displaystyle{=\Delta_k(z,z')}$};
		\end{tikzpicture}

	\vspace{1em}

		\begin{tikzpicture}[line width=1.5 pt, scale=1]
			\draw[vector] (-.5,0) -- (1.25,0);
			\node[right] at (1.5,0) {$\displaystyle{=-i\eta^{\mu\nu}G_k(z,z')}$};
		\end{tikzpicture}

	\vspace{1em}
	
		\begin{tikzpicture}[line width=1.5 pt, scale=1]
			\draw[vectorscalar] (-.5,0) -- (1.25,0);
			\node[right] at (1.5,0) {$\displaystyle{= i\bar G_k(z,z')}$};
		\end{tikzpicture}

	\vspace{1em}

		\begin{tikzpicture}[line width=1.5 pt, scale=1]
			\draw[vector] (0,0) -- (-1.25,0);
			\node[right] at (.75,0) {$\displaystyle{= \epsilon^\mu(q) f^{(0)}_A}$};
			\draw[fill=black] (0,0) circle (.3cm);
			\draw[fill=white] (0,0) circle (.295cm);
			\begin{scope}
		    	\clip (0,0) circle (.3cm);
		    	\foreach \x in {-1.0,-.8,...,.4}
					\draw[line width=1 pt] (\x,-.3) -- (\x+.6,.3);
		  	\end{scope}
		\end{tikzpicture}

	\vspace{1em}

		\begin{tikzpicture}[line width=1.5 pt, scale=1]
			\draw[fermionbar] (0,0) -- (-1.25,0);
			\node[right] at (.75,0) {$\displaystyle{=\frac{f_c}{\sqrt{R'}} 
			\left(\frac{z}{R}\right)^2 \left(\frac{z}{R'}\right)^{-c}u(p)}$};
			\draw[fill=black] (0,0) circle (.3cm);
			\draw[fill=white] (0,0) circle (.295cm);
			\begin{scope}
		    	\clip (0,0) circle (.3cm);
		    	\foreach \x in {-1.0,-.8,...,.4}
					\draw[line width=1 pt] (\x,-.3) -- (\x+.6,.3);
		  	\end{scope}
		\end{tikzpicture}

	\vspace{1em}

		\begin{tikzpicture}[line width=1.5 pt, scale=1]
			\draw[fermionbar] (.3,0) -- (1.2,0);
			\node[right] at (1.75,0) {$\displaystyle{=\bar u(p')\frac{f_c}{\sqrt{R'}} 
			\left(\frac{z}{R}\right)^2 \left(\frac{z}{R'}\right)^{-c}}$};
			\draw[fill=black] (0,0) circle (.3cm);
			\draw[fill=white] (0,0) circle (.295cm);
			\begin{scope}
		    	\clip (0,0) circle (.3cm);
		    	\foreach \x in {-1.0,-.8,...,.4}
					\draw[line width=1 pt] (\x,-.3) -- (\x+.6,.3);
		  	\end{scope}
		\end{tikzpicture}

\end{minipage}

\vspace{1em}

\noindent The 5D Lagrangian parameters are related to the usual Standard Model parameters by
\begin{align}
	g_5^2 &= g_\text{SM}^2 R\ln{R'}/{R}\\
	e_5 f^{(0)}_A &= e_\text{SM}\\
	Y_5 &= RY,
\end{align}
where $Y$ represents an anarchic 4D Yukawa matrix that is related to the Standard Model Yukawa by (\ref{eq:RS:anarchy:zero:mode:Yukawa}). The $f_c$ fermion flavor functions are defined in (\ref{eq:flavor:function}). The vector propagator functions $G_k(z,z')$ and $\bar G_k(z,z')$ are explicitly derived in \cite{Randall:2001gb}, which also contains generic formulae for analogous functions for fields of general spin and additional gauge boson vertices. Using the dimensionless $x$ and $y$ variables defined in (\ref{eq:dimensionless:vars}) and assuming $z>z'$, the Euclidean space vector Green's functions are
\begin{align}
	G_k(z,z') &= \frac{(R')^2}{R} G_y(x,x') = \frac{(R')^2}{R} \frac{xx'}{y} \frac{T_{10}(x,y)T_{10}(x',wy)}{S_{00}(wy,y)},\label{eq:G:def}\\
	\bar G_k(z,z') &= \frac{(R')^2}{R} \bar G_y(x,x') = \frac{(R')^2}{R} \frac{xx'}{y} \frac{S_{00}(x,y)S_{00}(x',wy)}{S_{00}(wy,y)},\label{eq:Gbar:def}
\end{align}
where 
\begin{align}
	T_{ij}(x,y) &= I_i(x)K_j(y)+ I_j(y)K_i(x)\\
	S_{ij}(x,y) &= I_i(x)K_j(y) - I_j(y)K_i(x)
\end{align}
and $w=R/R'$.
For $z<z'$ the above formula is modified by $x\leftrightarrow x'$. The three gauge boson couplings are given by
\vspace{1em}
\begin{tikzpicture}[line width=1.5 pt, scale=1]
	\draw[vector] (0,0) -- (90:1);
	\draw[vector] (0,0) -- (210:1);
	\draw[vector] (0,0) -- (-30:1);
	\node at (90:1.4) {$A_\mu$};
	\node at (210:1.4) {$W^+_\nu$};
	\node at (-30:1.5) {$W^-_\rho$};
	\node[right] at (1.7,.25) {$\displaystyle{=ie_5 \frac{R}{z}  
	\left[(k-k^+)^\rho \eta^{\mu\nu}
	+ (k^--k)^\nu \eta^{\mu\rho} 
	+ (k^+-k^-)^\mu \eta^{\nu\rho}
	\right] 
	}$};
\end{tikzpicture}
\vspace{1em}

\begin{tikzpicture}[line width=1.5 pt, scale=1]
	\draw[vector] (0,0) -- (90:1);
	\draw[vectorscalar] (0,0) -- (210:1);
	\draw[vectorscalar] (0,0) -- (-30:1);
	\node at (90:1.4) {$A_\mu$};
	\node at (210:1.4) {$W^+_5$};
	\node at (-30:1.5) {$W^-_5$};
	\node[right] at (1.7,.25) {$\displaystyle{=ie_5 \frac{R}{z}  
	(k^- -k^+)^\mu
	}$};
\end{tikzpicture}
\hspace{1.5 cm}
\begin{tikzpicture}[line width=1.5 pt, scale=1]
	\draw[vector] (0,0) -- (90:1);
	\draw[vector] (0,0) -- (210:1);
	\draw[vectorscalar] (0,0) -- (-30:1);
	\node at (90:1.4) {$A_\mu$};
	\node at (210:1.4) {$W^+_\nu$};
	\node at (-30:1.5) {$W^-_5$};
	\node[right] at (1.7,.25) {$\displaystyle{= e_5 \frac{R}{z}  
	\eta^{\mu\nu} (\partial_z - \partial_z^+)
	}$};
\end{tikzpicture}

\vspace{1em}

\noindent Here we have used the convention where all momenta are labeled by the charge of the particle and are flowing into the vertex. The $A_\mu W_5^+ W^-_\nu$ vertex is given by $e_5 (R/z)\eta^{\mu\nu}(\partial_z^\mu - \partial_z)$.
The Euclidan space fermion propagator $\Delta_k(z,z')$ is given in (\ref{eq:warped:G:fromF:Euclidean}).

\section{Derivation of fermion propagators}
\label{app:propagator:derivation}

General formulae for the scalar function associated with bulk propagators of arbitrary-spin fields in RS can be found in \cite{Randall:2001gb}. The special case of bulk fermion propagators with endpoints on the UV brane is presented in \cite{Contino:2004vy}. The Green's function equation for the general RS fermion propagator can be solved directly from the Strum-Liouville equation, though this can obscure some of the intuition of the results.
Here we provide a pedagogical derivation of the 5D bulk fermion propagator in a flat and warped interval extra dimension. See also the discussion in Appendix~\ref{sec:5D:momentum:space} which relates this construction to the usual pure momentum space formalism.

\subsection{Flat 5D fermion propagator}
\label{app:flatGreensFunc}

First we derive the chiral fermion propagator in a flat interval extra dimension $z\in (0,L)$ as a model calculation for the warped fermion propagator which is presented in Appendix \ref{app:warpedGreensFunc}. A complete set of propagators for a flat 5D interval was derived in \cite{Puchwein:2003jq} using finite temperature field theory techniques.

We derive these results by directly solving the Green's function equations.
The propagator from a given point $x'$ to a another point $x$ is given by the two-point Green's function of the 5D Dirac operator,
\begin{align}
    \mathcal D \, \Delta(x,x') &\equiv \left(i\gamma^M \partial_M-m\right)\Delta(x,x') = i\delta^{(5)}(x-x'),\label{eq:GreensFunction:flat}
\end{align}
where $M$ runs over 5D indices.
We shall treat the noncompact dimensions in momentum space and the finite dimension is in position space. In this formalism, the Green's function equation is
\begin{align}
    \left(\slashed{p} + i\partial_5\gamma^5 - m\right)\Delta(p,z,z') &= i\delta(z-z'),
\end{align}
where we use $\gamma^5 = \text{diag}(i\mathbbm{1}_2,-i\mathbbm{1}_2)$.

This is a first-order differential equation with nontrivial Dirac structure.
To solve this equation we define a pseudo-conjugate Dirac operator (which is neither a complex nor Hermitian conjugate),
\begin{align}
    \bar{\mathcal D} = i\gamma^M\partial_M + m.\label{eq:flat:pseudoconjugate}
\end{align}
Using this to ``square'' the Dirac operator, we can swap the Dirac equation for a simpler Klein-Gordon equation that is second order and diagonal on the space of Weyl spinors,
\begin{align}
    \mathcal D\bar{\mathcal D}
    &
    = \begin{pmatrix}
            \partial_5^2 - \partial^2 - m^2 & \\
             & \partial_5^2 - \partial^2 - m^2
        \end{pmatrix}.\label{eq:flat:squared}
\end{align}
It is straightforward to solve for the Green's functions $F(p,z,z')$ of the $\mathcal{DD}^*$ operator in mixed position/momentum space,
\begin{align}
    \mathcal{D}\bar{\mathcal D} F(p,z,z') =
        \begin{pmatrix}
            \partial_5^2 + p^2 - m^2 &  \\
             & \partial_5^2 + p^2 - m^2
        \end{pmatrix}
        \begin{pmatrix}
            F_- & \\
             & F_+
        \end{pmatrix}
        &= i\delta(z-z').\label{eq:flat:F:Greens}
\end{align}
From these we can trivially construct a solution for the Green's function of  (\ref{eq:GreensFunction:flat}),
\begin{align}
    \Delta(p,z,z') &\equiv \bar{\mathcal D} F(p,z,z') =
    \begin{pmatrix}
        \left(-\partial_5 + m\right) F_- & \sigma^\mu p_\mu F_+\\
        \bar\sigma^\mu p_\mu F_- & \left(\partial_5 +m\right)F_+
    \end{pmatrix}.\label{eq:flat:G:fromF}
\end{align}
We solve this by separating $F_\pm(z)$ into pieces
\begin{align}
    F_\pm(p,z,z') =
    \begin{cases}
        F^<_\pm(p,z,z') & \text{if } z<z'\\
        F^>_\pm(p,z,z') & \text{if } z>z'
    \end{cases}\label{eq:F:less:greater}
\end{align}
and then solving the homogeneous Klein-Gordon equations for each $F^<$ and $F^>$. The general solution is
\begin{align}
    F_\pm^{<,>}(p,z,z') &= A_\pm^{<,>}\cos(\chi_p z) + B_\pm^{<,>}\sin(\chi_p z),\label{eq:flat:F:general}
\end{align}
where the eight coefficients $A_\pm^{<,>}$ and $B_\pm^{<,>}$ are determined by the boundary conditions at $0, L$ and $z'$. The factor $\chi_p$ is the magnitude of $p_5$ and is defined by
\begin{align}
\chi_p = \sqrt{p^2 - m^2}.\label{eq:chi_p}
\end{align}

We impose matching boundary conditions at $z=z'$. By integrating the Green's function equation (\ref{eq:flat:F:Greens}) over a sliver $z\in[z'-\epsilon,z'+\epsilon]$ we obtain the conditions
\begin{align}
    \partial_5 F_\pm^>(z') - \partial_5 F_\pm^<(z') &= i,\label{eq:flat:BC:matching:1}\\
    F_\pm^>(z') - F_\pm^<(z') &= 0.\label{eq:flat:BC:matching:2}
\end{align}
These are a total of four equations. The remaining four equations imposed at the branes impose the chirality of the fermion zero mode and are equivalent to treating the interval as an orbifold. We denote the propagator for the 5D fermion with a left-chiral (right-chiral) zero mode by $\Delta^L$ ($\Delta^R$). We impose that the Green's function vanishes if a ``wrong-chirality'' state propagates to either brane,
\begin{align}
    P_R \left.\Delta^L(p,z,z')\right|_{z=0,L} = P_R \bar{\mathcal D} \left.F^L(p,z,z')\right|_{z=0,L} &= 0, \label{eq:flat:BC:orbifold:1}\\
    P_L \left.\Delta^R(p,z,z')\right|_{z=0,L} = P_L \bar{\mathcal D} \left.F^R(p,z,z')\right|_{z=0,L} &= 0, \label{eq:flat:BC:orbifold:2}
\end{align}
where $P_{L,R} = \frac{1}{2}(1\mp i\gamma^5)$ are the usual 4D chiral projection operators. Note from (\ref{eq:flat:G:fromF}) that each of these equations is actually a set of two boundary conditions on each brane. For example, the left-handed boundary conditions may be written explicitly as
\begin{align}
    \left.F_-^L(p,z,z')\right|_{z=0,L} &= 0,\\
    \left.(\partial_5+m) F_+^L(p,z,z')\right|_{z=0,L}&=0,
\end{align}
where we have used that $p_\mu$ is arbitrary. It is well-known that only one boundary condition for a Dirac fermion needs to be imposed in order not to overconstrain the first-order Dirac equation since the bulk equations of motion convert boundary conditions for $\chi$ into boundary conditions for $\psi$~\cite{Csaki:2003sh}. In this case, however, we work with a \textit{second}-order Klein-Gordon equation that does not mix $\chi$ and $\psi$. Thus the appearance and necessity of two boundary conditions per brane for a chiral fermion is not surprising; we are only converting the single boundary condition on $\Delta(p,z,z')$ into two boundary conditions for $F(p,z,z')$.

\begin{table}[t]
\begin{centering}
   \begin{tabular}{lclcclcl}
    \hline
        $A^{L<}_+=\phantom{+}\si_p(L-z')\si_pL$ &\quad& $A^{L>}_+=\phantom{+}\si_pz'\co_pL$
        &\quad&\quad&
        $A^{R<}_+=\phantom{+}0$ &\quad& $A^{R>}_+=-\co_pz'\si_pL$\\
        $B^{L<}_+=\phantom{+}0$ &\quad& $B^{L>}_+=\phantom{+}\si_pz'\si_pL$
        &\quad&\quad&
        $B^{R<}_+=-\co_p(L-z')$ &\quad& $B^{R>}_+=-\co_pz'\co_pL$\\
        $A^{L<}_-=\phantom{+}0$ &\quad& $A^{L>}_-=-\co_pz'\si_pL$
        &\quad&\quad&
        $A^{R<}_-=\phantom{+}\si_p(L-z')$ &\quad& $A^{R>}_-=-\si_pz'\co_pL$ \\
        $B^{L<}_-=\phantom{+}\co_p(L-z')$ &\quad& $B^{L>}_-=-\co_pz'\co_pL$
        &\quad&\quad&
        $B^{R<}_-=\phantom{+}0$ &\quad& $B^{R>}_-=\phantom{+}\si_pz'\si_pL$ \\
    \hline
  \end{tabular}
  \caption{Flat case coefficients in (\ref{eq:flat:F:general}) upon solving with the boundary conditions (\ref{eq:flat:BC:matching:1}--\ref{eq:flat:BC:orbifold:2}). We have used the notation $\co_px = \cos\chi_px$ and $\si_px = \sin\chi_px$.}
  \label{table:flat:coefficients}
\end{centering}
\end{table}
Solving for the coefficients $A_\pm^{<,>}(p,z)$ and $B_\pm^{<,>}(p,z)$ for each type of fermion (left- or right-chiral zero modes) one finds the results in Table~\ref{table:flat:coefficients}.
Using trigonometric identities one may combine the $z<z'$ and $z>z'$ results to obtain\footnote{This result differs from that of \cite{Puchwein:2003jq} by a factor of 2 since that paper treats the compactified space as an orbifold over the entire $S^1$ rather than just an interval $[0, \pi R]$.}
\begin{align}
    F^X_\pm &= \frac{-i\cos \chi_p\left(L-|z-z'|\right) +  \gamma^5 \wp_X \cos \chi_p\left(L - (z+z')\right)}{2\chi_p \sin\chi_p L},
\end{align}
where $X =\{L,R\}$ with $\wp_L=+1$ and $\wp_R=-1$. The fermion Green's function can then be obtained trivially from (\ref{eq:flat:G:fromF}).

Let us remark that the leading UV behavior of a brane-to-brane propagator (where the $k_5\gamma^5$ term vanishes) goes like
\begin{align}
    \Delta \sim \frac{\slashed{k}}{\chi_k}.
\end{align}

\subsection{Warped 5D fermion propagator}
\label{app:warpedGreensFunc}

We now derive the chiral fermion propagator in a warped interval extra dimension following the same strategy as Appendix \ref{app:flatGreensFunc}. 
%
The Dirac operator is obtained from the variation of the Randall-Sundrum free fermion action,
\begin{align}
    S_{\text{RS}}(\text{fermion}) &= \int dx\int^{R'}_R dz\; \left(\frac{R}{z}\right)^4 \bar\Psi \left(i\gamma^M\partial_M - i\frac{2}{z}\gamma^5 - \frac{c}{z}\right)\Psi,
\end{align}
where $c=mR$ and we have integrated the left-acting derivatives by parts. The Dirac operator is a product of the $(R/z)^4$ prefactor coming from the AdS geometry and an operator $\mathcal D$ given by
\begin{align}
\mathcal D &= i\gamma^M\partial_M - i\frac{2}{z}\gamma^5 - \frac{c}{z}.
\end{align}
We would like to find the mixed position/momentum space two-point Green's function satisfying
\begin{align}
    (R/z)^4\,\mathcal{D}\, \Delta(p,z,z') &= i\delta(z-z').\label{eq:warped:greens}
\end{align}
Following (\ref{eq:flat:pseudoconjugate}) we define a pseudo-conjugate Dirac operator
\begin{align}
    \bar{\mathcal D} = i\gamma^M\partial_M - i\frac{2}{z}\gamma^5 +\frac{c}{z}\label{eq:warped:Dstar}
\end{align}
and `square' $\mathcal D$ into a diagonal second-order operator,
\begin{align}
    \mathcal{D}\bar{\mathcal D} =
    \begin{pmatrix}
        \mathcal{D}\bar{\mathcal D}_{\phantom{*}-} & 0\\
        0 & \mathcal{D}\bar{\mathcal D}_{\phantom{*}+}
    \end{pmatrix}
    \quad\quad\quad\quad\quad\quad
    \mathcal{D}\bar{\mathcal D}_{\phantom{*}\pm}  = \partial^2 - \partial_5^2 + \frac 4z \partial_5 + \frac{c^2\pm c -6}{z^2}.
\end{align}
Next we follow (\ref{eq:flat:F:Greens}) and solve for the Green's function of this squared operator in mixed position/momentum space where $\partial^2 \to -p^2$,
\begin{align}
    -(R/z)^4\,
    \mathcal{D}\bar{\mathcal D}F(p,z,z')
    =
    -\left(\frac{R}{z}\right)^4
    \begin{pmatrix}
        \mathcal{D}\bar{\mathcal D}_{\phantom{*}-} & \\
         &  \mathcal{D}\bar{\mathcal D}_{\phantom{*}+}
    \end{pmatrix}
    \begin{pmatrix}
        F_- & \\
         & F_+
    \end{pmatrix}
    =
    i\delta(z-z'). \label{eq:warped:KG:greens}
\end{align}
The solution to the Dirac Green's function equation (\ref{eq:warped:greens}) is then given by $\Delta(p,z,z')=\bar{\mathcal D} F(p,z,z')$.
We shall separate $F(p,z,z')$ into solutions for the cases $z>z'$ and $z<z'$ following (\ref{eq:F:less:greater}). The general solution to the homogeneous equation (\ref{eq:warped:KG:greens}) with $z\neq z'$ is
\begin{align}
    F_\pm^{<,>}(p,z,z') &= A_\pm^{<,>}\, z^{\frac 52} J_{c\pm\frac 12}(pz) + B_\pm^{<,>}\, z^{\frac 52} Y_{c\pm\frac 12}(pz),\label{eq:Fpm}
\end{align}
where $J_n$ and $Y_n$ are Bessel functions of the first and second kinds, $A_\pm^{<,>}$ and $B_\pm^{<,>}$ are coefficients to be determined by boundary conditions,
and $p$ is the analog of $\chi_p$ defined by $p=\sqrt{p_\mu p^\mu}$.
Note that this differs from (\ref{eq:chi_p}) since there is no explicit bulk mass dependence. In (\ref{eq:Fpm}) the bulk masses enter only in the order of the Bessel functions as $(c\pm\frac 12)$.

The matching boundary conditions at $z=z'$ are given by  (\ref{eq:flat:BC:matching:1}) and (\ref{eq:flat:BC:matching:2}) modified by a factor of $(R/z')^4$ from (\ref{eq:warped:KG:greens}),
\begin{align}
    \partial_5 F_\pm^>(z') - \partial_5 F_\pm^<(z') &= 
    i(R/z')^{-4},\label{eq:warped:BC:matching:1}\\
    F_\pm^>(z') - F_\pm^<(z') &= 0.\label{eq:warped:BC:matching:2}
\end{align}
The chiral boundary conditions are the same as in the flat case, (\ref{eq:flat:BC:orbifold:1}) and (\ref{eq:flat:BC:orbifold:2}) with the appropriate insertion of (\ref{eq:warped:Dstar}).

We may now solve for the $A$ and $B$ coefficients. It is useful to write these in terms of common factors that appear in their expressions. To this end, let us define the prefactors
\begin{align}
    \alpha_L &= \frac{i\pi}{2R^4}\frac{1}{S^-_c(pR,pR')} & \alpha_R &= \frac{i\pi}{2R^4}\frac{1}{S^+_c(pR,pR')}\label{eq:warped:pre:coefficients}
\end{align}
and a set of antisymmetric functions
\begin{align}
    S^\pm_c(x,y) &= J_{c\pm \frac{1}{2}}(x)Y_{c\pm\frac{1}{2}}(y)- J_{c\pm \frac{1}{2}}(y)Y_{c\pm \frac{1}{2}}(x)\label{eq:auxiliary:pm}\\
    \tilde S^{\pm}_c(x,y)&= J_{c\pm\frac{1}{2}}(x)Y_{c\mp\frac{1}{2}}(y)- J_{c\mp \frac{1}{2}}(y)Y_{c\pm \frac{1}{2}}(x)\label{eq:auxiliary:tilde:pm}
\end{align}
With these definitions the coefficients for the left- and right-handed $F$ functions  are given in Table~\ref{table:warped:coefficients}.
\begin{table}[t]
\begin{centering}
   \begin{tabular}{lcl}
    \hline
        $A^{L<}_+ =-\alpha_L z'^{\frac 52} Y_{c-\frac 12} \left(pR\right) {\tilde S^+_c(pz',pR')}$
        &\quad\quad\quad\quad\quad&
        $A^{R<}_+ =-\alpha_R z'^{\frac 52} Y_{c+\frac 12} \left(pR\right) {S^+_c(pz',pR')} $ \\
        $B^{L<}_+ =\phantom{+}\alpha_L z'^{\frac 52} J_{c-\frac 12} \left(pR\right) {\tilde S^+_c(pz',pR')} $
        &\quad\quad\quad\quad\quad&
        $   B^{R<}_+ =\phantom{+}\alpha_R z'^{\frac 52} J_{c+\frac 12} \left(pR\right) {S^+_c(pz',pR')}$ \\
        $A^{L<}_- =- \alpha_L z'^{\frac 52} Y_{c-\frac 12} \left(pR\right) {S^-_c(pz',pR')}$
        &\quad\quad\quad\quad\quad&
        $A^{R<}_- =- \alpha_R z'^{\frac 52} Y_{c+\frac 12} \left(pR\right) {\tilde S^-_c(pz',pR')} $ \\
        $B^{L<}_- =\phantom{+}\alpha_L z'^{\frac 52} J_{c-\frac 12} \left(pR\right) {S^-_c(pz',pR')}$
        &\quad\quad\quad\quad\quad&
        $B^{R<}_- =\phantom{+}\alpha_R z'^{\frac 52} J_{c+\frac 12} \left(pR\right) {\tilde S^-_c(pz',pR')}$
\\
    \hline
  \end{tabular}
  \caption{Left-handed RS fermion propagator coefficients: the $z>z'$ coefficients are obtained by swapping $R\leftrightarrow R'$ in the arguments of the functions,  leaving the $\alpha_{L,R}$ constant.}
  \label{table:warped:coefficients}
\end{centering}
\end{table}
The $F_\pm^{L,R}$ functions may thus be written out succinctly for $z\leq z'$ as
\begin{align}
    F^{L<}_+ &= \alpha_L \left(zz'\right)^{5/2}
    \tilde S^+_c\left(pz',p R'\right) \tilde S^-_c(p R,p z)\label{eq:FLlp}
    \\
    F^{L<}_- &= \alpha_L \left(zz'\right)^{5/2}
    S^-_c\left(pz',p R'\right) S^-_c(p R,p z)\label{eq:FLlm}
    \\
    F^{R<}_+ &= \alpha_R \left(zz'\right)^{5/2}
    S^+_c\left(pz',p R'\right) S^+_c(p R,p z)\label{eq:FRlp}
    \\
    F^{R<}_- &= \alpha_R \left(zz'\right)^{5/2}
    \tilde S^-_c\left(pz',p R'\right) \tilde S^+_c(p R,p z)\label{eq:FRlm}
\end{align}
The expressions for $z>z'$ are obtained by making the replacement $\{ R\leftrightarrow R'\}$ in the arguments of the $S_c$ functions. We now use the notation in (\ref{eq:F:less:greater}) and drop the $<,>$ superscripts. From these the fermion Green's function can be obtained trivially from the analog of (\ref{eq:flat:G:fromF}),
\begin{align}
    \Delta(p,z,z') &\equiv \bar{\mathcal D}F(p,z,z') =
    \begin{pmatrix}
    D_-F_- & \sigma^\mu p_\mu F_+ \\ 
    \bar\sigma^\mu p_\mu F_- & D_+F_+
    \end{pmatrix}, \quad\quad\quad D_\pm \equiv \pm\left(\partial_5-\frac 2z\right)+\frac cz.\label{eq:warped:G:fromF}
\end{align}
Note that in the UV limit ($\chi_p\gg 1/R$) the Bessel functions reduce to phase-shifted trigonometric functions so that we indeed recover the flat 5D propagators.

\subsection{Euclidean warped 5D fermion propagator}
\label{app:warpedGreensFunc:Euc}

Finally, it is convenient to write the Wick-rotated form of the fermion propagators since these will provide the relevant Feynman rules in loop diagrams such as $\mu\to e \gamma$. We shall write out the scalar $F$ functions in a convenient form that we use throughout the rest of this document. The derivation is identical to that outlined above with the replacement $p^2 = -p_E^2$ (i.e.\ $\partial = i\partial_E$) in the Green's function equation so that we shall simply state the results. The Euclidean scalar functions are written in terms of the modified Bessel functions $I$ and $K$ which behave like exponentials in the UV. Let us define the auxiliary functions
\begin{align}
	S_c(x_\pm,x'_\pm) &= I_{c\pm 1/2}(x) K_{c\pm 1/2}(x') - I_{c\pm 1/2}(x') K_{c\pm 1/2}(x)\\
	S_c(x_\pm,x'_\mp) &= I_{c\pm 1/2}(x) K_{c\mp 1/2}(x') - I_{c\mp 1/2}(x') K_{c\pm 1/2}(x)\\
	T_c(x_\pm,x'_\mp) &= I_{c\pm 1/2}(x) K_{c\mp 1/2}(x') + I_{c\mp 1/2}(x') K_{c\pm 1/2}(x).
\end{align}
Since we would like to write dimensionless loop integrals, let us define the dimensionless variables $y\equiv k_E R'$ and $x=k_Ez$, which are the natural quantities which appear as arguments of the Bessel functions. We write the warp factor as $w=(R/R')$. It is convenient to pull out overall factors to write the $F$ functions as
\begin{align}
	F_\pm(k_E,z,z')=iw^{-4}R'\tilde F_{\pm,y}^{xx'}.
\end{align}
The Euclidean scalar functions for $x>x'$ (i.e.\ $z>z'$) are given by
\begin{align}
\tilde F^L_{-} &=\phantom{+} \frac{(xx')^{5/2}}{y^5} \frac{S_{c_L}(x_-,y_-) S_{c_L}(x'_-,wy_-)}{S_{c_L}(y_-,wy_-)}
&
\tilde F^L_{+} &=- \frac{(xx')^{5/2}}{y^5} \frac{T_{c_L}(x_+,y_-) T_{c_L}(x'_+,wy_-)}{S_{c_L}(y_-,wy_-)}
\label{eq:tilde:F:1}
\\
\tilde F^R_{-} &=- \frac{(xx')^{5/2}}{y^5} \frac{T_{c_R}(x_-,y_+) T_{c_R}(x'_-,wy_+)}{S_{c_R}(y_+,wy_+)}
&
\tilde F^R_{+} &=\phantom{+} \frac{(xx')^{5/2}}{y^5} \frac{S_{c_R}(x_+,y_+) S_{c_R}(x'_+,wy_+)}{S_{c_R}(y_+,wy_+)}.
\label{eq:tilde:F:2}
\end{align}
The functions for $x<x'$ are given by replacing $x\leftrightarrow x'$ in the above formulas. With these definitions the Euclidean fermion propagator given by the analog of (\ref{eq:warped:G:fromF}),
\begin{align}
    \Delta(k_E,x,x') &\equiv i\frac{R'}{w^4}\bar{\mathcal D}\tilde F_y^{xx'} =
    \begin{pmatrix}
    y\tilde D_+\tilde F_- & \sigma^\mu y_\mu \tilde F_+ \\ 
    \bar\sigma^\mu y_\mu \tilde F_- & y\tilde D_- \tilde F_+
    \end{pmatrix}, \quad\quad\quad \tilde D_\pm \equiv \pm\left(\partial_x-\frac 2x\right)+\frac cx.\label{eq:warped:G:fromF:Euclidean}
\end{align}

\section{Finiteness of the brane-localized neutral Higgs diagram}
\label{app:Finiteness}

As explained in Section~\ref{sec:heuristic:brane:5D}, the finiteness of the one-loop result and logarithmic divergence at two-loop order becomes opaque to na\"ive 5D power counting arguments when the Higgs is brane-localized. Additional cancellations of leading-order terms in loop momentum are required to sensibly interpolate between the superficial degree of divergence of the bulk and brane-localized scenarios. For the charged Higgs this cancellation mechanism came from an $M_W^2$ insertion, which led to an additional $1/k^2$ factor relative to the bulk field. Here we shall elucidate the finiteness of the single-mass-insertion brane-localized neutral scalar loop. 

At one-loop order this finiteness can be seen explicitly by the cancellation between the neutral Higgs and the neutral Goldstone. However, there is an additional chiral cancellation that occurs between the two diagrams associated a single intermediate neutral boson. Indeed, because the Higgs and neutral Goldstone do not appear to completely cancel at two-loop order, this additional cancellation is necessary for the power-counting arguments given in Section~\ref{sec:heuristic:twoloop}. 

We highlight this cancellation in two ways. The pure momentum space calculation highlights the role of the chiral boundary conditions, while the mixed position/momentum space calculation shows an explicit cancellation while including the full scalar structure the amplitude.

\subsection{Momentum space}

Here we shall see that 4D Lorentz invariance combined with the chiral boundary conditions forces the UV divergence of the two diagrams in Fig.~\ref{fig:1MIH0} to cancel.

We first note that the propagators to the photon vertex each have an endpoint in the bulk. This implies that the leading-order contributions to these propagators in the UV limit are proportional to the uncompactified flat-space 5D propagators,
\begin{align}
\Delta =
\begin{pmatrix}
    \Delta_{\psi\chi} & \Delta_{\psi\psi}\\
    \Delta_{\chi\chi} & \Delta_{\chi\psi}
\end{pmatrix}
\sim
\frac{1}{k^2-k_5^2}
\begin{pmatrix}
    ik_5 & k_\mu\sigma^\mu\\
    k_\mu\bar\sigma^\mu & -ik_5
\end{pmatrix}
=\frac{k_\mu\gamma^\mu + k_5\gamma^5}{k^2-k_5^2},
\label{eq:flat:5D:2x2:propagator}
\end{align}
where we have written $\Delta_{\psi\chi}$ to mean the propagation of a left-handed Weyl spinor $\chi$ into a right-handed spinor $\psi$. The terms along the diagonal come from $k_5\gamma^5$ and represent the chirality-flipping part of the propagator. The  boundary conditions require the wrong-chirality modes, the SU(2) doublet $\psi_L$ and SU(2) singlet $\chi_R$, to vanish on the IR brane.
Thus, the fermion may propagate to the wrong-chirality spinor in the bulk only if it propagates back to the correct-chirality spinor when it returns to the brane.
For an internal left-handed Weyl fermion $\chi_L$, the portion of the amplitude coming from the photon emission takes the form
\begin{align}
    \Delta_{\chi\chi}\sigma^\mu\Delta_{\chi\chi} + \Delta_{\chi\psi}\bar\sigma^\mu\Delta_{\psi\chi} \sim \left(k_\alpha \bar\sigma^\alpha\right)\sigma^\mu \left(k_\beta \bar\sigma^\beta\right) + (k_5)^2 \bar\sigma^\mu.
\end{align}
 Combining with the analogous expression for a right-handed Weyl fermion in the loop, the relevant part of the photon emission amplitude can be written as
\begin{align}
    \frac{\slashed{k}\gamma^\mu\slashed{k} + (k_5)^2\gamma^\mu}{(k^2-k_5^2)^2},
\end{align}
where these terms correspond to a fermion of the correct and incorrect chirality propagating into the brane. The second term can be simplified using
\begin{align}
    \int dk_5\,\frac{(k_5)^2}{(k^2-k_5^2)^2} =
    \int dk_5\,\frac{-k^2}{(k^2-k_5^2)^2},\label{eq:heuristic:k5int}
\end{align}
which can be confirmed by Wick rotating both sides, $k^2 \to -k_E^2$, and performing the $dk_5$ integral explicitly.
Now it is easy to see that the divergent contributions from the diagrams in Fig.~\ref{fig:1MIH0} cancel. The boundary conditions force brane-to-brane propagators to go like $\slashed{k}$ with no $\gamma^5$ part. Thus we may write the internal fermion structure of the amplitudes as
\begin{align}
    \mathcal M_{(a)} + \mathcal M_{(b)} \sim \slashed{k}\left(\slashed{k}\gamma^\mu\slashed{k} - k^2\gamma^\mu\right) + \left(\slashed{k}\gamma^\mu\slashed{k} - kß^2\gamma^\mu\right)\slashed{k} = 0.\label{eq:chiral:cancellation}
\end{align}
The key minus sign between the two terms in the photon emission comes from the chiral boundary conditions that force the second term to pick up the relative sign between the two diagonal blocks of $\gamma^5$. 

Let us remark that it is crucial that the denominator in (\ref{eq:heuristic:k5int}) contains exactly two propagators or else the equality would not hold.
One might be concerned that the brane-to-brane propagator should also contribute an additional factor of $(k^2-k_5^2)$ to the denominator (the $k_5\gamma^5$ term vanishes in the numerator from boundary conditions).
Such a factor is indeed present in the full calculation, but because 5D Lorentz invariance is broken on the brane, $k_5$ is not conserved there and this factor actually includes a \textit{different}, uncorrelated fifth momentum component, $\tilde k_5$, which can be taken the be independent of the $dk_5$ integral. This is a manifestation of the principles in Appendix~\ref{app:5D:EFT}. As a check, one can perform the $d\tilde k_5$ integral for this brane-to-brane propagator and obtain the same $\slashed{k}/|k|$ UV behavior found in the careful derivation performed in Appendix \ref{app:flatGreensFunc}.

\subsection{Position/momentum space}

In Appendix \ref{app:flatGreensFunc} we derived the flat-space bulk fermion propagator,
\begin{align}
    \Delta(p,x_5,x_5') &= \left(\slashed{p}-i\gamma^5\partial_5+m\right)\frac{-i\cos \chi_p\left(L-|x_5-x_5'|\right) +  \gamma^5 \wp(X) \cos \chi_p\left(L - (x_5+x_5')\right)}{2\chi_p \sin\chi_p L},
\end{align}
where the zero mode chirality is given by $X =\{L,R\}$ with $\wp(L)=+1$ and $\wp(R)=-1$. We then argued at the end of Appendix \ref{app:warpedGreensFunc} that the propagators in a warped extra dimension reduce to this case up to overall phases. Thus we expect the amplitudes to have the same UV behavior up to finite factors. The relevant flat-space one-loop diagrams contributing to the operator (\ref{eq:5Doperator}) are shown in Fig.~\ref{fig:1MIH0}. We start with Fig.~\ref{fig:1MIH0:a} and assume that the decay is from $\mu_L$ to $e_R$. The loop propagators with $(x_5,x_5')=(L,z)$, $(z,L)$ and $(L,L)$ can be written as
\begin{eqnarray}
 \Delta(k',L,z)&=& -i\frac{\slashed{k'}\cos\chi_{k'}z-i\gamma^5\chi_{k'}\sin\chi_{k'}z}{\chi_{k'}\sin\chi_{k'}L}P_R
\\
 \Delta(k,z,L)&=& -i\frac{\slashed{k}\cos\chi_{k}z+i\gamma^5\chi_{k}\sin\chi_{k}z}{\chi_{k}\sin\chi_{k}L}
P_R
\\
 \Delta(k,L,L)&=& -i\frac{\slashed{k}\cos\chi_{k}L}{\chi_{k}\sin\chi_{k}L}
P_R,
\end{eqnarray}
where $k'=k+q$.
We have used the chiral boundary conditions to simplify $\Delta(k,L,L)$. Since we are interested in the UV behavior we have dropped the terms proportional to the bulk mass $m$ from the internal propagators because these are finite. Combining the propagators together and doing the same calculation for Fig.~\ref{fig:1MIH0:b}, the amplitudes become
\begin{align}
    {\mathcal M}^\mu_{(a)} =&  \int \frac{d^4k}{(2\pi)^4} \, dz\; \;\bar u(p')\left\{ \frac{\slashed{k'}\,\gamma^{\mu}\,\slashed k \,f(k,z)+\chi_{k}\chi_{k'}\,\gamma^{\mu}\,g(k,z)}{\chi_k\chi_{k'}\,[(p+k)^2-m_H^2]}\,\right\}\frac{\slashed k\,\cot\chi_kL}{\chi_k} u(p)\label{eq:flat:divergent:a}\\
    {\mathcal M}^\mu_{(b)} =&  \int \frac{d^4k}{(2\pi)^4} \, dz\; \;\bar u(p')\frac{\slashed{k'}\,\cot\chi_{k'}L}{\chi_{k'}}\left\{ \frac{\slashed{k'}\,\gamma^{\mu}\,\slashed k \,f(k,z)+\chi_{k}\chi_{k'}\,\gamma^{\mu}\,g(k,z)}{\chi_k\chi_{k'}\,[(p+k)^2-m_H^2]}\,\right\}u(p)\label{eq:flat:divergent:b}
\end{align}
where we have written
\begin{align}
    f(k,z) &= -\frac{\cos(\chi_{k+q}z)\cos(\chi_kz)}{\sin\chi_{k+q}L\sin\chi_kL}\\
    g(k,z) &= -\frac{\sin(\chi_{k+q}z)\sin(\chi_kz)}{\sin\chi_{k+q}L\sin\chi_kL}.
\end{align}
Note that all of the $z$ dependence is manifestly contained in sines and cosines. Further we have neglected the flavor-dependence of the $\chi_k$ factors since these also come from the bulk masses via (\ref{eq:chi_p}) and are negligible in the UV.

Upon Wick rotation the trigonometric functions become hyperbolic functions which are exponentials in the Euclidean momentum,
\begin{align}
    \cos\chi_kz \to \cosh(\chi_{k_E}z) &= \frac12\left(e^{\chi_{k_E}z}+e^{-\chi_{k_E}z}\right)\\
    \sin\chi_kz \to i\sinh(\chi_{k_E}z) &= \frac i2\left(e^{\chi_{k_E}z}-e^{-\chi_{k_E}z}\right).
\end{align}
We may now replace the trigonometric functions with the appropriate Euclidean exponentials. Since we are concerned with the UV behavior, we may drop terms which are exponentially suppressed for large $k$ over the entire range of $z$. The remaining terms are simple exponentials and can be integrated over the interval. One finds that the trigonometric terms in (\ref{eq:flat:divergent:a}) and (\ref{eq:flat:divergent:b}) yield the expression
\begin{align}
    \frac{i}{\chi_{k_E+q}+\chi_{k_E}} \to \frac{-1}{\chi_{k+q}+\chi_k},
\end{align}
where on the right we have reversed our Wick rotation to obtain a Minkowski space expression for the terms which are not exponentially suppressed in Euclidean momentum. After doing this, the leading order term in $\cot\chi L$ in (\ref{eq:flat:divergent:a}) and (\ref{eq:flat:divergent:b}) equals $i^{-1}$ and the terms in the braces become
\begin{equation}
\left\{ \frac{(\slashed k+\slashed q)\,\gamma^{\mu}\,\slashed k \,-\chi_{k+q}\chi_k\,\gamma^{\mu}\,}{\chi_k\chi_{k+q}\,(\chi_k+\chi_{k+q})\,[(p+k)^2-m_H^2]}\,\right\},
\end{equation}
which gives the numerator of (\ref{eq:chiral:cancellation}).

In terms of these quantities the potentially divergent amplitudes can be written as
\begin{align}
    {\mathcal M}^\mu_{(a)} =&  \int \frac{d^4k}{(2\pi)^4}\; \frac{1}{(\chi_{k+q}+\chi_{k})[(p+k)^2-m_H^2] } \bar u(p)\left\{\frac{(\slashed k+\slashed q)}{\chi_{k+q}}\gamma^\mu -\gamma^\mu\frac{\slashed{k}}{\chi_k}\right\}u(p+q)\\
    {\mathcal M}^\mu_{(b)} =&  \int \frac{d^4k}{(2\pi)^4}\; \frac{1}{(\chi_{k+q}+\chi_{k})[(p+k)^2-m_H^2]} \bar u(p)\left\{\gamma^\mu\frac{\slashed{k}}{\chi_k}-\frac{(\slashed k+\slashed q)}{\chi_{k+q}}\gamma^\mu\right\}u(p+q),
\end{align}
therefore these two terms cancel each other in the UV and the operator (\ref{eq:5Doperator}) is finite.

Higher mass insertions do not spoil this cancellation since these are associated with internal brane-to-brane propagators whose UV limit goes like $\Delta(k)\sim\slashed{k}/\chi_k$. The chiral structure of the effective operator (\ref{eq:5Doperator}) requires that only diagrams with an odd number of mass insertions contribute. Using the UV limit $\Delta(k)^2 \to 1$ one notes that the divergence structure reduces to the case above.


\bibliographystyle{utphys}
\bibliography{RSMuEGamBib}

\end{document}